\documentclass[aps, prl, twocolumn, a4paper, 10pt, showpacs,reprint,superscriptaddress,nofootinbib]{revtex4-2}

\usepackage[utf8]{inputenc}
\usepackage{amsmath}
\usepackage{amssymb}
\usepackage{amsfonts}
\usepackage{amstext}
\usepackage{amssymb}
\usepackage{amsthm}
\usepackage{bbold}
\usepackage{mathtools}
\usepackage{physics}
\usepackage{graphicx}
\usepackage{dcolumn}
\usepackage{bm}
\usepackage{braket}
\usepackage{color}
\usepackage[english]{babel}
\usepackage[svgnames]{xcolor}
\usepackage{setspace}
\usepackage{diagbox}
\usepackage{placeins}
\usepackage{algorithm}
\usepackage{algpseudocode}
\usepackage[inline]{enumitem}
\usepackage{enumerate}
\usepackage[normalem]{ulem}
\usepackage{tikz}
\usepackage[symbol]{footmisc}

\usepackage{cancel}
\DeclareUnicodeCharacter{2212}{\ensuremath{-}}

\definecolor{deeppurple}{rgb}{0.7, 0, 0.8}

\allowdisplaybreaks
\advance\parskip 0.4pt

\newcommand{\change}[1]{\textcolor{black}{#1}}

\newcommand{\pia}[1]{\textcolor{DarkGreen}{P.S.: #1}}

\newcommand{\greta}[1]{\textcolor{orange}{G.R..: #1}}
\usepackage{hyperref}
\hypersetup{
	colorlinks=true,
	linkcolor=blue,
	filecolor=blue,
	citecolor = red,      
	urlcolor=cyan,
}

\usepackage{orcidlink}

\begin{document}

\renewcommand{\figureautorefname}{Fig.}
\renewcommand{\sectionautorefname}{Appendix}
\renewcommand{\subsectionautorefname}{Appendix}

\renewcommand{\equationautorefname}{Eq.}
\renewcommand{\thesection}{\Roman{section}}
\renewcommand{\thesubsection}{\Alph{subsection}}

\title{
Tensor-Programmable Quantum Circuits for Solving Differential Equations}
\date{\today}
\author{Pia Siegl$^{\ast,\dag}$
\orcidlink{0000-0003-2249-8121}}

\affiliation{Institute for Quantum Physics, University of Hamburg, Luruper Chaussee 149, 22761 Hamburg, Germany}
\affiliation{Institute of Software Methods for Product Virtualization, German Aerospace Center (DLR), Nöthnitzer Straße 46b, 01187 Dresden, Germany}
\author{Greta Sophie Reese$^{\ast,\S}$
\orcidlink{0009-0001-9135-7499}}
\affiliation{Institute for Quantum Physics, University of Hamburg, Luruper Chaussee 149, 22761 Hamburg, Germany}
\affiliation{Center for Optical Quantum Technologies, University of Hamburg, 22761 Hamburg, Germany}
\affiliation{The Hamburg Centre for Ultrafast Imaging, Hamburg, Germany}
\author{Tomohiro Hashizume \orcidlink{0000-0002-7154-5417}}
\affiliation{Institute for Quantum Physics, University of Hamburg, Luruper Chaussee 149, 22761 Hamburg, Germany}
\affiliation{The Hamburg Centre for Ultrafast Imaging, Hamburg, Germany}
\author{Nis-Luca van Hülst \orcidlink{0009-0004-9893-3614}}
\affiliation{Institute for Quantum Physics, University of Hamburg, Luruper Chaussee 149, 22761 Hamburg, Germany}
\author{Dieter Jaksch \orcidlink{0000-0002-9704-3941}}
\affiliation{Institute for Quantum Physics, University of Hamburg, Luruper Chaussee 149, 22761 Hamburg, Germany}
\affiliation{The Hamburg Centre for Ultrafast Imaging, Hamburg, Germany}
\affiliation{Clarendon Laboratory, University of Oxford, Parks Road, Oxford OX1 3PU, UK}

\def\thefootnote{$\ast$}\footnotetext{These authors contributed equally to this work}
\def\thefootnote{$\dag$}\footnotetext{Contact author: pia.siegl@dlr.de}
\def\thefootnote{$\S$}\footnotetext{Contact author: greta.reese@uni-hamburg.de}

\begin{abstract}
We present a quantum solver for partial differential equations based on a flexible matrix product operator representation. Utilizing mid-circuit measurements and a state-dependent norm correction, this scheme overcomes the restriction of unitary operators.
Hence, it allows for the direct implementation of a broad class of differential equations governing the dynamics of classical and quantum systems. The capabilities of the framework are demonstrated \change{for linear and non-linear partial differential equations}  \change{using the  example of the linearized Euler equations with absorbing boundaries} \change{and the nonlinear Burgers' equation}. 
\change{For a turbulence data set, we demonstrate potential advantages of the quantum tensor scheme over its classical counterparts.}
\end{abstract}

\maketitle
\section{Introduction}
Solving partial differential equations (PDEs) is a core task in many research and industry areas, ranging from the financial sector \cite{Black1973,Lee2012} and material science \cite{Bian2016,Wu2025}
to computational fluid dynamics \cite{Courant1967, ferziger2002:CMFD, WU2022}.
Despite the enormous amount of resources nowadays available in classical computing, solving PDEs remains a challenge. 
One example is computational fluid dynamics, where resolving all relevant spatial scales quickly demands billions of data points \cite{Slotnick2014}, and approximations and the use of models become mandatory \cite{Menter1994, Wilcox1998, Pope2000, Sagaut2005,Germano1991}.

Quantum computers offer an efficient representation of classical data, as the number of qubits needed for amplitude encoding scales logarithmically with the number of data points \cite{Long2001,Gourianov2022,Jaksch2023, Givi2020}. 
To solve PDEs with quantum computers different approaches have been proposed: (i) algebraic quantum linear solvers as the Harrow-Hassidim-Lloyd (HHL) \cite{Harrow2009} algorithm and its extensions \cite{Ambainis2010-arxiv, Childs2017, Penuel2024,Lloyd2020}; (ii) 
specific PDEs were solved efficiently with discrete time-stepping schemes  \cite{Brearly2024, over2024_diffusion}; (iii) variational quantum algorithms (VQAs) \cite{Peruzzo2014, Kandala2017, Cerezo2022}, that rely on a hybrid scheme combining parameterized quantum circuits and a classical parameter optimization.

Despite the rapid advancement in quantum hardware and error correction \cite{Acharya2024, Reichardt2024-arxiv, Vandam2024-arxiv} and the promises for near term devices with significant numbers of logical qubits \cite{QueraRoadmap, IBMRoadmap}, quantum linear solvers are expected to stay expensive or even unfeasible due to the large demand in resources \cite{Penuel2024} {and their unfavorable scaling with the stiffness of the problem \cite{Harrow2009}.
VQAs instead are characterized by shallow circuit structures, are predicted to exhibit beneficial scaling 
\cite{Penuel2024,Lubasch2020,Gourianov2022} and were successfully applied in various areas \cite{Schilling2024, Lubasch2020,Huggins2020, Kandala2017}. 
While noisy hardware can limit the accuracy of VQAs \cite{Umer2025}, noisy optimization is possible \change{\cite{Pool2024, Yuguo2024,Rosenberg2022}}  and strategies like circuit recompilation \cite{Jaderberg2020,Jaderberg2022} can significantly reduce the sensitivity to noise \change{ \cite{Kim2023}}.

\begin{figure*}
    \centering
\includegraphics[width=\linewidth]{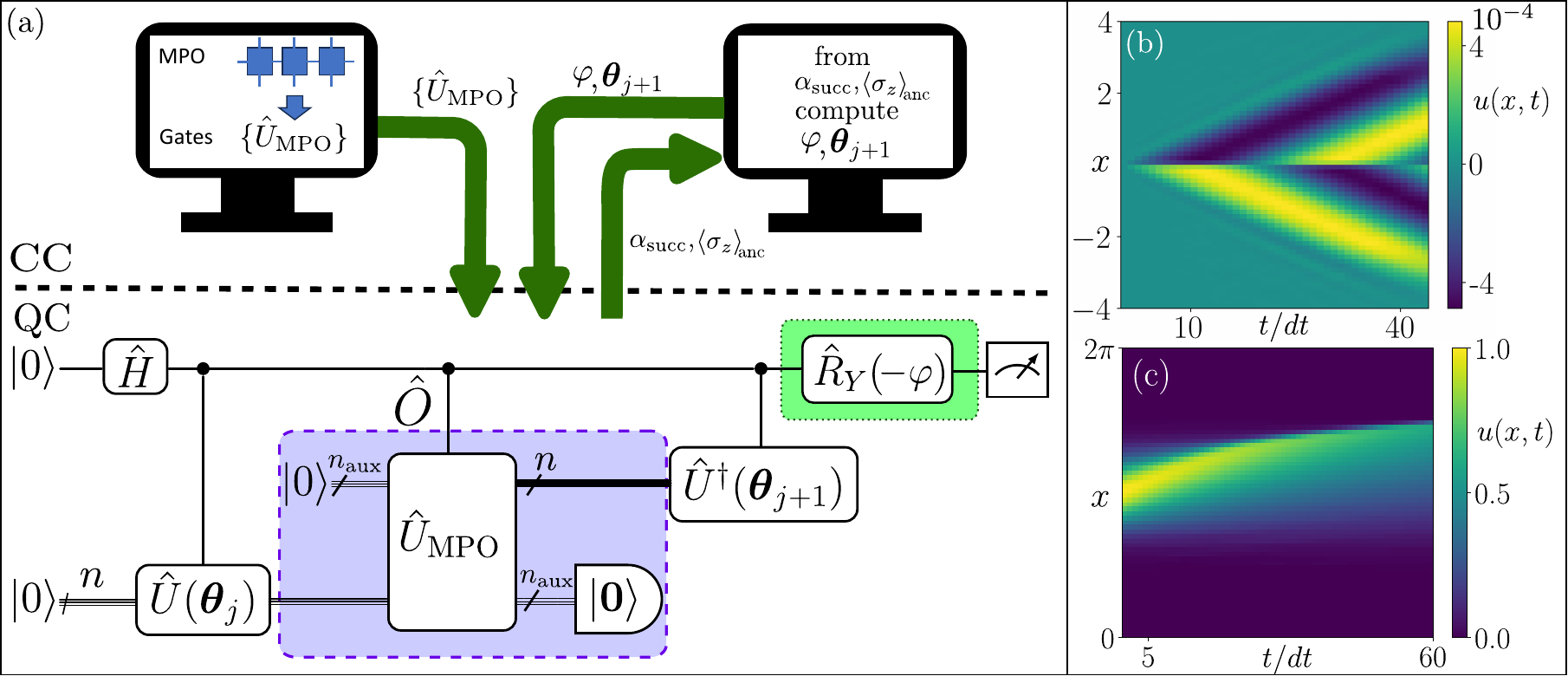}
\caption{(a) Hybrid quantum-classical routine to solve \change{linear} PDEs iteratively in a variational manner. 
The computation of the unitary gates $\hat{U}_{\text{MPO}}$ representing the operator $\hat{O}$ as well as the optimization routine take place on a classical computer (CC, upper part). The overlap $\bra{0}\hat{U}^{\dagger}(\boldsymbol{\theta}_{j+1})\hat{O}\hat{U}(\boldsymbol{\theta}_j)\ket{0}$ that determines the cost function, necessary to compute the solution at the next iteration step, is computed on the quantum computer (QC, lower part) using an adapted Hadamard test.
The angles $\boldsymbol{\theta}_j$ describe a previous iteration step, while $\boldsymbol{\theta}_{j+1}$ are to be determined during the classical optimization process.
This procedure allows to map classical differential solvers to the quantum computer. 
(b) Solution of the Euler equations 
with our quantum differential solver (QDS) for 45 time steps $dt$ encoded into 6 ansatz qubits. Depicted is the discretized 
velocity $u(x,t)$ evolution over time which arises is induced by a periodic pressure point source at $x=0$. 
\change{
(c) Solution of the non-linear Burgers' equations. Depicted is the discretized velocity field $u(x,t)$ over time, where initial Gauss peak evolves into a shock wave.}
A detailed description of all system and training parameters is given in
\autoref{subsec:Parameter}.
 }
\label{fig:circuit-schemes}
\end{figure*}
In addition to quantum computing, so called quantum inspired methods are under increasing attention as differential solvers.
A prominent example are Matrix Product States (MPS) \cite{Oseledets2011, Orus2014} which exhibit a low-rank representation of many functions \cite{Oseledets2013} and have proven successful in solving PDEs on classical hardware, showing potential to compete with conventional solvers and delivering remarkable results across various applications \cite{Gourianov2022,Ye2022,Kiffner2023, Kornev2023-arxiv,Ye2024, Peddinti2024, Hölscher2025, Gourianov2024}.

MPS methods show great promise when combined with VQAs.
While the efficiency of MPS is limited to solutions with bounded entanglement \cite{Schachenmeyer2013,Gourianov2022}, quantum circuits  \change{were shown to feature} an exponential reduction in the number of variables parameterizing the solution \cite{Lubasch2020} \change{for certain use cases}.
Further, MPS algorithms scale at least polynomially better when ported to quantum computers \cite{Lubasch2020,Gourianov2022}, offering at least the same speed-up as Grover's algorithm for unstructured search \cite{Grover1996} and defining an upper bound of the circuit depths for state encoding.
In combination with known methods to encode MPS with quantum circuits \cite{Ran2020, Malz2024, Smith2024}, transferring Matrix Product Operators (MPOs)  \cite{Nibbi2024, Termanova2024} is an important step to fully translate MPS-based algorithms onto quantum circuits.

Mapping classical PDEs on quantum computers demands mimicking the effect of non-linear and non-unitary dynamics by linear and unitary quantum operations.
Lubasch~\textit{et al.} significantly advanced the field of quantum differential solvers (QDS) by introducing a VQA that solves non-linear PDEs \cite{Lubasch2020} relevant in classical and quantum physics. 
This strategy offers an efficient classical parametrization of the solution which allows to circumvent the read-out problem for each time-step \change{and enables the computation of different observables like velocity moments, coarse-grained solutions or spatial means without rerunning the full time evolution \cite{Goldack2025,Uchida2024, Lubasch2020}.} 
 While it was used to solve core fluid dynamic problems as the Burgers' equation \cite{Jaksch2023, Jesus2025prep} and was extended \cite{Sarma2024} to various boundary conditions \cite{Over2024}
and space-time methods \cite{Pool2024}, it is limited to 
PDEs that directly map onto known quantum operators. As an additional drawback, cost functions are build up from numerous contributions where each requires a quantum circuit that needs to be measured individually. \change{Furthermore, the cost function is not bounded, making it difficult to estimate the training progress and accuracy without comparing with a classical solution.}
Going to generic PDEs requires an entirely different approach that we will 
\change{introduce in the following and apply to a linear and a non-linear PDE.}

In this \change{article}, we introduce a \change{tensor-programmable variational quantum algorithm} which utilizes the operator representation as MPO-based quantum circuits and has several advantages over previous approaches.
First, it allows for the incorporation of non-unitary operators, extending the range of PDEs and solution techniques on quantum computers.
Second, the number of quantum circuits $M$, required to build up the cost function, can be significantly reduced compared to previously introduced schemes. \change{In general, all terms of the PDE can be summarized within one quantum circuit, allowing to infer the convergence of the training progress directly from one expectation value.}
Furthermore, it opens the path to a broadly-applicable and modular scheme for solving problems in a wide range of scientific and industrial fields.

\change{This paper is structured as follows. Section~\ref{sec:methods} introduces the tensor-programmable quantum scheme, starting with a general overview to then address the details on the operator mapping in subsection~\ref{sec:methods-opMat} and iterations steps in~\ref{sec:methods-itSteps}. Furthermore it explains the necessary norm correction in subsection~\ref{sec:methods-norm} and introduces an adaption of the Hadamard test and a convergence measure for the optimization procedure in subsection~\ref{sec:methods-adH}. Subsection~\ref{sec:methods-nonLin} extends the introduced scheme to non-linear differential equations.
In section~\ref{sec:application}, we apply the introduced scheme to the linearized Euler equations (\ref{sec:application-euler}) and the non-linear Burgers' equation (\ref{sec:application-burgers}), while a third use-case, the linear advection-diffusion equation is provided in the Appendix~\ref{sec:app-AdvDiff} to show the successful application when using a larger number of qubits. Section~\ref{sec:ScalingComp} presents a comparison of this scheme with classical MPS based algorithms at the example of a turbulent dataset and considers the scaling of the operator. A conclusion is given in section~\ref{sec:conclusion}}.

\section{Methods} \label{sec:methods}
\change{In this section, we explain the tensor-programmable quantum scheme for linear and non-linear PDEs.
For simplicity, we first give a general overview of the method for linear PDEs, which is depicted in \autoref{fig:circuit-schemes}, together with the simulation results. We deepen the discussions in the following subsections and extend the scheme to non-linear PDEs in subsection~\ref{sec:methods-nonLin}. }

The solution $\phi(\bm{x},j)$ on a discretized grid at iteration step $j$ is amplitude encoded into a quantum register composed of $n$ qubits \cite{Schuld2021}. This state is generated by a quantum gate $\hat{U}(\boldsymbol{\theta}_j)$ (see \autoref{fig:circuit-schemes}~(a)) that is classically parametrized by a real vector $\boldsymbol{\theta}_j$ and an additional real number $\theta_j^0$ setting the norm of $\phi(\bm{x},j)$.

Here, we focus on uniform discretizations in one spatial dimension for simplicity. The extension to higher dimensions and non-uniform grids is conceptually straightforward.
In one dimension the state on the quantum register is given by $\theta_j ^0 \ket{\psi_j}=
\sum_{l=0}^{2^n-1}\phi(x, j )\ket{x_b}$,
where $x_b$ is the binary form of $x$, with $\ket{x_b}$ 
representing the computational basis states of the $n$ qubit quantum register.
Therefore, the vector $\boldsymbol{\theta}_j$ provides a classical representation of the solution which is exponentially compressed for restricted circuit depths \cite{Lubasch2020}.
\change{To encode the initial state into a quantum state, methods like the efficient approximate encoding  of MPS via shallow quantum circuits \cite{Ben-Dov2024}
can be used if the initial state is not trivial.}

The evolution of the system by one step is characterized by an operator $\hat{O}$ with $\theta^0_{j+1}\ket{\psi_{j+1}} = \hat{O}\theta^0_j\ket{\psi_j}$. 
We determine $\boldsymbol{\theta}_{j+1}$ by solving a problem dependent cost function $\mathcal{C}$ that is proportional to the 
overlap $\mathcal{C}\propto-\bra{0}\hat{U}^{\dagger}(\boldsymbol{\theta}_{j+1})\hat{O}\ket{\psi_j}$. 
The overlap is measured via an adapted Hadamard test (\autoref{fig:circuit-schemes}~(a), green box) by evaluating $\braket{\sigma_z}_{\text{anc}}$ of a global ancilla qubit at the end of the quantum circuit. 
The overlap is fed back to a classical computer that variationally updates the parameter vector $\boldsymbol{\theta}_{j+1}$ until a pre-defined convergence criterion is reached. 

The operator $\hat{O}$ can summarize the terms of the PDE in one quantum circuit, resulting in a single cost term or split them into several contributions.
The most suited strategy should be chosen in dependence on the PDE and the available quantum hardware resources.
The operator is in general non-unitary and implemented probabilistically (\autoref{fig:circuit-schemes}~(a), purple box). 
The success probability $\alpha_{\text{succ}}$ of the operator application is fed back to the classical computer to compute a norm correction, necessary to obtain the new normalization constant $\theta^0_{j+1}$.

\subsection{\change{Operator mapping}}\label{sec:methods-opMat}
To realize the non-unitary operator $\hat{O}$, we utilize the operator representation in terms of MPOs. Many discretized differential operators exhibit a low-rank MPO representation with small bond dimension $\zeta$ \cite{Kazeev2012, Oseledets2010, Kiffner2023}, including derivatives of higher order accuracy and various boundary conditions. \change{A collection of relevant MPOs for the considered PDEs is given in \autoref{subsec:MPO_representation}}.
The classical MPO is translated into a set of unitaries $\hat{U}_{\text{MPO}}$ with an algorithm proposed by Termanova~\textit{et al.} \cite{Termanova2024}, which is outlined in 
\change{\autoref{subsec:MPO-to-UNITARIES}}.
The algorithm introduces a new MPO consisting out of isometric tensors to approximate the original operator up to a multiplicative constant $c_{\text{MPO}}$ \change{and with relative approximation error $E$ \cite{Termanova2024}}.
This isometric MPO requires a larger bond dimension $Z>\zeta$, compensating for the reduced degrees of freedom due to the isometric constraints. This bond dimension defines an auxiliary qubit register of size ${n_{\text{aux}}}=\log_2(Z)$ \cite{Termanova2024}.
The isometric MPO is converted into unitary gates $\hat{U}_{\text{MPO}}$, which are part of the quantum circuit in \autoref{fig:circuit-schemes}~(a). Subsequent
mid-circuit measurements and postselection are employed to ensure that the operation corresponds to the actual MPO. Only those instances are kept, where the auxiliary register is measured in the state $\ket{\bm{0}}_{\text{aux}}$.


Tracking the number of successful and total runs determines the success probability $\alpha_{\text{succ}}$ of the postselection during the cost function evaluation. No further quantum circuit is required. \change{An alternative method to estimate $\alpha_{\text{succ}}$ from the training parameter $\varphi$ is describe later in this manuscript.}
Importantly, $\alpha_{\text{succ}}$ was shown to have favorable magnitude and scaling for various examples \cite{Termanova2024}, \change{ which can be extended to all relevant differential operators in this work as shown in \autoref{subsec:scaling_succ_prob}.}
In contrast to other approaches \cite{Zhao2023}, there is no exponential decay of the overall success probability of the algorithm with the number of iteration steps $j$.

\subsection{\change{Computing iteration steps}}\label{sec:methods-itSteps}
We identify the parameters defining the next iteration step $j+1$ by solving a problem dependent cost function $\mathcal{C}$.
Using the parameters from the previous iteration step as the initial guess for the optimization simplifies the process, as they are typically close to the solution of the current iteration step. This closeness 
\change{leads to a significantly improved trainability even in the presence of shots and larger qubit numbers \cite{Puig2025} (c.f. \autoref{subsec:cost-landscape})} 
\subsection{\change{Norm correction}}\label{sec:methods-norm}
As the operators are in general non-unitary the computation of the normalization constant $\theta^0_{j+1}$ requires to incorporate several correction factors.
One is the constant $c_{\text{MPO}}$, computed in the creation of the isometric MPO.
Furthermore, there is the  state and operator  dependent norm constant $f_{\hat{O},j}$ which accounts for the difference stemming from casting a non norm-conserving operator into a norm conserving form on the quantum computer. The necessary correction for each iteration step  $j$ can be computed from $\alpha_{\text{succ}}$ as
\change{
\begin{equation}
f_{\hat{O},j} =\frac{1+\alpha_{\text{succ}}}{2\sin(\varphi)-(\sqrt{\alpha_{\text{succ}}}-\frac{1}{\sqrt{\alpha_{\text{succ}}}})\cos(\varphi)} 
\end{equation}
}
where $\varphi\in(0,\pi/2]$ is the rotation angle of the $\hat{R}_Y$-gate on the global ancilla qubit (cf.~\autoref{fig:circuit-schemes}~(a)) which 
\change{has an optimal value} for each iteration step $j$.
Then, the effect of the operator $\hat{O}$ can be computed using
$\hat{O}\theta_j^0\ket{\psi_j}\ket{\textbf{0}}_{\mathrm{aux}}=c_{\text{MPO}}f_{\hat{O},j}\change{P_{\ket{0}_{\mathrm{aux}}\bra{0}_{\mathrm{aux}}}}\hat{U}_{\text{MPO}}\theta_j^0\ket{\psi_j} \change{\ket{\textbf{0}}_{\mathrm{aux}}}$, allowing for a correct estimation of $\theta_{j+1}^0$.
\change{Here, $\ket{}_{\mathrm{aux}}$, denotes the auxillary qubit register for the operator application and $P_{\ket{0}_{\mathrm{aux}}\bra{0}_{\mathrm{aux}}}$ denotes the projector of these qubits on $\ket{0}$.}

The overlap $\bra{0}\hat{U}^{\dagger}(\boldsymbol{\theta}_{j+1})\hat{O}\ket{\psi_j}$ in the cost function $\mathcal{C}$ can be computed using the measurement result $\braket{\sigma_z}_\text{anc}$ of a Hadamard test \cite{Lubasch2020}.
There, a global ancilla qubit controls the applications of the ansätze $\hat{U}^{\dagger}(\boldsymbol{\theta}_{j+1})$ and $\hat{U}(\boldsymbol{\theta}_{j})$ and the operator $\hat{U}_{\text{MPO}}$, and is measured in the computational bases at the end of the circuit (cf. \autoref{fig:circuit-schemes}~(a)).

\begin{figure*}
    \centering
\includegraphics[width=\linewidth]{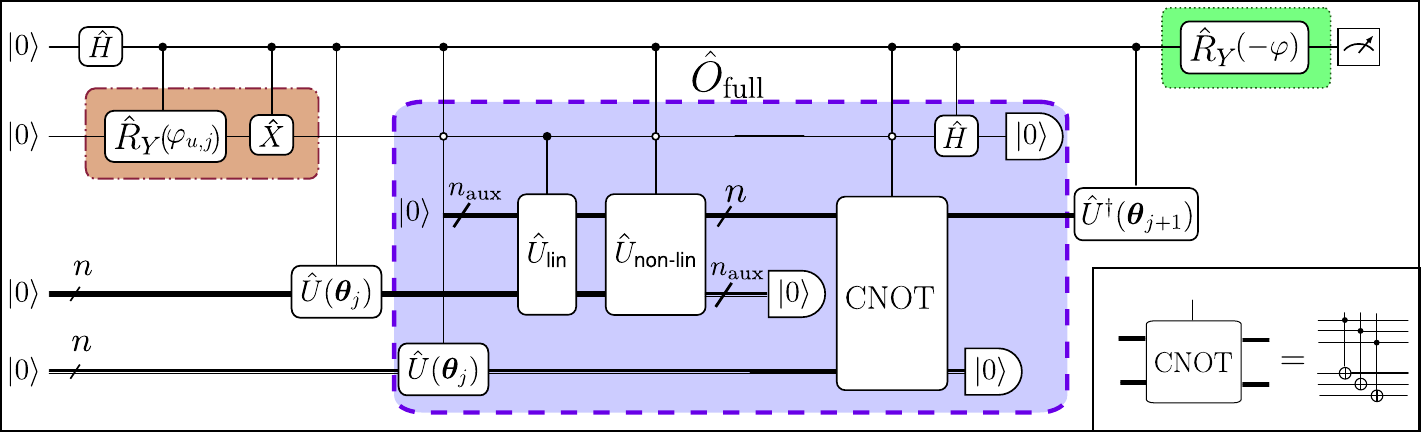}
\caption{\change{Quantum circuit to encode one time-step of a non-linear PDE. Linear and non-linear operators are separated into two MPOs and an additional auxillary qubit is introduced allowing to create a weighted superposition of the linear and non-linear contributions.
Double control of the global ancilla qubit and the extra auxiliary qubit can be avoided by applying $\mathrm{CNOT}$ gates before and after the non-linear operators, where the global ancilla acts as control and the extra auxiliary qubit as target. Then, the control on the global auxillary qubit can be omitted. The concrete decomposition of the $\mathrm{CNOT}$ gate is shown for the example of $n=3$.}}
\label{fig:circuit-non-lin}
\end{figure*}

\subsection{\change{Adapted Hadamard test and convergence measure}}\label{sec:methods-adH}
If the success probability $\alpha_{\text{succ}}$ is smaller than one, 
the probabilistic application of the MPO has a negative impact on the Hadamard test.
It stems from an increased contribution of $\ket{0}_\text{anc}$, which is always successful compared to $\ket{1}_\text{anc}$, where runs may be discarded, compared right after the application of the operator.
This imbalance causes an increased variance of $\braket{\sigma_z}_\text{anc}$, raising the  number of shots required to determine the cost term with a given accuracy. This non-optimal behavior can be mitigated with an adaption of the standard Hadamard test as shown in \autoref{fig:circuit-schemes}~(a) (green box). We substitute \change{the second Hadamard gate by an angle-dependent rotation gate $\hat{R}_Y(\change{-}\varphi)$}. 
The rotation angle \change{that maximizes $\braket{\sigma_z}_\text{anc}$} depends on $\alpha_{\text{succ}}$ (c.f. \autoref{subsec:adaptedH}).
For $\alpha_{\rm{succ}}=1$ the optimal angle $\varphi=\pi/2$ restores the Hadamard gate.

\change{When all operators of the PDE are summarized within one quantum circuit and the parameters $\boldsymbol{\theta}_{j+1}$ are well trained, there is an optimal 
\begin{equation} \label{eq:phi-alpha}
  \varphi_{\rm{opt}}=2\arctan(\sqrt{\alpha_{\text{succ}}}),  
\end{equation}
that results in $\braket{\sigma_z}_\text{anc}=1$. 
While optimizing $\boldsymbol{\theta}_{j+1}$, the training progress can be tracked with the fidelity  $\mathcal{F}=||\braket{\mathbf{0}|\hat U^{\dagger}(\boldsymbol{\theta}_{j+1})P_{\ket{0}_{\textrm{aux}}\bra{0}_{\textrm{aux}}} \hat U_{\textrm{MPO}}|\psi_{j}}||^2$.
Combining all operators in one quantum circuit allows to infer the fidelity  directly from $\braket{\sigma_z}_\text{anc}$ according to (c.f. \autoref{subsec:adaptedH} for derivation details) 
\begin{equation}
\mathcal{F} = \frac{(\alpha_{\text{succ}}\braket{\sigma_z}_{\textrm{anc}}+\alpha_{\text{succ}}\cos(\varphi_{\textrm{opt}})+\braket{\sigma_z}_{\textrm{anc}}-\cos(\varphi_{\textrm{opt}}))^2}{4\alpha_{\text{succ}}\sin^2(\varphi_{\textrm{opt}})}.
\end{equation}
Then, the fidelity can act as a convergence measure that reflects the status of the training without the need for verification with a classical solution\change{. This is a significant advantage compared with previous approaches \cite{Lubasch2020, Over2024}, where each term is treated individually and this knowledge on the solutions convergence can not be easily inferred.}
The parameter $\varphi_{\rm{opt}}$ can be either computed from the estimated success probability, by counting the number of discarded shots during the training process, or can be trained itself. If $\varphi$ is trained and the problem is described by one quantum circuit, $\alpha_{\text{succ}}$ can be computed from Eq.~(\ref{eq:phi-alpha}). As all trainable parameters are placed after the postselection procedure, the parameter-shift rule  \cite{Mitarai2018, Schuld2018} or coherent gradient approximations \cite{Kandala2017,Spall1998} apply.} 
\subsection{\change{Extension to non-linear differential equations}}\label{sec:methods-nonLin}
\change{To make the quantum-tensor scheme applicable beyond linear PDEs, we consider differential equations of the form 
\begin{equation}
u^{j+1}=(\hat O_{\rm{lin}} + u^j\hat O_{\rm{non-lin}})u^j = \hat{O}_{\rm{full}}u^j,
\end{equation}
where $\hat{O}_{\rm{lin}}$ and $\hat{O}_{\rm{non-lin}}$ are both linear operators. Prominent examples of such PDEs are the Burgers' equation and the Navier-Stokes equations.
We can not summarize $\hat O_{\rm{lin}}$ and $\hat O_{\rm{non-lin}}$ into one MPO, as the non-linear term has a direct dependence on $u^j$, which would require to compute new unitaries for each iteration step. To preserve the advantages of the combined circuit for  non-linear problems, we introduce an extended circuit shown in \autoref{fig:circuit-non-lin}. A second weighting ancilla qubit allows the simultaneous application of the linear and non-linear terms. For the non-linear term we cast $\hat O_{\rm{non-lin}}$ in the unitary $\hat U_{\rm{non-lin}}$, while the non-linearity itself is computed with $\mathrm{CNOT}$ gates as introduced by Lubasch in \cite{Lubasch2020}.
The angle $\varphi_u$ in the rotation gate $\hat R_{Y}(\varphi_u)$  on the weighting ancilla qubit is responsible for the correct weighting of the linear and the non-linear term. For explicit Euler time stepping, where the leading linear term is the identity matrix, 
\begin{equation}
\varphi_{u,j}=2\arctan{\left(\frac{\theta_j^0\abs{c_{\rm{MPO,non-lin}}}}{\abs{c_{\rm{MPO,lin}}}}\right)},
\end{equation}
where $c_{\rm{MPO,lin}}$ and  $c_{\rm{MPO,non-lin}}$ are the multiplicative factors from the unitary creation of  $\hat U_{\rm{lin}}$ and $\hat U_{\rm{\rm{non-lin}}}$ respectively.
To compute the norm of the next time step, we compute $f_{\hat{O},j}$ as before.
Then, the effect of the total operator $\hat{O}_{\rm{full}}$ can be computed as $\hat{O}_{\rm{full}}\theta^0_j\ket{\psi_j}\ket{\textbf{0}}_{\mathrm{aux}} = \sqrt{2}~c_{\rm{MPO,lin}}f_{\hat{O},j}r_{u}P_{\ket{0}_{\mathrm{aux}}\bra{0}_{\mathrm{aux}}}\hat U_{\rm{full}}\theta^0_j\ket{\psi_j}\ket{\textbf{0}}_{\mathrm{aux}}$, where $\hat U_{\rm{full}}$ summarizes the linear and non-linear operator actions and $r_{u} = k_u/k_u^{QC}$ stems from the difference between the real ratio
$k_u =\theta^0_j\abs{c_{\rm{MPO,non-lin}}}/\abs{c_{\rm{{MPO,lin}}}}$ 
and its actual application $k_u^{QC}=\sin(1/2 \varphi_{u,j})$
on the quantum computer.} 
\section{Application to Differential Equations}\label{sec:application}
\change{
In the following, we show the successful application of this new approach to two use cases. We will first discuss the linearized Euler equation, to demonstrate the efficient incorporation of complex operators and time stepping schemes. 
Next, we consider the non-linear Burgers' equation to demonstrate the newly introduced quantum circuit for non-linear problems and the working principle of the convergence estimation from the expectation value. A third use-case of the advection-diffusion equation demonstrating the usability with larger qubit numbers is presented in \autoref{sec:app-AdvDiff}.
For both use cases, the ansatz functions are encoded with a brickwall ansatz for the circuit $\hat{U}(\boldsymbol{\theta}_j)$ as depicted in \autoref{fig:building-blocks}~(a) and all operators are casted into highly accurate unitary approximations with  bond dimension $Z=16$ and relative approximation error below $E=5\times10^{-10}$.
\change{For all use-cases the reinitialization of the new training weights from the previous iteration step leads to a sine-shape training landscape for each training parameter even in the presence of shot noise number
 and hence} facilitates the optimization of the generally non-convex cost function. \change{The training landscape remains pronounced for increasing system size, ensuring the trainability even for large qubit number. Details are given in \autoref{subsec:cost-landscape}.}}


\subsection{\change{Linear Euler equations}}\label{sec:application-euler}
\change{The $1D$ linear Euler equations describes } the evolution of velocity $u(x,t)$ and pressure $p(x,t)$ of an inviscid flow \cite{Chorin1979} over time $t$. To this aim, we employ a noise-free quantum computing simulator and the explicit 4th order Runge-Kutta (RK4) method to compute the solution of the next time step $dt$.
We consider the particular situation of a periodic pressure point source with constant amplitude $A_0$ and angular frequency $\omega$ in the center of the domain $f(x,t) = A_0\delta (x)\sin(\omega t)$. 
Furthermore non-reflective boundary conditions are applied using the sponge layer method \cite{Peric2015}, where a damping zone near the boundary is introduced via the sponge function $\gamma(x)$ (cf. \autoref{subsec:MPO_representation}).

The coupled system of equations reads
\begin{equation}
   \begin{array}{l}
    \label{eq:Euler_eqs}
    \dfrac{\partial p}{\partial t} = 
    - \bar{\rho}c^{2}\left(\dfrac{\partial u}{\partial x} \right) + f(x,t) - \gamma(x)p, \\
    \dfrac{\partial u}{\partial t} = - \dfrac{1}{\bar{\rho}}\dfrac{\partial p}{\partial x} - \gamma(x)u,
    \end{array}
\end{equation}
with the constants density $\bar{\rho}$ and the speed of sound $c$.
We implement $8$\textsuperscript{th}-order accurate finite differences with Dirichlet boundary conditions and a staggered grid to avoid checkerboard oscillations  \cite{harlow1965}. 
For computational simplicity we 
separate the fields into two ansatz circuits. 
Our scheme still significantly reduces qubit and circuit count compared to previously introduced methods \cite{Lubasch2020,Over2024, Sarma2024}, where additional circuits for each order of accuracy and the boundary implementations are needed. Instead, we can represent all operators acting on one field with one circuit of bounded depth.
Furthermore, the sponge operator
does not require an additional qubit register.
The resulting cost functions, using RK4, are given in \autoref{subsec:Cost}. 

\begin{figure*}[bt]
    \centering
    \includegraphics[width=\linewidth]{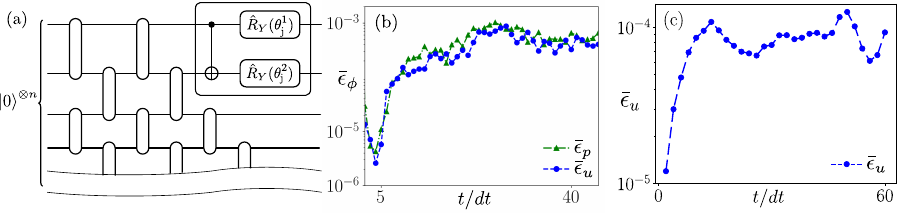}
    \caption{
     (a) Brickwall ansatz, with variational parameters. Each 2-qubit block is composed of two $\hat{R}_Y(\theta_i)$-gates and one $ \mathrm{CNOT}$-gate. The ansatz can be used for a variable number of layers $L$, with each layer consisting of one column of 2-qubit blocks. 
     (b) Evolution of the relative error 
     $\bar{\epsilon}_{\phi}(t)$ (cf. \autoref{eq:rel_error}) for the Euler's equation.
     Here, $\phi(x,t)$ corresponds to the discretized solutions $u(x,t)$ and $p(x,t)$ of the Euler equation.
     To represent the solutions, a brickwall ansatz with $6$ qubits and $14$ layers is used. 
     \change{(c) Evolution of the relative error  $\bar{\epsilon}_{u}(t)$ for the Burgers' equation. To represent the solutions, a brickwall ansatz with $6$ qubits and $10$ layers is used. 
     }
     }
	\label{fig:building-blocks}
\end{figure*}

\begin{figure}[bt]
    \centering
\includegraphics[width=\columnwidth]{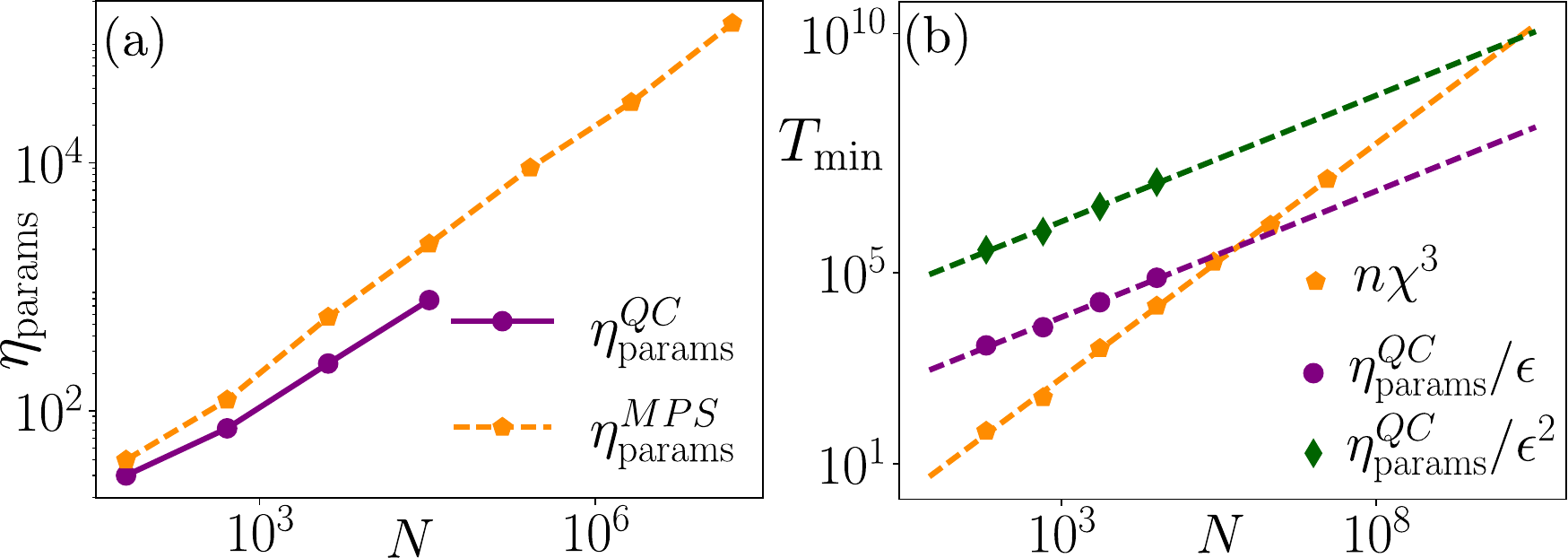}
\caption{\change{Representation capabilities of the brickwall and the MPS ansatz for a three-dimensional turbulent flow defined on $N$ grid points. (a) Number of parameters to represent an increasing section of an isotropic flow field of total size $N_{\textrm{tot}}=1024^3$ approximated up to an accuracy of $\bar{\epsilon}_v=0.01$.
(b) Estimate of the minimal required cost $T_{\textrm{min}}$ computed for the given number of parameters from panel (a) and for an allowed maximal the sampling error $\epsilon=0.01$ The two minimal costs for the quantum scheme give the best (purple, circle) and the worst (green, diamont) case scenario for the cost contribution of the sampling error. The dashed lines corresponds to a liner interpolation in the log-log-scale.}}

\label{fig:layer-and-chi}
\end{figure}

Initially, the velocity and the pressure field are zero in the whole domain. Due to the pressure point source, an increasing pressure peak forms during the first time steps, leading to a non-zero contribution in the velocity field.
This peak propagates towards the boundaries while additional peaks are formed by the source as shown in \autoref{fig:circuit-schemes}~(b).
To assess the quality of the solution, we use the normalized fidelity $\mathcal{F}^{n}=|(\phi(x,t)^{QDS}, \phi(x,t)^{cl})|^2/\norm{\phi(x,t)^{QDS}}^2\norm{\phi(x,t)^{cl}}^2$ \cite{nielsen_chuang2010}, as a measure of closeness between the solution computed with our QDS ($\phi(x,t)^{QDS}$) and a classical solver ($\phi(x,t)^{cl}$), with ($\cdot, \cdot$) being the inner product. The relative error is defined as 
\begin{equation} \label{eq:rel_error}
\change{\bar{\epsilon}_{\phi}(t)=1-\mathcal{F}^n}
\end{equation}
and depicted in \autoref{fig:building-blocks}~(b).
During the time evolution the maximal relative error is $0.1\%$, which is better than the error that would be introduced by current quantum hardware. 

\subsection{\change{Non-linear Burgers' equation}}\label{sec:application-burgers}
\change{The non-linear weakly-viscous Burgers' equation is defined as
\begin{equation}
    \frac{\partial u}{\partial t} = -u  \frac{\partial u}{\partial x} + \nu \frac{\partial^2 u}{\partial x^2},
\end{equation}
}
\change{
where $u$ is the velocity and $\nu$ the viscosity of the fluid. We implement a first-order accurate backward derivative $\frac{\partial}{\partial x}$ and a second-order accurate central derivative $\frac{\partial^2}{\partial x^2}$ with periodic boundary conditions with $x\in[0, 2\pi)$.}

\change{
We consider the particular situation of a Gauss peak as initial conditions $u(x,t=0)=\exp(\frac{-(x-\pi)^2}{\sigma})$ with $\sigma=0.5$ and choose $\nu$ and $dt$, such that a shock evolves within a reasonable simulation time. The detailed simulation parameters and the used cost function are listed in \autoref{subsec:Parameter} and \autoref{subsec:Cost}.
The velocity evolution is shown if \autoref{fig:circuit-schemes}~(c), where the initial Gauss peak evolves into a shock wave. When compared to the classical solution, the relative error $\bar\epsilon_u$ reaches values around $10^{-4}$, showing a good agreement over the whole time evolution.}
\change{We observe that the scaling of $\alpha_{\rm{succ}}$ of the non-linear multiplication over system size behaves similarly as reported by Lubasch~\textit{et al.} \cite{Lubasch2020} for the expectation value in their non-linear use-case. More details are given in  \autoref{subsec:scaling_succ_prob}.} 


\section{Scaling Analysis}\label{sec:ScalingComp}
\change{In addition to the simulation itself, we are interested in the scaling behavior of the quantum ansatz. To this aim we first focus on a comparison with MPS-based algorithms, which are the classical counterpart of the  tensor-programmable quantum scheme \cite{Termanova2024, Haghshenas2022}. Second, we consider the cost arising from the operator application.}
\subsection{\change{Comparison with MPS methods}}
\change{For the scaling comparison, we focus on the cost caused by the field encoding, which is normally dominant, as the operators feature a low-rank MPO representation (c.f. \autoref{subsec:MPO_representation}).}
\change{We are mainly interested in two performance indicators, namely the memory consumption and the computational complexity. The memory consumption is dominated by number of parameters necessary to represent the field per iteration step which scales as $O(n \chi^2)$ for the classical MPS and as $O(\eta_{\rm{params}})$ for the quantum ansatz, where $\eta_{\rm{params}}$ is the number of parameters $\mathbf{\boldsymbol{\theta}}_j$ in the brickwall ansatz.}

\change{
Regarding the computational complexity, it is well known, that MPS algorithms scale at least as $n\chi^3$ \cite{Gourianov2022} for linear operations. For the non-linear Hadamard product, stable simulations with a scaling of $n\chi^4$ where reported \cite{Gourianov2022}.
The quantum algorithm scales mainly with the number of parameters necessary to represent the solution \cite{Haghshenas2022} as $O(\eta_{\rm{params}})$. Additionally, sampling the expectation value $\braket{\sigma_z}_{\textrm{anc}}$ adds an additional cost $O(1/\epsilon^2)$ depending on the allowed sampling error $\epsilon$. 
There are methods that can reduce this cost to values close to $O(1/\epsilon)$ \cite{Wang2019, Grinko2021}. 
When all operators are combined in one quantum circuit and the expectation value approaches $\braket{\sigma}_{\textrm{anc}}=1$, the sampling cost approaches $1/\epsilon$ even without additional modifications. 
Given the large shot overhead caused by the scaling of the sampling error, the quantum tensor network method can clearly not outperform the classical algorithms for small $\chi$. However, for large-scale simulation where larger $\chi$ are required, this constant shot-overhead decreases in importance, and the first factor, the number of required parameters becomes decisive.}

\change{We consider the highly relevant use-case of a three-dimensional turbulent flow field, namely a velocity field of an isotropic turbulent flow on $N_{\textrm{tot}}=1024^3$ grid points provided by the Johns Hopkins Turbulence Database at a Taylor-scale Reynolds number of $Re_{\tau}\approx433$ \cite{John_Hopkins_Turb, John_Hopkins_Turb2, John_Hopkins_Turb3}.
To reduce the problem into numerically feasible sizes, we consider a three-dimensional box of size $N=2^{3n_{\textrm{d}}}$ at the center of the domain, where each spatial direction is resolved using $n_{\textrm{d}}$ qubits. Increasing $n_{\textrm{d}}$ allows to  study the scaling behavior with $N$ and to provide an estimate for the full flow field.}

\change{Figure~\ref{fig:layer-and-chi}~(a) depicts the number of parameters necessary for the MPS and the quantum circuit to represent the flow field of different box-sizes, as well as their ratio. 
For the MPS, we consider the maximal bond dimension $\chi$ required to represent the $y$-component of the velocity field with a fidelity $\bar{\epsilon}_v<0.01$. 
For the quantum circuit, we train a brickwall ansatz to represent the field with increasing layer number until $\bar{\epsilon}_v<0.01$. For this, we have replace the $\mathrm{CNOT}$ gates in the brickwall ansatz with $CZ$ gates, as this allows to initialize additional low-depth layers as an identity.
The reduction in parameters when comparing the quantum circuit and the MPS representation increases with the system size and the bond dimension. We expect this advantage to stem from the better entangling properties of the quantum circuit. Importantly, while the shown MPS-representation was created using successive singular value decompositions  \cite{Schollw_ck_2011} and is hence already optimal with respect to the $l_2$-norm, the quantum circuit representation does not possess this optimality. Different circuit architectures \cite{Haghshenas2022} and techniques as incremental structural learning \cite{Jaderberg2020, Jaderberg2022} can potentially be exploited to reduce the number of parameters further, which would also improve the scaling properties.
Additionally, \cite{Brandao2016} has provided an theoretical upper bound when representing each unitary as a $\kappa$-design, where the circuit depth scales only polynomially in the number of qubits, and the choosen $\kappa$ adds (potentially large) constant factor.\\} 
\change{In \autoref{fig:layer-and-chi}~(b), we look at the minimal cost $T_{\textrm{min}}$ of the MPS ($T^{\textrm{MPS}}_{\textrm{min}}= n\chi^3$) and the quantum approach ($T^{\textrm{QC}}_{\textrm{min}}= \eta^{\textrm{QC}}_{\textrm{params}}/\epsilon^{(2)}$ ), given the computed number of parameters and assuming a maximal sampling error of the size of the target fidelity $\epsilon=0.01$. 
While $T_{\textrm{min}}$ is lower for the MPS approach for small qubit numbers, the quantum scheme shows its potential for larger scale simulations.\\}
\change{Choosing a suitable training strategy for the quantum use-case is mandatory to preserve the potential advantage. Gradient based methods, which rely on the exact computation of each gradient, add an additional scaling factor $O(\eta_{\textrm{params}})$, which would annihilate any advantage. However, strategies to estimate the gradient coherently, as the simultaneous perturbation stochastic approximation (SPSA) can fix the number of circuit evaluations to a number independent of the number of trainable parameters \cite{Kandala2017,Spall1998}, and is still capable to train the solution to optimal parameters 
(c.f. \autoref{subsec:cost-landscape}).\\}
\change{Clearly, this is a simplified scaling analysis. It neglects the potential additional cost of $\chi^4$ of the MPS scheme in the presence of non-linear operations and the impact of the success probability of the quantum circuit. Both depend on the chosen differential equation and its implementation details.  
For small success probabilities, the discussed expected theoretical advantage might be shifted to larger system sizes
or might require additional strategies as e.g. amplitude amplification, phase estimation or their combination as utilized by Goswani~\textit{et al.} \cite{Goswami2025-arxiv} to be maintained.
Furthermore, we have assumed that both the classical and the quantum tensor network algorithm involve a comparable scaling of the number of training steps with system size. Importantly, both approaches provide the fidelity $\mathcal{F}$ as convergence measure which allows to adapt the training hyperparameters during the training procedure and improving the convergence.
A significant part of the cost in the quantum algorithm is caused by the number of required shots. As this process is inherently parallelizable, practical benefits might occur even before theoretical advantages are reached.}
\subsection{\change{Scaling of the operator}}
In previous schemes \cite{Lubasch2020, Over2024}  the required number of auxillary qubits used to encode potentials and finite difference derivatives scales linear with the number of ansatz qubits $n$. 
In comparison, in our  approach
the number of auxillary qubits depends solely on $Z$, which is
expected to be independent of $n$.
Additionally, the number of quantum circuits $M$ is considerably reduced.
This beneficial scaling in qubit and circuit number is achieved with the same scaling of the circuit depth.
The scaling of the number of 2-qubit gates for the operator application, $N_{\rm{2q,op}}$, is upper bounded \cite{Shende2004} by $N_{\rm{2q,op}}=k n Z^2$, with a universal proportionality constant $k$.
\section{Conclusion}\label{sec:conclusion}
This paper demonstrates how generic VQA can be programmed using MPOs, allowing for the seamless integration of non-unitary operators. Importantly, higher order differential operators and various boundary conditions can be incorporated with little or no additional cost. \change{Tensor network algorithms were effectively implemented for many large-scale simulations but can face challenges in the presence of complex data. Using the example of a turbulent flow field, we can show the additional compression capabilities of the quantum circuit. This leads us to expect successful scaling of the tensor-programmable quantum circuits, particularly when exploring additional parameter reduction techniques \cite{Jaderberg2020, Jaderberg2022}. This potential is further enhanced by employing advanced optimization strategies, such as multigrid and local optimization, which have proven extremely successful in classical tensor network schemes \cite{Gourianov2022, Lubasch2018} and are also accessible for VQAs \cite{Pool2024,Slattery2022}.}

\change{Being applicable to linear and non-linear PDEs,} the presented scheme contains all building blocks for solving PDEs critical for science and industry, e.g., the Navier-Stokes equations.
This will become especially valuable once quantum hardware reaches the required capacity for industry relevant use cases, a milestone that, according to projections from companies like IBM and QuEra, could be achieved by 2029 \cite{QueraRoadmap,IBMRoadmap}.
Furthermore, the flexible operator representation of this scheme would enable interfacing between quantum algorithms and existing classical software packages.

\change{The tensor-programmable quantum algorithms allows to encode all operators of a PDE into a single quantum circuit. This bears two main advantages. First, it increases the expectation value of the global ancilla qubit. Second, this expectation value is directly connected to the fidelity between the trained and the correct solution and can hence act as a global measure of convergence. This allows to estimate the quality of the solution without expensive comparisons to classical solution and yields the potential for adaptive optimization strategies, where hyperparameters and circuit depths are improved within the training loop.}
Finally, \change{these improvements} can potentially be augmented by utilizing the potential of phase estimation techniques \cite{Kitaev1995,Fomichev2023} to improve the precision of the cost function measurements.
\change{In the presence of low success probabilities additional techniques as amplitude amplification \cite{Goswami2025-arxiv} should be considered, to decrease the number of required shots.}
These questions will be investigated in future studies.\\

\section*{Acknowledgment}
We thank Theofanis Panagos for helpful discussions on the use case \change{and M. Lautaro Hickmann for helpful discussions on circuit optimization routines.}
P.S. acknowledges financial support by the DLR-Quantum-Fellowship Program.
G.S.R., N.v.H. and D.J. acknowledge funding from the European Union’s Horizon Europe
research and innovation program (HORIZON-CL4-2021-DIGITAL-EMERGING-02-10) under grant agreement No. 101080085 QCFD.\\
G.S.R. is supported by the Cluster of Excellence ''CUI: Advanced Imaging of Matter'' of the Deutsche Forschungsgemeinschaft (DFG) - EXC 2056 - project ID 390715994.
T.H. and D.J. acknowledge funding by the Cluster of Excellence ``Advanced Imaging of Matter'' of the Deutsche Forschungsgemeinschaft (DFG) – EXC 2056 - project ID 390715994.\\
D.J. acknowledges support from the Hamburg Quantum Computing Initiative (HQIC) project EFRE. The project is co-financed by ERDF of the European Union and by ``Fonds of the Hamburg Ministry of Science, Research, Equalities and Districts (BWFGB)''.\\
Further the authors would like to thank Airbus and the BMW Group for providing the use case in the context of the Airbus-BMW Group Quantum Computing Challenge 2024. 
\section*{Data Availability}
The data that support the findings of this article are openly available \cite{dataset}.
\FloatBarrier
\bibliography{MPOToQCforDiff.bib}

\begin{thebibliography}{98}%
\makeatletter
\providecommand \@ifxundefined [1]{%
 \@ifx{#1\undefined}
}%
\providecommand \@ifnum [1]{%
 \ifnum #1\expandafter \@firstoftwo
 \else \expandafter \@secondoftwo
 \fi
}%
\providecommand \@ifx [1]{%
 \ifx #1\expandafter \@firstoftwo
 \else \expandafter \@secondoftwo
 \fi
}%
\providecommand \natexlab [1]{#1}%
\providecommand \enquote  [1]{``#1''}%
\providecommand \bibnamefont  [1]{#1}%
\providecommand \bibfnamefont [1]{#1}%
\providecommand \citenamefont [1]{#1}%
\providecommand \href@noop [0]{\@secondoftwo}%
\providecommand \href [0]{\begingroup \@sanitize@url \@href}%
\providecommand \@href[1]{\@@startlink{#1}\@@href}%
\providecommand \@@href[1]{\endgroup#1\@@endlink}%
\providecommand \@sanitize@url [0]{\catcode `\\12\catcode `\$12\catcode
  `\&12\catcode `\#12\catcode `\^12\catcode `\_12\catcode `\%12\relax}%
\providecommand \@@startlink[1]{}%
\providecommand \@@endlink[0]{}%
\providecommand \url  [0]{\begingroup\@sanitize@url \@url }%
\providecommand \@url [1]{\endgroup\@href {#1}{\urlprefix }}%
\providecommand \urlprefix  [0]{URL }%
\providecommand \Eprint [0]{\href }%
\providecommand \doibase [0]{https://doi.org/}%
\providecommand \selectlanguage [0]{\@gobble}%
\providecommand \bibinfo  [0]{\@secondoftwo}%
\providecommand \bibfield  [0]{\@secondoftwo}%
\providecommand \translation [1]{[#1]}%
\providecommand \BibitemOpen [0]{}%
\providecommand \bibitemStop [0]{}%
\providecommand \bibitemNoStop [0]{.\EOS\space}%
\providecommand \EOS [0]{\spacefactor3000\relax}%
\providecommand \BibitemShut  [1]{\csname bibitem#1\endcsname}%
\let\auto@bib@innerbib\@empty
\bibitem [{\citenamefont {Black}\ and\ \citenamefont
  {Scholes}(1973)}]{Black1973}%
  \BibitemOpen
  \bibfield  {author} {\bibinfo {author} {\bibfnamefont {F.}~\bibnamefont
  {Black}}\ and\ \bibinfo {author} {\bibfnamefont {M.}~\bibnamefont
  {Scholes}},\ }\bibfield  {title} {\bibinfo {title} {The pricing of options
  and corporate liabilities},\ }\href
  {https://api.semanticscholar.org/CorpusID:154552078} {\bibfield  {journal}
  {\bibinfo  {journal} {J. Political Econ.}\ }\textbf {\bibinfo {volume}
  {81}},\ \bibinfo {pages} {637} (\bibinfo {year} {1973})}\BibitemShut
  {NoStop}%
\bibitem [{\citenamefont {Lee}\ and\ \citenamefont {Sun}(2012)}]{Lee2012}%
  \BibitemOpen
  \bibfield  {author} {\bibinfo {author} {\bibfnamefont {S.~T.}\ \bibnamefont
  {Lee}}\ and\ \bibinfo {author} {\bibfnamefont {H.-W.}\ \bibnamefont {Sun}},\
  }\bibfield  {title} {\bibinfo {title} {Fourth-order compact scheme with local
  mesh refinement for option pricing in jump-diffusion model},\ }\href
  {https://doi.org/10.1002/num.20677} {\bibfield  {journal} {\bibinfo
  {journal} {Numer. Methods Partial Differ. Equ.}\ }\textbf {\bibinfo {volume}
  {28}},\ \bibinfo {pages} {1079} (\bibinfo {year} {2012})}\BibitemShut
  {NoStop}%
\bibitem [{\citenamefont {Bian}\ \emph {et~al.}(2016)\citenamefont {Bian},
  \citenamefont {Kim},\ and\ \citenamefont {Karniadakis}}]{Bian2016}%
  \BibitemOpen
  \bibfield  {author} {\bibinfo {author} {\bibfnamefont {X.}~\bibnamefont
  {Bian}}, \bibinfo {author} {\bibfnamefont {C.}~\bibnamefont {Kim}},\ and\
  \bibinfo {author} {\bibfnamefont {G.~E.}\ \bibnamefont {Karniadakis}},\
  }\bibfield  {title} {\bibinfo {title} {111 years of brownian motion},\ }\href
  {http://dx.doi.org/10.1039/C6SM01153E} {\bibfield  {journal} {\bibinfo
  {journal} {Soft Matter}\ }\textbf {\bibinfo {volume} {12}},\ \bibinfo {pages}
  {6331} (\bibinfo {year} {2016})}\BibitemShut {NoStop}%
\bibitem [{\citenamefont {Wu}\ \emph {et~al.}(2025)\citenamefont {Wu},
  \citenamefont {Zhang},\ and\ \citenamefont {Mao}}]{Wu2025}%
  \BibitemOpen
  \bibfield  {author} {\bibinfo {author} {\bibfnamefont {X.}~\bibnamefont
  {Wu}}, \bibinfo {author} {\bibfnamefont {Y.}~\bibnamefont {Zhang}},\ and\
  \bibinfo {author} {\bibfnamefont {S.}~\bibnamefont {Mao}},\ }\bibfield
  {title} {\bibinfo {title} {Learning the physics-consistent material behavior
  from measurable data via pde-constrained optimization},\ }\href
  {https://doi.org/https://doi.org/10.1016/j.cma.2025.117748} {\bibfield
  {journal} {\bibinfo  {journal} {Comput. Methods Appl. Mech. Eng.}\ }\textbf
  {\bibinfo {volume} {437}},\ \bibinfo {pages} {117748} (\bibinfo {year}
  {2025})}\BibitemShut {NoStop}%
\bibitem [{\citenamefont {Courant}\ \emph {et~al.}(1967)\citenamefont
  {Courant}, \citenamefont {Friedrichs},\ and\ \citenamefont
  {Lewy}}]{Courant1967}%
  \BibitemOpen
  \bibfield  {author} {\bibinfo {author} {\bibfnamefont {R.}~\bibnamefont
  {Courant}}, \bibinfo {author} {\bibfnamefont {K.}~\bibnamefont
  {Friedrichs}},\ and\ \bibinfo {author} {\bibfnamefont {H.}~\bibnamefont
  {Lewy}},\ }\bibfield  {title} {\bibinfo {title} {On the partial difference
  equations of mathematical physics},\ }\href
  {https://doi.org/10.1147/rd.112.0215} {\bibfield  {journal} {\bibinfo
  {journal} {IBM J. Res. Dev.}\ }\textbf {\bibinfo {volume} {11}},\ \bibinfo
  {pages} {215} (\bibinfo {year} {1967})}\BibitemShut {NoStop}%
\bibitem [{\citenamefont {Ferziger}\ and\ \citenamefont
  {Peri\'{c}}(2002)}]{ferziger2002:CMFD}%
  \BibitemOpen
  \bibfield  {author} {\bibinfo {author} {\bibfnamefont {J.~H.}\ \bibnamefont
  {Ferziger}}\ and\ \bibinfo {author} {\bibfnamefont {M.}~\bibnamefont
  {Peri\'{c}}},\ }\href {https://doi.org/10.1007/978-3-642-56026-2} {\emph
  {\bibinfo {title} {{Computational Methods for Fluid Dynamics}}}},\ \bibinfo
  {edition} {3rd}\ ed.\ (\bibinfo  {publisher} {Springer},\ \bibinfo {address}
  {Berlin},\ \bibinfo {year} {2002})\BibitemShut {NoStop}%
\bibitem [{\citenamefont {Wu}(2022)}]{WU2022}%
  \BibitemOpen
  \bibfield  {author} {\bibinfo {author} {\bibfnamefont {P.}~\bibnamefont
  {Wu}},\ }\bibfield  {title} {\bibinfo {title} {Recent advances in the
  application of computational fluid dynamics in the development of rotary
  blood pumps},\ }\href
  {https://www.sciencedirect.com/science/article/pii/S2590093522000649}
  {\bibfield  {journal} {\bibinfo  {journal} {Med. Nov. Technol. Devices}\
  }\textbf {\bibinfo {volume} {16}},\ \bibinfo {pages} {100177} (\bibinfo
  {year} {2022})}\BibitemShut {NoStop}%
\bibitem [{\citenamefont {Slotnick}\ \emph {et~al.}(2014)\citenamefont
  {Slotnick}, \citenamefont {Khodadoust}, \citenamefont {Alonso}, \citenamefont
  {Darmofal}, \citenamefont {Gropp}, \citenamefont {Lurie},\ and\ \citenamefont
  {Mavriplis}}]{Slotnick2014}%
  \BibitemOpen
  \bibfield  {author} {\bibinfo {author} {\bibfnamefont {J.~P.}\ \bibnamefont
  {Slotnick}}, \bibinfo {author} {\bibfnamefont {A.}~\bibnamefont
  {Khodadoust}}, \bibinfo {author} {\bibfnamefont {J.~J.}\ \bibnamefont
  {Alonso}}, \bibinfo {author} {\bibfnamefont {D.~L.}\ \bibnamefont
  {Darmofal}}, \bibinfo {author} {\bibfnamefont {W.}~\bibnamefont {Gropp}},
  \bibinfo {author} {\bibfnamefont {E.~A.}\ \bibnamefont {Lurie}},\ and\
  \bibinfo {author} {\bibfnamefont {D.~J.}\ \bibnamefont {Mavriplis}},\
  }\bibfield  {title} {\bibinfo {title} {Cfd vision 2030 study: A path to
  revolutionary computational aerosciences}\ }(\bibinfo {year}
  {2014})\BibitemShut {NoStop}%
\bibitem [{\citenamefont {Menter}(1994)}]{Menter1994}%
  \BibitemOpen
  \bibfield  {author} {\bibinfo {author} {\bibfnamefont {F.~R.}\ \bibnamefont
  {Menter}},\ }\bibfield  {title} {\bibinfo {title} {Two-equation
  eddy-viscosity turbulence models for engineering applications},\ }\href
  {https://doi.org/10.2514/3.12149} {\bibfield  {journal} {\bibinfo  {journal}
  {AIAA J.}\ }\textbf {\bibinfo {volume} {32}},\ \bibinfo {pages} {1598}
  (\bibinfo {year} {1994})}\BibitemShut {NoStop}%
\bibitem [{\citenamefont {Wilcox}(1998)}]{Wilcox1998}%
  \BibitemOpen
  \bibfield  {author} {\bibinfo {author} {\bibfnamefont {D.}~\bibnamefont
  {Wilcox}},\ }\bibfield  {title} {\bibinfo {title} {Turbulence modeling for
  cfd},\ }\href@noop {} {\bibfield  {journal} {\bibinfo  {journal} {DCW
  industries, La Canada}\ } (\bibinfo {year} {1998})}\BibitemShut {NoStop}%
\bibitem [{\citenamefont {Pope}(2000)}]{Pope2000}%
  \BibitemOpen
  \bibfield  {author} {\bibinfo {author} {\bibfnamefont {S.~B.}\ \bibnamefont
  {Pope}},\ }\href@noop {} {\emph {\bibinfo {title} {Turbulent Flows}}}\
  (\bibinfo  {publisher} {Cambridge University Press},\ \bibinfo {year}
  {2000})\BibitemShut {NoStop}%
\bibitem [{\citenamefont {Sagaut}(2005)}]{Sagaut2005}%
  \BibitemOpen
  \bibfield  {author} {\bibinfo {author} {\bibfnamefont {P.}~\bibnamefont
  {Sagaut}},\ }\href@noop {} {\emph {\bibinfo {title} {Large eddy simulation
  for incompressible flows: an introduction}}}\ (\bibinfo  {publisher}
  {Springer Science \& Business Media},\ \bibinfo {year} {2005})\BibitemShut
  {NoStop}%
\bibitem [{\citenamefont {Germano}\ \emph {et~al.}(1991)\citenamefont
  {Germano}, \citenamefont {Piomelli}, \citenamefont {Moin},\ and\
  \citenamefont {Cabot}}]{Germano1991}%
  \BibitemOpen
  \bibfield  {author} {\bibinfo {author} {\bibfnamefont {M.}~\bibnamefont
  {Germano}}, \bibinfo {author} {\bibfnamefont {U.}~\bibnamefont {Piomelli}},
  \bibinfo {author} {\bibfnamefont {P.}~\bibnamefont {Moin}},\ and\ \bibinfo
  {author} {\bibfnamefont {W.~H.}\ \bibnamefont {Cabot}},\ }\bibfield  {title}
  {\bibinfo {title} {A dynamic subgrid-scale eddy viscosity model},\ }\href
  {https://doi.org/10.1063/1.857955} {\bibfield  {journal} {\bibinfo  {journal}
  {Phys. Fluids A: Fluid Dyn.}\ }\textbf {\bibinfo {volume} {3}},\ \bibinfo
  {pages} {1760} (\bibinfo {year} {1991})}\BibitemShut {NoStop}%
\bibitem [{\citenamefont {Long}\ and\ \citenamefont {Sun}(2001)}]{Long2001}%
  \BibitemOpen
  \bibfield  {author} {\bibinfo {author} {\bibfnamefont {G.-L.}\ \bibnamefont
  {Long}}\ and\ \bibinfo {author} {\bibfnamefont {Y.}~\bibnamefont {Sun}},\
  }\bibfield  {title} {\bibinfo {title} {Efficient scheme for initializing a
  quantum register with an arbitrary superposed state},\ }\href
  {https://doi.org/10.1103/PhysRevA.64.014303} {\bibfield  {journal} {\bibinfo
  {journal} {Phys. Rev. A}\ }\textbf {\bibinfo {volume} {64}},\ \bibinfo
  {pages} {014303} (\bibinfo {year} {2001})}\BibitemShut {NoStop}%
\bibitem [{\citenamefont {Gourianov}\ \emph {et~al.}(2022)\citenamefont
  {Gourianov}, \citenamefont {Lubasch}, \citenamefont {Dolgov}, \citenamefont
  {{van den Berg}}, \citenamefont {Babaee}, \citenamefont {Givi}, \citenamefont
  {Kiffner},\ and\ \citenamefont {Jaksch}}]{Gourianov2022}%
  \BibitemOpen
  \bibfield  {author} {\bibinfo {author} {\bibfnamefont {N.}~\bibnamefont
  {Gourianov}}, \bibinfo {author} {\bibfnamefont {M.}~\bibnamefont {Lubasch}},
  \bibinfo {author} {\bibfnamefont {S.}~\bibnamefont {Dolgov}}, \bibinfo
  {author} {\bibfnamefont {Q.~Y.}\ \bibnamefont {{van den Berg}}}, \bibinfo
  {author} {\bibfnamefont {H.}~\bibnamefont {Babaee}}, \bibinfo {author}
  {\bibfnamefont {P.}~\bibnamefont {Givi}}, \bibinfo {author} {\bibfnamefont
  {M.}~\bibnamefont {Kiffner}},\ and\ \bibinfo {author} {\bibfnamefont
  {D.}~\bibnamefont {Jaksch}},\ }\bibfield  {title} {\bibinfo {title} {A
  quantum-inspired approach to exploit turbulence structures},\ }\href
  {https://doi.org/10.1038/s43588-021-00181-1} {\bibfield  {journal} {\bibinfo
  {journal} {Nat. Comput. Sci.}\ }\textbf {\bibinfo {volume} {2}},\ \bibinfo
  {pages} {30} (\bibinfo {year} {2022})}\BibitemShut {NoStop}%
\bibitem [{\citenamefont {Jaksch}\ \emph {et~al.}(2023)\citenamefont {Jaksch},
  \citenamefont {Givi}, \citenamefont {Daley},\ and\ \citenamefont
  {Rung}}]{Jaksch2023}%
  \BibitemOpen
  \bibfield  {author} {\bibinfo {author} {\bibfnamefont {D.}~\bibnamefont
  {Jaksch}}, \bibinfo {author} {\bibfnamefont {P.}~\bibnamefont {Givi}},
  \bibinfo {author} {\bibfnamefont {A.~J.}\ \bibnamefont {Daley}},\ and\
  \bibinfo {author} {\bibfnamefont {T.}~\bibnamefont {Rung}},\ }\bibfield
  {title} {\bibinfo {title} {Variational quantum algorithms for computational
  fluid dynamics},\ }\href {https://doi.org/10.2514/1.J062426} {\bibfield
  {journal} {\bibinfo  {journal} {AIAA J.}\ }\textbf {\bibinfo {volume} {61}},\
  \bibinfo {pages} {1885} (\bibinfo {year} {2023})}\BibitemShut {NoStop}%
\bibitem [{\citenamefont {Givi}\ \emph {et~al.}(2020)\citenamefont {Givi},
  \citenamefont {Daley}, \citenamefont {Mavriplis},\ and\ \citenamefont
  {Malik}}]{Givi2020}%
  \BibitemOpen
  \bibfield  {author} {\bibinfo {author} {\bibfnamefont {P.}~\bibnamefont
  {Givi}}, \bibinfo {author} {\bibfnamefont {A.~J.}\ \bibnamefont {Daley}},
  \bibinfo {author} {\bibfnamefont {D.}~\bibnamefont {Mavriplis}},\ and\
  \bibinfo {author} {\bibfnamefont {M.}~\bibnamefont {Malik}},\ }\bibfield
  {title} {\bibinfo {title} {Quantum speedup for aeroscience and engineering},\
  }\href {https://doi.org/10.2514/1.J059183} {\bibfield  {journal} {\bibinfo
  {journal} {AIAA J.}\ }\textbf {\bibinfo {volume} {58}},\ \bibinfo {pages}
  {3715} (\bibinfo {year} {2020})}\BibitemShut {NoStop}%
\bibitem [{\citenamefont {Harrow}\ \emph {et~al.}(2009)\citenamefont {Harrow},
  \citenamefont {Hassidim},\ and\ \citenamefont {Lloyd}}]{Harrow2009}%
  \BibitemOpen
  \bibfield  {author} {\bibinfo {author} {\bibfnamefont {A.~W.}\ \bibnamefont
  {Harrow}}, \bibinfo {author} {\bibfnamefont {A.}~\bibnamefont {Hassidim}},\
  and\ \bibinfo {author} {\bibfnamefont {S.}~\bibnamefont {Lloyd}},\ }\bibfield
   {title} {\bibinfo {title} {Quantum algorithm for linear systems of
  equations},\ }\href {https://doi.org/10.1103/PhysRevLett.103.150502}
  {\bibfield  {journal} {\bibinfo  {journal} {Phys. Rev. Lett.}\ }\textbf
  {\bibinfo {volume} {103}},\ \bibinfo {pages} {150502} (\bibinfo {year}
  {2009})}\BibitemShut {NoStop}%
\bibitem [{\citenamefont {Ambainis}(2010)}]{Ambainis2010-arxiv}%
  \BibitemOpen
  \bibfield  {author} {\bibinfo {author} {\bibfnamefont {A.}~\bibnamefont
  {Ambainis}},\ }\href@noop {} {\bibinfo {title} {Variable time amplitude
  amplification and a faster quantum algorithm for solving systems of linear
  equations}} (\bibinfo {year} {2010}),\ \Eprint
  {https://arxiv.org/abs/1010.4458} {arXiv:1010.4458} \BibitemShut {NoStop}%
\bibitem [{\citenamefont {Childs}\ \emph {et~al.}(2017)\citenamefont {Childs},
  \citenamefont {Kothari},\ and\ \citenamefont {Somma}}]{Childs2017}%
  \BibitemOpen
  \bibfield  {author} {\bibinfo {author} {\bibfnamefont {A.~M.}\ \bibnamefont
  {Childs}}, \bibinfo {author} {\bibfnamefont {R.}~\bibnamefont {Kothari}},\
  and\ \bibinfo {author} {\bibfnamefont {R.~D.}\ \bibnamefont {Somma}},\
  }\bibfield  {title} {\bibinfo {title} {Quantum algorithm for systems of
  linear equations with exponentially improved dependence on precision},\
  }\href {https://doi.org/10.1137/16M1087072} {\bibfield  {journal} {\bibinfo
  {journal} {SIAM J. Comput.}\ }\textbf {\bibinfo {volume} {46}},\ \bibinfo
  {pages} {1920} (\bibinfo {year} {2017})}\BibitemShut {NoStop}%
\bibitem [{\citenamefont {Penuel}\ \emph {et~al.}(2024)\citenamefont {Penuel},
  \citenamefont {Katabarwa}, \citenamefont {Johnson}, \citenamefont {Farquhar},
  \citenamefont {Cao},\ and\ \citenamefont {Garrett}}]{Penuel2024}%
  \BibitemOpen
  \bibfield  {author} {\bibinfo {author} {\bibfnamefont {J.}~\bibnamefont
  {Penuel}}, \bibinfo {author} {\bibfnamefont {A.}~\bibnamefont {Katabarwa}},
  \bibinfo {author} {\bibfnamefont {P.~D.}\ \bibnamefont {Johnson}}, \bibinfo
  {author} {\bibfnamefont {C.}~\bibnamefont {Farquhar}}, \bibinfo {author}
  {\bibfnamefont {Y.}~\bibnamefont {Cao}},\ and\ \bibinfo {author}
  {\bibfnamefont {M.~C.}\ \bibnamefont {Garrett}},\ }\href@noop {} {\bibinfo
  {title} {Feasibility of accelerating incompressible computational fluid
  dynamics simulations with fault-tolerant quantum computers}} (\bibinfo {year}
  {2024}),\ \Eprint {https://arxiv.org/abs/2406.06323} {arXiv:2406.06323}
  \BibitemShut {NoStop}%
\bibitem [{\citenamefont {Lloyd}\ \emph {et~al.}(2020)\citenamefont {Lloyd},
  \citenamefont {Palma}, \citenamefont {Gokler}, \citenamefont {Kiani},
  \citenamefont {Liu}, \citenamefont {Marvian}, \citenamefont {Tennie},\ and\
  \citenamefont {Palmer}}]{Lloyd2020}%
  \BibitemOpen
  \bibfield  {author} {\bibinfo {author} {\bibfnamefont {S.}~\bibnamefont
  {Lloyd}}, \bibinfo {author} {\bibfnamefont {G.~D.}\ \bibnamefont {Palma}},
  \bibinfo {author} {\bibfnamefont {C.}~\bibnamefont {Gokler}}, \bibinfo
  {author} {\bibfnamefont {B.}~\bibnamefont {Kiani}}, \bibinfo {author}
  {\bibfnamefont {Z.-W.}\ \bibnamefont {Liu}}, \bibinfo {author} {\bibfnamefont
  {M.}~\bibnamefont {Marvian}}, \bibinfo {author} {\bibfnamefont
  {F.}~\bibnamefont {Tennie}},\ and\ \bibinfo {author} {\bibfnamefont
  {T.}~\bibnamefont {Palmer}},\ }\href@noop {} {\bibinfo {title} {Quantum
  algorithm for nonlinear differential equations}} (\bibinfo {year} {2020}),\
  \Eprint {https://arxiv.org/abs/2011.06571} {arXiv:2011.06571} \BibitemShut
  {NoStop}%
\bibitem [{\citenamefont {Brearley}\ and\ \citenamefont
  {Laizet}(2024)}]{Brearly2024}%
  \BibitemOpen
  \bibfield  {author} {\bibinfo {author} {\bibfnamefont {P.}~\bibnamefont
  {Brearley}}\ and\ \bibinfo {author} {\bibfnamefont {S.}~\bibnamefont
  {Laizet}},\ }\bibfield  {title} {\bibinfo {title} {Quantum algorithm for
  solving the advection equation using hamiltonian simulation},\ }\href
  {https://link.aps.org/doi/10.1103/PhysRevA.110.012430} {\bibfield  {journal}
  {\bibinfo  {journal} {Phys. Rev. A}\ }\textbf {\bibinfo {volume} {110}},\
  \bibinfo {pages} {012430} (\bibinfo {year} {2024})}\BibitemShut {NoStop}%
\bibitem [{\citenamefont {Over}\ \emph {et~al.}(2024)\citenamefont {Over},
  \citenamefont {Bengoechea}, \citenamefont {Brearley}, \citenamefont
  {Laizet},\ and\ \citenamefont {Rung}}]{over2024_diffusion}%
  \BibitemOpen
  \bibfield  {author} {\bibinfo {author} {\bibfnamefont {P.}~\bibnamefont
  {Over}}, \bibinfo {author} {\bibfnamefont {S.}~\bibnamefont {Bengoechea}},
  \bibinfo {author} {\bibfnamefont {P.}~\bibnamefont {Brearley}}, \bibinfo
  {author} {\bibfnamefont {S.}~\bibnamefont {Laizet}},\ and\ \bibinfo {author}
  {\bibfnamefont {T.}~\bibnamefont {Rung}},\ }\href@noop {} {\bibinfo {title}
  {Quantum algorithm for the advection-diffusion equation with optimal success
  probability}} (\bibinfo {year} {2024}),\ \Eprint
  {https://arxiv.org/abs/2410.07909} {arXiv:2410.07909} \BibitemShut {NoStop}%
\bibitem [{\citenamefont {Peruzzo}\ \emph {et~al.}(2014)\citenamefont
  {Peruzzo}, \citenamefont {McClean}, \citenamefont {Shadbolt}, \citenamefont
  {Yung}, \citenamefont {Zhou}, \citenamefont {Love}, \citenamefont
  {Aspuru-Guzik},\ and\ \citenamefont {O’Brien}}]{Peruzzo2014}%
  \BibitemOpen
  \bibfield  {author} {\bibinfo {author} {\bibfnamefont {A.}~\bibnamefont
  {Peruzzo}}, \bibinfo {author} {\bibfnamefont {J.}~\bibnamefont {McClean}},
  \bibinfo {author} {\bibfnamefont {P.}~\bibnamefont {Shadbolt}}, \bibinfo
  {author} {\bibfnamefont {M.-H.}\ \bibnamefont {Yung}}, \bibinfo {author}
  {\bibfnamefont {X.-Q.}\ \bibnamefont {Zhou}}, \bibinfo {author}
  {\bibfnamefont {P.~J.}\ \bibnamefont {Love}}, \bibinfo {author}
  {\bibfnamefont {A.}~\bibnamefont {Aspuru-Guzik}},\ and\ \bibinfo {author}
  {\bibfnamefont {J.~L.}\ \bibnamefont {O’Brien}},\ }\bibfield  {title}
  {\bibinfo {title} {A variational eigenvalue solver on a photonic quantum
  processor},\ }\href {https://doi.org/10.1038/ncomms5213} {\bibfield
  {journal} {\bibinfo  {journal} {Nat. Commun.}\ }\textbf {\bibinfo {volume}
  {5}} (\bibinfo {year} {2014})}\BibitemShut {NoStop}%
\bibitem [{\citenamefont {Kandala}\ \emph {et~al.}(2017)\citenamefont
  {Kandala}, \citenamefont {Mezzacapo}, \citenamefont {Temme}, \citenamefont
  {Takita}, \citenamefont {Brink}, \citenamefont {Chow},\ and\ \citenamefont
  {Gambetta}}]{Kandala2017}%
  \BibitemOpen
  \bibfield  {author} {\bibinfo {author} {\bibfnamefont {A.}~\bibnamefont
  {Kandala}}, \bibinfo {author} {\bibfnamefont {A.}~\bibnamefont {Mezzacapo}},
  \bibinfo {author} {\bibfnamefont {K.}~\bibnamefont {Temme}}, \bibinfo
  {author} {\bibfnamefont {M.}~\bibnamefont {Takita}}, \bibinfo {author}
  {\bibfnamefont {M.}~\bibnamefont {Brink}}, \bibinfo {author} {\bibfnamefont
  {J.~M.}\ \bibnamefont {Chow}},\ and\ \bibinfo {author} {\bibfnamefont
  {J.~M.}\ \bibnamefont {Gambetta}},\ }\bibfield  {title} {\bibinfo {title}
  {Hardware-efficient variational quantum eigensolver for small molecules and
  quantum magnets},\ }\href {https://api.semanticscholar.org/CorpusID:4390182}
  {\bibfield  {journal} {\bibinfo  {journal} {Nature}\ }\textbf {\bibinfo
  {volume} {549}},\ \bibinfo {pages} {242} (\bibinfo {year}
  {2017})}\BibitemShut {NoStop}%
\bibitem [{\citenamefont {Cerezo}\ \emph {et~al.}(2022)\citenamefont {Cerezo},
  \citenamefont {Sharma}, \citenamefont {Arrasmith},\ and\ \citenamefont
  {Coles}}]{Cerezo2022}%
  \BibitemOpen
  \bibfield  {author} {\bibinfo {author} {\bibfnamefont {M.}~\bibnamefont
  {Cerezo}}, \bibinfo {author} {\bibfnamefont {K.}~\bibnamefont {Sharma}},
  \bibinfo {author} {\bibfnamefont {A.}~\bibnamefont {Arrasmith}},\ and\
  \bibinfo {author} {\bibfnamefont {P.}~\bibnamefont {Coles}},\ }\bibfield
  {title} {\bibinfo {title} {Variational quantum state eigensolver},\ }\href
  {https://doi.org/10.1038/s41534-022-00611-6} {\bibfield  {journal} {\bibinfo
  {journal} {Npj Quantum Inf.}\ }\textbf {\bibinfo {volume} {8}},\ \bibinfo
  {pages} {113} (\bibinfo {year} {2022})}\BibitemShut {NoStop}%
\bibitem [{\citenamefont {AI}\ and\ \citenamefont
  {Collaborators}(2025)}]{Acharya2024}%
  \BibitemOpen
  \bibfield  {author} {\bibinfo {author} {\bibfnamefont {G.~Q.}\ \bibnamefont
  {AI}}\ and\ \bibinfo {author} {\bibnamefont {Collaborators}},\ }\bibfield
  {title} {\bibinfo {title} {Quantum error correction below the surface code
  threshold},\ }\href@noop {} {\bibfield  {journal} {\bibinfo  {journal}
  {Nature}\ }\textbf {\bibinfo {volume} {638}},\ \bibinfo {pages} {920–926}
  (\bibinfo {year} {2025})}\BibitemShut {NoStop}%
\bibitem [{\citenamefont {Reichardt}\ \emph {et~al.}(2024)\citenamefont
  {Reichardt}, \citenamefont {Aasen}, \citenamefont {Chao}, \citenamefont
  {Chernoguzov}, \citenamefont {van Dam}, \citenamefont {Gaebler},
  \citenamefont {Gresh}, \citenamefont {Lucchetti}, \citenamefont {Mills},
  \citenamefont {Moses}, \citenamefont {Neyenhuis}, \citenamefont {Paetznick},
  \citenamefont {Paz}, \citenamefont {Siegfried}, \citenamefont {da~Silva},
  \citenamefont {Svore}, \citenamefont {Wang},\ and\ \citenamefont
  {Zanner}}]{Reichardt2024-arxiv}%
  \BibitemOpen
  \bibfield  {author} {\bibinfo {author} {\bibfnamefont {B.~W.}\ \bibnamefont
  {Reichardt}}, \bibinfo {author} {\bibfnamefont {D.}~\bibnamefont {Aasen}},
  \bibinfo {author} {\bibfnamefont {R.}~\bibnamefont {Chao}}, \bibinfo {author}
  {\bibfnamefont {A.}~\bibnamefont {Chernoguzov}}, \bibinfo {author}
  {\bibfnamefont {W.}~\bibnamefont {van Dam}}, \bibinfo {author} {\bibfnamefont
  {J.~P.}\ \bibnamefont {Gaebler}}, \bibinfo {author} {\bibfnamefont
  {D.}~\bibnamefont {Gresh}}, \bibinfo {author} {\bibfnamefont
  {D.}~\bibnamefont {Lucchetti}}, \bibinfo {author} {\bibfnamefont
  {M.}~\bibnamefont {Mills}}, \bibinfo {author} {\bibfnamefont {S.~A.}\
  \bibnamefont {Moses}}, \bibinfo {author} {\bibfnamefont {B.}~\bibnamefont
  {Neyenhuis}}, \bibinfo {author} {\bibfnamefont {A.}~\bibnamefont
  {Paetznick}}, \bibinfo {author} {\bibfnamefont {A.}~\bibnamefont {Paz}},
  \bibinfo {author} {\bibfnamefont {P.~E.}\ \bibnamefont {Siegfried}}, \bibinfo
  {author} {\bibfnamefont {M.~P.}\ \bibnamefont {da~Silva}}, \bibinfo {author}
  {\bibfnamefont {K.~M.}\ \bibnamefont {Svore}}, \bibinfo {author}
  {\bibfnamefont {Z.}~\bibnamefont {Wang}},\ and\ \bibinfo {author}
  {\bibfnamefont {M.}~\bibnamefont {Zanner}},\ }\href@noop {} {\bibinfo {title}
  {Demonstration of quantum computation and error correction with a tesseract
  code}} (\bibinfo {year} {2024}),\ \Eprint {https://arxiv.org/abs/2409.04628}
  {arXiv:2409.04628} \BibitemShut {NoStop}%
\bibitem [{\citenamefont {van Dam}\ \emph {et~al.}(2024)\citenamefont {van
  Dam}, \citenamefont {Liu}, \citenamefont {Low}, \citenamefont {Paetznick},
  \citenamefont {Paz}, \citenamefont {Silva}, \citenamefont {Sundaram},
  \citenamefont {Svore},\ and\ \citenamefont {Troyer}}]{Vandam2024-arxiv}%
  \BibitemOpen
  \bibfield  {author} {\bibinfo {author} {\bibfnamefont {W.}~\bibnamefont {van
  Dam}}, \bibinfo {author} {\bibfnamefont {H.}~\bibnamefont {Liu}}, \bibinfo
  {author} {\bibfnamefont {G.~H.}\ \bibnamefont {Low}}, \bibinfo {author}
  {\bibfnamefont {A.}~\bibnamefont {Paetznick}}, \bibinfo {author}
  {\bibfnamefont {A.}~\bibnamefont {Paz}}, \bibinfo {author} {\bibfnamefont
  {M.}~\bibnamefont {Silva}}, \bibinfo {author} {\bibfnamefont
  {A.}~\bibnamefont {Sundaram}}, \bibinfo {author} {\bibfnamefont
  {K.}~\bibnamefont {Svore}},\ and\ \bibinfo {author} {\bibfnamefont
  {M.}~\bibnamefont {Troyer}},\ }\href@noop {} {\bibinfo {title} {End-to-end
  quantum simulation of a chemical system}} (\bibinfo {year} {2024}),\ \Eprint
  {https://arxiv.org/abs/2409.05835} {arXiv:2409.05835} \BibitemShut {NoStop}%
\bibitem [{Que()}]{QueraRoadmap}%
  \BibitemOpen
  \href@noop {} {\bibinfo {title} {Quera computing releases a groundbreaking
  roadmap for advanced error-corrected quantum computers, pioneering the next
  frontier in quantum innovation}},\ \bibinfo {howpublished}
  {\url{https://www.quera.com/press-releases/quera-computing-releases-a-groundbreaking-roadmap-for-advanced-error-corrected-quantum-computers-pioneering-the-next-frontier-in-quantum-innovation-0}},\
  \bibinfo {note} {visited on 2025-10-06}\BibitemShut {NoStop}%
\bibitem [{IBM()}]{IBMRoadmap}%
  \BibitemOpen
  \href@noop {} {\bibinfo {title} {Ibm technology atlas quantum roadmap}},\
  \bibinfo {howpublished} {\url{https://www.ibm.com/roadmaps/quantum.pdf}},\
  \bibinfo {note} {visited on 2024-10-10}\BibitemShut {NoStop}%
\bibitem [{\citenamefont {Lubasch}\ \emph {et~al.}(2020)\citenamefont
  {Lubasch}, \citenamefont {Joo}, \citenamefont {Moinier}, \citenamefont
  {Kiffner},\ and\ \citenamefont {Jaksch}}]{Lubasch2020}%
  \BibitemOpen
  \bibfield  {author} {\bibinfo {author} {\bibfnamefont {M.}~\bibnamefont
  {Lubasch}}, \bibinfo {author} {\bibfnamefont {J.}~\bibnamefont {Joo}},
  \bibinfo {author} {\bibfnamefont {P.}~\bibnamefont {Moinier}}, \bibinfo
  {author} {\bibfnamefont {M.}~\bibnamefont {Kiffner}},\ and\ \bibinfo {author}
  {\bibfnamefont {D.}~\bibnamefont {Jaksch}},\ }\bibfield  {title} {\bibinfo
  {title} {Variational quantum algorithms for nonlinear problems},\ }\href
  {https://doi.org/10.1103/PhysRevA.101.010301} {\bibfield  {journal} {\bibinfo
   {journal} {Phys. Rev. A}\ }\textbf {\bibinfo {volume} {101}},\ \bibinfo
  {pages} {010301} (\bibinfo {year} {2020})}\BibitemShut {NoStop}%
\bibitem [{\citenamefont {Schilling}\ \emph {et~al.}(2024)\citenamefont
  {Schilling}, \citenamefont {Preti}, \citenamefont {M\"uller}, \citenamefont
  {Calarco},\ and\ \citenamefont {Motzoi}}]{Schilling2024}%
  \BibitemOpen
  \bibfield  {author} {\bibinfo {author} {\bibfnamefont {M.}~\bibnamefont
  {Schilling}}, \bibinfo {author} {\bibfnamefont {F.}~\bibnamefont {Preti}},
  \bibinfo {author} {\bibfnamefont {M.~M.}\ \bibnamefont {M\"uller}}, \bibinfo
  {author} {\bibfnamefont {T.}~\bibnamefont {Calarco}},\ and\ \bibinfo {author}
  {\bibfnamefont {F.}~\bibnamefont {Motzoi}},\ }\bibfield  {title} {\bibinfo
  {title} {Exponentiation of parametric hamiltonians via unitary
  interpolation},\ }\href
  {https://link.aps.org/doi/10.1103/PhysRevResearch.6.043278} {\bibfield
  {journal} {\bibinfo  {journal} {Phys. Rev. Res.}\ }\textbf {\bibinfo {volume}
  {6}},\ \bibinfo {pages} {043278} (\bibinfo {year} {2024})}\BibitemShut
  {NoStop}%
\bibitem [{\citenamefont {Huggins}\ \emph {et~al.}(2020)\citenamefont
  {Huggins}, \citenamefont {Lee}, \citenamefont {Baek}, \citenamefont
  {O’Gorman},\ and\ \citenamefont {Whaley}}]{Huggins2020}%
  \BibitemOpen
  \bibfield  {author} {\bibinfo {author} {\bibfnamefont {W.~J.}\ \bibnamefont
  {Huggins}}, \bibinfo {author} {\bibfnamefont {J.}~\bibnamefont {Lee}},
  \bibinfo {author} {\bibfnamefont {U.}~\bibnamefont {Baek}}, \bibinfo {author}
  {\bibfnamefont {B.}~\bibnamefont {O’Gorman}},\ and\ \bibinfo {author}
  {\bibfnamefont {K.~B.}\ \bibnamefont {Whaley}},\ }\bibfield  {title}
  {\bibinfo {title} {A non-orthogonal variational quantum eigensolver},\ }\href
  {https://dx.doi.org/10.1088/1367-2630/ab867b} {\bibfield  {journal} {\bibinfo
   {journal} {New J. Phys.}\ }\textbf {\bibinfo {volume} {22}},\ \bibinfo
  {pages} {073009} (\bibinfo {year} {2020})}\BibitemShut {NoStop}%
\bibitem [{\citenamefont {Umer}\ \emph {et~al.}(2025)\citenamefont {Umer},
  \citenamefont {Mastorakis}, \citenamefont {Evangelou},\ and\ \citenamefont
  {Angelakis}}]{Umer2025}%
  \BibitemOpen
  \bibfield  {author} {\bibinfo {author} {\bibfnamefont {M.}~\bibnamefont
  {Umer}}, \bibinfo {author} {\bibfnamefont {E.}~\bibnamefont {Mastorakis}},
  \bibinfo {author} {\bibfnamefont {S.}~\bibnamefont {Evangelou}},\ and\
  \bibinfo {author} {\bibfnamefont {D.~G.}\ \bibnamefont {Angelakis}},\
  }\bibfield  {title} {\bibinfo {title} {Probing the limits of variational
  quantum algorithms for nonlinear ground states on real quantum hardware: The
  effects of noise},\ }\href
  {https://link.aps.org/doi/10.1103/PhysRevA.111.012626} {\bibfield  {journal}
  {\bibinfo  {journal} {Phys. Rev. A}\ }\textbf {\bibinfo {volume} {111}},\
  \bibinfo {pages} {012626} (\bibinfo {year} {2025})}\BibitemShut {NoStop}%
\bibitem [{\citenamefont {Pool}\ \emph {et~al.}(2024)\citenamefont {Pool},
  \citenamefont {Somoza}, \citenamefont {Mc~Keever}, \citenamefont {Lubasch},\
  and\ \citenamefont {Horstmann}}]{Pool2024}%
  \BibitemOpen
  \bibfield  {author} {\bibinfo {author} {\bibfnamefont {A.~J.}\ \bibnamefont
  {Pool}}, \bibinfo {author} {\bibfnamefont {A.~D.}\ \bibnamefont {Somoza}},
  \bibinfo {author} {\bibfnamefont {C.}~\bibnamefont {Mc~Keever}}, \bibinfo
  {author} {\bibfnamefont {M.}~\bibnamefont {Lubasch}},\ and\ \bibinfo {author}
  {\bibfnamefont {B.}~\bibnamefont {Horstmann}},\ }\bibfield  {title} {\bibinfo
  {title} {Nonlinear dynamics as a ground-state solution on quantum
  computers},\ }\href
  {https://link.aps.org/doi/10.1103/PhysRevResearch.6.033257} {\bibfield
  {journal} {\bibinfo  {journal} {Phys. Rev. Res.}\ }\textbf {\bibinfo {volume}
  {6}},\ \bibinfo {pages} {033257} (\bibinfo {year} {2024})}\BibitemShut
  {NoStop}%
\bibitem [{\citenamefont {Shao}\ \emph {et~al.}(2024)\citenamefont {Shao},
  \citenamefont {Wei}, \citenamefont {Cheng},\ and\ \citenamefont
  {Liu}}]{Yuguo2024}%
  \BibitemOpen
  \bibfield  {author} {\bibinfo {author} {\bibfnamefont {Y.}~\bibnamefont
  {Shao}}, \bibinfo {author} {\bibfnamefont {F.}~\bibnamefont {Wei}}, \bibinfo
  {author} {\bibfnamefont {S.}~\bibnamefont {Cheng}},\ and\ \bibinfo {author}
  {\bibfnamefont {Z.}~\bibnamefont {Liu}},\ }\bibfield  {title} {\bibinfo
  {title} {Simulating noisy variational quantum algorithms: A polynomial
  approach},\ }\href {https://doi.org/10.1103/PhysRevLett.133.120603}
  {\bibfield  {journal} {\bibinfo  {journal} {Phys. Rev. Lett.}\ }\textbf
  {\bibinfo {volume} {133}},\ \bibinfo {pages} {120603} (\bibinfo {year}
  {2024})}\BibitemShut {NoStop}%
\bibitem [{\citenamefont {Rosenberg}\ \emph {et~al.}(2022)\citenamefont
  {Rosenberg}, \citenamefont {Ginsparg},\ and\ \citenamefont
  {McMahon}}]{Rosenberg2022}%
  \BibitemOpen
  \bibfield  {author} {\bibinfo {author} {\bibfnamefont {E.}~\bibnamefont
  {Rosenberg}}, \bibinfo {author} {\bibfnamefont {P.}~\bibnamefont
  {Ginsparg}},\ and\ \bibinfo {author} {\bibfnamefont {P.~L.}\ \bibnamefont
  {McMahon}},\ }\bibfield  {title} {\bibinfo {title} {Experimental error
  mitigation using linear rescaling for variational quantum eigensolving with
  up to 20 qubits},\ }\href {https://doi.org/10.1088/2058-9565/ac3b37}
  {\bibfield  {journal} {\bibinfo  {journal} {Quantum Science and Technology}\
  }\textbf {\bibinfo {volume} {7}},\ \bibinfo {pages} {015024} (\bibinfo {year}
  {2022})}\BibitemShut {NoStop}%
\bibitem [{\citenamefont {Jaderberg}\ \emph {et~al.}(2020)\citenamefont
  {Jaderberg}, \citenamefont {Agarwal}, \citenamefont {Leonhardt},
  \citenamefont {Kiffner},\ and\ \citenamefont {Jaksch}}]{Jaderberg2020}%
  \BibitemOpen
  \bibfield  {author} {\bibinfo {author} {\bibfnamefont {B.}~\bibnamefont
  {Jaderberg}}, \bibinfo {author} {\bibfnamefont {A.}~\bibnamefont {Agarwal}},
  \bibinfo {author} {\bibfnamefont {K.}~\bibnamefont {Leonhardt}}, \bibinfo
  {author} {\bibfnamefont {M.}~\bibnamefont {Kiffner}},\ and\ \bibinfo {author}
  {\bibfnamefont {D.}~\bibnamefont {Jaksch}},\ }\bibfield  {title} {\bibinfo
  {title} {Minimum hardware requirements for hybrid quantum–classical dmft},\
  }\href {https://dx.doi.org/10.1088/2058-9565/ab972b} {\bibfield  {journal}
  {\bibinfo  {journal} {Quantum Sci. Technol.}\ }\textbf {\bibinfo {volume}
  {5}},\ \bibinfo {pages} {034015} (\bibinfo {year} {2020})}\BibitemShut
  {NoStop}%
\bibitem [{\citenamefont {Jaderberg}\ \emph {et~al.}(2022)\citenamefont
  {Jaderberg}, \citenamefont {Eisfeld}, \citenamefont {Jaksch},\ and\
  \citenamefont {Mostame}}]{Jaderberg2022}%
  \BibitemOpen
  \bibfield  {author} {\bibinfo {author} {\bibfnamefont {B.}~\bibnamefont
  {Jaderberg}}, \bibinfo {author} {\bibfnamefont {A.}~\bibnamefont {Eisfeld}},
  \bibinfo {author} {\bibfnamefont {D.}~\bibnamefont {Jaksch}},\ and\ \bibinfo
  {author} {\bibfnamefont {S.}~\bibnamefont {Mostame}},\ }\bibfield  {title}
  {\bibinfo {title} {Recompilation-enhanced simulation of electron–phonon
  dynamics on ibm quantum computers},\ }\href
  {https://dx.doi.org/10.1088/1367-2630/ac8a69} {\bibfield  {journal} {\bibinfo
   {journal} {New J. Phys.}\ }\textbf {\bibinfo {volume} {24}},\ \bibinfo
  {pages} {093017} (\bibinfo {year} {2022})}\BibitemShut {NoStop}%
\bibitem [{\citenamefont {Kim}\ \emph {et~al.}(2023)\citenamefont {Kim},
  \citenamefont {Eddins}, \citenamefont {Anand}, \citenamefont {Wei},
  \citenamefont {van~den Berg}, \citenamefont {Rosenblatt}, \citenamefont
  {Nayfeh}, \citenamefont {Wu}, \citenamefont {Zaletel}, \citenamefont
  {Temme},\ and\ \citenamefont {Kandala}}]{Kim2023}%
  \BibitemOpen
  \bibfield  {author} {\bibinfo {author} {\bibfnamefont {Y.}~\bibnamefont
  {Kim}}, \bibinfo {author} {\bibfnamefont {A.}~\bibnamefont {Eddins}},
  \bibinfo {author} {\bibfnamefont {S.}~\bibnamefont {Anand}}, \bibinfo
  {author} {\bibfnamefont {K.~X.}\ \bibnamefont {Wei}}, \bibinfo {author}
  {\bibfnamefont {E.}~\bibnamefont {van~den Berg}}, \bibinfo {author}
  {\bibfnamefont {S.}~\bibnamefont {Rosenblatt}}, \bibinfo {author}
  {\bibfnamefont {H.}~\bibnamefont {Nayfeh}}, \bibinfo {author} {\bibfnamefont
  {Y.}~\bibnamefont {Wu}}, \bibinfo {author} {\bibfnamefont {M.}~\bibnamefont
  {Zaletel}}, \bibinfo {author} {\bibfnamefont {K.}~\bibnamefont {Temme}},\
  and\ \bibinfo {author} {\bibfnamefont {A.}~\bibnamefont {Kandala}},\
  }\bibfield  {title} {\bibinfo {title} {Evidence for the utility of quantum
  computing before fault tolerance},\ }\href
  {https://doi.org/10.1038/s41586-023-06096-3} {\bibfield  {journal} {\bibinfo
  {journal} {Nature}\ }\textbf {\bibinfo {volume} {618}},\ \bibinfo {pages}
  {500} (\bibinfo {year} {2023})}\BibitemShut {NoStop}%
\bibitem [{\citenamefont {Oseledets}(2011)}]{Oseledets2011}%
  \BibitemOpen
  \bibfield  {author} {\bibinfo {author} {\bibfnamefont {I.~V.}\ \bibnamefont
  {Oseledets}},\ }\bibfield  {title} {\bibinfo {title} {Tensor-train
  decomposition},\ }\href {https://doi.org/10.1137/090752286} {\bibfield
  {journal} {\bibinfo  {journal} {SIAM J. Sci. Comput.}\ }\textbf {\bibinfo
  {volume} {33}},\ \bibinfo {pages} {2295} (\bibinfo {year}
  {2011})}\BibitemShut {NoStop}%
\bibitem [{\citenamefont {Orús}(2014)}]{Orus2014}%
  \BibitemOpen
  \bibfield  {author} {\bibinfo {author} {\bibfnamefont {R.}~\bibnamefont
  {Orús}},\ }\bibfield  {title} {\bibinfo {title} {A practical introduction to
  tensor networks: Matrix product states and projected entangled pair states},\
  }\href {https://www.sciencedirect.com/science/article/pii/S0003491614001596}
  {\bibfield  {journal} {\bibinfo  {journal} {Ann. Phys.}\ }\textbf {\bibinfo
  {volume} {349}},\ \bibinfo {pages} {117} (\bibinfo {year}
  {2014})}\BibitemShut {NoStop}%
\bibitem [{\citenamefont {Oseledets}(2013)}]{Oseledets2013}%
  \BibitemOpen
  \bibfield  {author} {\bibinfo {author} {\bibfnamefont {I.~V.}\ \bibnamefont
  {Oseledets}},\ }\bibfield  {title} {\bibinfo {title} {Constructive
  representation of functions in low-rank tensor formats},\ }\href
  {https://doi.org/10.1007/s00365-012-9175-x} {\bibfield  {journal} {\bibinfo
  {journal} {Constructive Approximation}\ }\textbf {\bibinfo {volume} {37}},\
  \bibinfo {pages} {1} (\bibinfo {year} {2013})}\BibitemShut {NoStop}%
\bibitem [{\citenamefont {Ye}\ and\ \citenamefont {Loureiro}(2022)}]{Ye2022}%
  \BibitemOpen
  \bibfield  {author} {\bibinfo {author} {\bibfnamefont {E.}~\bibnamefont
  {Ye}}\ and\ \bibinfo {author} {\bibfnamefont {N.~F.~G.}\ \bibnamefont
  {Loureiro}},\ }\bibfield  {title} {\bibinfo {title} {Quantum-inspired method
  for solving the vlasov-poisson equations},\ }\href
  {https://link.aps.org/doi/10.1103/PhysRevE.106.035208} {\bibfield  {journal}
  {\bibinfo  {journal} {Phys. Rev. E}\ }\textbf {\bibinfo {volume} {106}},\
  \bibinfo {pages} {035208} (\bibinfo {year} {2022})}\BibitemShut {NoStop}%
\bibitem [{\citenamefont {Kiffner}\ and\ \citenamefont
  {Jaksch}(2023)}]{Kiffner2023}%
  \BibitemOpen
  \bibfield  {author} {\bibinfo {author} {\bibfnamefont {M.}~\bibnamefont
  {Kiffner}}\ and\ \bibinfo {author} {\bibfnamefont {D.}~\bibnamefont
  {Jaksch}},\ }\bibfield  {title} {\bibinfo {title} {Tensor network reduced
  order models for wall-bounded flows},\ }\href
  {https://doi.org/10.1103/PhysRevFluids.8.124101} {\bibfield  {journal}
  {\bibinfo  {journal} {Phys. Rev. Fluids}\ }\textbf {\bibinfo {volume} {8}},\
  \bibinfo {pages} {124101} (\bibinfo {year} {2023})}\BibitemShut {NoStop}%
\bibitem [{\citenamefont {Kornev}\ \emph {et~al.}(2023)\citenamefont {Kornev},
  \citenamefont {Dolgov}, \citenamefont {Pinto}, \citenamefont {Pflitsch},
  \citenamefont {Perelshtein},\ and\ \citenamefont
  {Melnikov}}]{Kornev2023-arxiv}%
  \BibitemOpen
  \bibfield  {author} {\bibinfo {author} {\bibfnamefont {E.}~\bibnamefont
  {Kornev}}, \bibinfo {author} {\bibfnamefont {S.}~\bibnamefont {Dolgov}},
  \bibinfo {author} {\bibfnamefont {K.}~\bibnamefont {Pinto}}, \bibinfo
  {author} {\bibfnamefont {M.}~\bibnamefont {Pflitsch}}, \bibinfo {author}
  {\bibfnamefont {M.}~\bibnamefont {Perelshtein}},\ and\ \bibinfo {author}
  {\bibfnamefont {A.}~\bibnamefont {Melnikov}},\ }\href@noop {} {\bibinfo
  {title} {Numerical solution of the incompressible navier-stokes equations for
  chemical mixers via quantum-inspired tensor train finite element method}}
  (\bibinfo {year} {2023}),\ \Eprint {https://arxiv.org/abs/2305.10784}
  {arXiv:2305.10784} \BibitemShut {NoStop}%
\bibitem [{\citenamefont {Ye}\ and\ \citenamefont {Loureiro}(2024)}]{Ye2024}%
  \BibitemOpen
  \bibfield  {author} {\bibinfo {author} {\bibfnamefont {E.}~\bibnamefont
  {Ye}}\ and\ \bibinfo {author} {\bibfnamefont {N.~F.}\ \bibnamefont
  {Loureiro}},\ }\bibfield  {title} {\bibinfo {title} {{Quantized tensor
  networks for solving the Vlasov\textendash{}Maxwell equations}},\ }\href
  {https://doi.org/10.1017/S0022377824000503} {\bibfield  {journal} {\bibinfo
  {journal} {J. Plasma Phys.}\ }\textbf {\bibinfo {volume} {90}},\ \bibinfo
  {pages} {805900301} (\bibinfo {year} {2024})}\BibitemShut {NoStop}%
\bibitem [{\citenamefont {Peddinti}\ \emph {et~al.}(2024)\citenamefont
  {Peddinti}, \citenamefont {Pisoni}, \citenamefont {Marini}, \citenamefont
  {Lott}, \citenamefont {Argentieri}, \citenamefont {Tiunov},\ and\
  \citenamefont {Aolita}}]{Peddinti2024}%
  \BibitemOpen
  \bibfield  {author} {\bibinfo {author} {\bibfnamefont {R.~D.}\ \bibnamefont
  {Peddinti}}, \bibinfo {author} {\bibfnamefont {S.}~\bibnamefont {Pisoni}},
  \bibinfo {author} {\bibfnamefont {A.}~\bibnamefont {Marini}}, \bibinfo
  {author} {\bibfnamefont {P.}~\bibnamefont {Lott}}, \bibinfo {author}
  {\bibfnamefont {H.}~\bibnamefont {Argentieri}}, \bibinfo {author}
  {\bibfnamefont {E.}~\bibnamefont {Tiunov}},\ and\ \bibinfo {author}
  {\bibfnamefont {L.}~\bibnamefont {Aolita}},\ }\bibfield  {title} {\bibinfo
  {title} {Quantum-inspired framework for computational fluid dynamics},\
  }\href {https://doi.org/10.1038/s42005-024-01623-8} {\bibfield  {journal}
  {\bibinfo  {journal} {Commun. Phys.}\ }\textbf {\bibinfo {volume} {7}},\
  \bibinfo {pages} {135} (\bibinfo {year} {2024})}\BibitemShut {NoStop}%
\bibitem [{\citenamefont {H\"olscher}\ \emph {et~al.}(2025)\citenamefont
  {H\"olscher}, \citenamefont {Rao}, \citenamefont {M\"uller}, \citenamefont
  {Klepsch}, \citenamefont {Luckow}, \citenamefont {Stollenwerk},\ and\
  \citenamefont {Wilhelm}}]{Hölscher2025}%
  \BibitemOpen
  \bibfield  {author} {\bibinfo {author} {\bibfnamefont {L.}~\bibnamefont
  {H\"olscher}}, \bibinfo {author} {\bibfnamefont {P.}~\bibnamefont {Rao}},
  \bibinfo {author} {\bibfnamefont {L.}~\bibnamefont {M\"uller}}, \bibinfo
  {author} {\bibfnamefont {J.}~\bibnamefont {Klepsch}}, \bibinfo {author}
  {\bibfnamefont {A.}~\bibnamefont {Luckow}}, \bibinfo {author} {\bibfnamefont
  {T.}~\bibnamefont {Stollenwerk}},\ and\ \bibinfo {author} {\bibfnamefont
  {F.~K.}\ \bibnamefont {Wilhelm}},\ }\bibfield  {title} {\bibinfo {title}
  {Quantum-inspired fluid simulation of two-dimensional turbulence with gpu
  acceleration},\ }\href
  {https://link.aps.org/doi/10.1103/PhysRevResearch.7.013112} {\bibfield
  {journal} {\bibinfo  {journal} {Phys. Rev. Res.}\ }\textbf {\bibinfo {volume}
  {7}},\ \bibinfo {pages} {013112} (\bibinfo {year} {2025})}\BibitemShut
  {NoStop}%
\bibitem [{\citenamefont {Gourianov}\ \emph {et~al.}(2025)\citenamefont
  {Gourianov}, \citenamefont {Givi}, \citenamefont {Jaksch},\ and\
  \citenamefont {Pope}}]{Gourianov2024}%
  \BibitemOpen
  \bibfield  {author} {\bibinfo {author} {\bibfnamefont {N.}~\bibnamefont
  {Gourianov}}, \bibinfo {author} {\bibfnamefont {P.}~\bibnamefont {Givi}},
  \bibinfo {author} {\bibfnamefont {D.}~\bibnamefont {Jaksch}},\ and\ \bibinfo
  {author} {\bibfnamefont {S.~B.}\ \bibnamefont {Pope}},\ }\bibfield  {title}
  {\bibinfo {title} {Tensor networks enable the calculation of turbulence
  probability distributions},\ }\href
  {https://www.science.org/doi/abs/10.1126/sciadv.ads5990} {\bibfield
  {journal} {\bibinfo  {journal} {Sci. Adv.}\ }\textbf {\bibinfo {volume}
  {11}},\ \bibinfo {pages} {eads5990} (\bibinfo {year} {2025})}\BibitemShut
  {NoStop}%
\bibitem [{\citenamefont {Schachenmayer}\ \emph {et~al.}(2013)\citenamefont
  {Schachenmayer}, \citenamefont {Lanyon}, \citenamefont {Roos},\ and\
  \citenamefont {Daley}}]{Schachenmeyer2013}%
  \BibitemOpen
  \bibfield  {author} {\bibinfo {author} {\bibfnamefont {J.}~\bibnamefont
  {Schachenmayer}}, \bibinfo {author} {\bibfnamefont {B.~P.}\ \bibnamefont
  {Lanyon}}, \bibinfo {author} {\bibfnamefont {C.~F.}\ \bibnamefont {Roos}},\
  and\ \bibinfo {author} {\bibfnamefont {A.~J.}\ \bibnamefont {Daley}},\
  }\bibfield  {title} {\bibinfo {title} {Entanglement growth in quench dynamics
  with variable range interactions},\ }\href
  {https://doi.org/10.1103/PhysRevX.3.031015} {\bibfield  {journal} {\bibinfo
  {journal} {Phys. Rev. X}\ }\textbf {\bibinfo {volume} {3}},\ \bibinfo {pages}
  {031015} (\bibinfo {year} {2013})}\BibitemShut {NoStop}%
\bibitem [{\citenamefont {Grover}(1996)}]{Grover1996}%
  \BibitemOpen
  \bibfield  {author} {\bibinfo {author} {\bibfnamefont {L.~K.}\ \bibnamefont
  {Grover}},\ }\bibfield  {title} {\bibinfo {title} {A fast quantum mechanical
  algorithm for database search},\ }in\ \href
  {https://doi.org/10.1145/237814.237866} {\emph {\bibinfo {booktitle}
  {Proceedings of the Twenty-Eighth Annual ACM Symposium on Theory of
  Computing}}},\ \bibinfo {series and number} {STOC '96}\ (\bibinfo
  {publisher} {Association for Computing Machinery},\ \bibinfo {address} {New
  York, NY, USA},\ \bibinfo {year} {1996})\ p.\ \bibinfo {pages}
  {212–219}\BibitemShut {NoStop}%
\bibitem [{\citenamefont {Ran}(2020)}]{Ran2020}%
  \BibitemOpen
  \bibfield  {author} {\bibinfo {author} {\bibfnamefont {S.-J.}\ \bibnamefont
  {Ran}},\ }\bibfield  {title} {\bibinfo {title} {Encoding of matrix product
  states into quantum circuits of one- and two-qubit gates},\ }\href
  {https://link.aps.org/doi/10.1103/PhysRevA.101.032310} {\bibfield  {journal}
  {\bibinfo  {journal} {Phys. Rev. A}\ }\textbf {\bibinfo {volume} {101}},\
  \bibinfo {pages} {032310} (\bibinfo {year} {2020})}\BibitemShut {NoStop}%
\bibitem [{\citenamefont {Malz}\ \emph {et~al.}(2024)\citenamefont {Malz},
  \citenamefont {Styliaris}, \citenamefont {Wei},\ and\ \citenamefont
  {Cirac}}]{Malz2024}%
  \BibitemOpen
  \bibfield  {author} {\bibinfo {author} {\bibfnamefont {D.}~\bibnamefont
  {Malz}}, \bibinfo {author} {\bibfnamefont {G.}~\bibnamefont {Styliaris}},
  \bibinfo {author} {\bibfnamefont {Z.-Y.}\ \bibnamefont {Wei}},\ and\ \bibinfo
  {author} {\bibfnamefont {J.~I.}\ \bibnamefont {Cirac}},\ }\bibfield  {title}
  {\bibinfo {title} {Preparation of matrix product states with log-depth
  quantum circuits},\ }\href {https://doi.org/10.1103/PhysRevLett.132.040404}
  {\bibfield  {journal} {\bibinfo  {journal} {Phys. Rev. Lett.}\ }\textbf
  {\bibinfo {volume} {132}},\ \bibinfo {pages} {040404} (\bibinfo {year}
  {2024})}\BibitemShut {NoStop}%
\bibitem [{\citenamefont {Smith}\ \emph {et~al.}(2024)\citenamefont {Smith},
  \citenamefont {Khan}, \citenamefont {Clark}, \citenamefont {Girvin},\ and\
  \citenamefont {Wei}}]{Smith2024}%
  \BibitemOpen
  \bibfield  {author} {\bibinfo {author} {\bibfnamefont {K.~C.}\ \bibnamefont
  {Smith}}, \bibinfo {author} {\bibfnamefont {A.}~\bibnamefont {Khan}},
  \bibinfo {author} {\bibfnamefont {B.~K.}\ \bibnamefont {Clark}}, \bibinfo
  {author} {\bibfnamefont {S.}~\bibnamefont {Girvin}},\ and\ \bibinfo {author}
  {\bibfnamefont {T.-C.}\ \bibnamefont {Wei}},\ }\bibfield  {title} {\bibinfo
  {title} {Constant-depth preparation of matrix product states with adaptive
  quantum circuits},\ }\href
  {https://link.aps.org/doi/10.1103/PRXQuantum.5.030344} {\bibfield  {journal}
  {\bibinfo  {journal} {PRX Quantum}\ }\textbf {\bibinfo {volume} {5}},\
  \bibinfo {pages} {030344} (\bibinfo {year} {2024})}\BibitemShut {NoStop}%
\bibitem [{\citenamefont {Nibbi}\ and\ \citenamefont
  {Mendl}(2024)}]{Nibbi2024}%
  \BibitemOpen
  \bibfield  {author} {\bibinfo {author} {\bibfnamefont {M.}~\bibnamefont
  {Nibbi}}\ and\ \bibinfo {author} {\bibfnamefont {C.~B.}\ \bibnamefont
  {Mendl}},\ }\bibfield  {title} {\bibinfo {title} {Block encoding of matrix
  product operators},\ }\href
  {https://link.aps.org/doi/10.1103/PhysRevA.110.042427} {\bibfield  {journal}
  {\bibinfo  {journal} {Phys. Rev. A}\ }\textbf {\bibinfo {volume} {110}},\
  \bibinfo {pages} {042427} (\bibinfo {year} {2024})}\BibitemShut {NoStop}%
\bibitem [{\citenamefont {Termanova}\ \emph {et~al.}(2024)\citenamefont
  {Termanova}, \citenamefont {Melnikov}, \citenamefont {Mamenchikov},
  \citenamefont {Belokonev}, \citenamefont {Dolgov}, \citenamefont
  {Berezutskii}, \citenamefont {Ellerbrock}, \citenamefont {Mansell},\ and\
  \citenamefont {Perelshtein}}]{Termanova2024}%
  \BibitemOpen
  \bibfield  {author} {\bibinfo {author} {\bibfnamefont {A.}~\bibnamefont
  {Termanova}}, \bibinfo {author} {\bibfnamefont {A.}~\bibnamefont {Melnikov}},
  \bibinfo {author} {\bibfnamefont {E.}~\bibnamefont {Mamenchikov}}, \bibinfo
  {author} {\bibfnamefont {N.}~\bibnamefont {Belokonev}}, \bibinfo {author}
  {\bibfnamefont {S.}~\bibnamefont {Dolgov}}, \bibinfo {author} {\bibfnamefont
  {A.}~\bibnamefont {Berezutskii}}, \bibinfo {author} {\bibfnamefont
  {R.}~\bibnamefont {Ellerbrock}}, \bibinfo {author} {\bibfnamefont
  {C.}~\bibnamefont {Mansell}},\ and\ \bibinfo {author} {\bibfnamefont
  {M.}~\bibnamefont {Perelshtein}},\ }\bibfield  {title} {\bibinfo {title}
  {Tensor quantum programming},\ }\href
  {https://iopscience.iop.org/article/10.1088/1367-2630/ad985b/meta} {\bibfield
   {journal} {\bibinfo  {journal} {New J. Phys.}\ }\textbf {\bibinfo {volume}
  {26}},\ \bibinfo {pages} {123019} (\bibinfo {year} {2024})}\BibitemShut
  {NoStop}%
\bibitem [{\citenamefont {Goldack}\ \emph {et~al.}(2025)\citenamefont
  {Goldack}, \citenamefont {Atia}, \citenamefont {Alberton},\ and\
  \citenamefont {Jansen}}]{Goldack2025}%
  \BibitemOpen
  \bibfield  {author} {\bibinfo {author} {\bibfnamefont {M.}~\bibnamefont
  {Goldack}}, \bibinfo {author} {\bibfnamefont {Y.}~\bibnamefont {Atia}},
  \bibinfo {author} {\bibfnamefont {O.}~\bibnamefont {Alberton}},\ and\
  \bibinfo {author} {\bibfnamefont {K.}~\bibnamefont {Jansen}},\ }\href@noop {}
  {\bibinfo {title} {Computing statistical properties of velocity fields on
  current quantum hardware}} (\bibinfo {year} {2025})\BibitemShut {NoStop}%
\bibitem [{\citenamefont {Uchida}\ \emph {et~al.}(2024)\citenamefont {Uchida},
  \citenamefont {Miyamoto}, \citenamefont {Yamazaki}, \citenamefont
  {Fujisawa},\ and\ \citenamefont {Yoshida}}]{Uchida2024}%
  \BibitemOpen
  \bibfield  {author} {\bibinfo {author} {\bibfnamefont {F.}~\bibnamefont
  {Uchida}}, \bibinfo {author} {\bibfnamefont {K.}~\bibnamefont {Miyamoto}},
  \bibinfo {author} {\bibfnamefont {S.}~\bibnamefont {Yamazaki}}, \bibinfo
  {author} {\bibfnamefont {K.}~\bibnamefont {Fujisawa}},\ and\ \bibinfo
  {author} {\bibfnamefont {N.}~\bibnamefont {Yoshida}},\ }\href
  {https://arxiv.org/abs/2412.17206} {\bibinfo {title} {Quantum simulation of
  burgers turbulence: Nonlinear transformation and direct evaluation of
  statistical quantities}} (\bibinfo {year} {2024}),\ \Eprint
  {https://arxiv.org/abs/2412.17206} {arXiv:2412.17206 [quant-ph]} \BibitemShut
  {NoStop}%
\bibitem [{\citenamefont {Jesus~et al.}(2025)}]{Jesus2025prep}%
  \BibitemOpen
  \bibfield  {author} {\bibinfo {author} {\bibfnamefont {J.~D.}\ \bibnamefont
  {Jesus~et al.}},\ }\href@noop {} {\bibinfo {title} {in preparation}}
  (\bibinfo {year} {2025})\BibitemShut {NoStop}%
\bibitem [{\citenamefont {Sarma}\ \emph {et~al.}(2024)\citenamefont {Sarma},
  \citenamefont {Watts}, \citenamefont {Moosa}, \citenamefont {Liu},\ and\
  \citenamefont {McMahon}}]{Sarma2024}%
  \BibitemOpen
  \bibfield  {author} {\bibinfo {author} {\bibfnamefont {A.}~\bibnamefont
  {Sarma}}, \bibinfo {author} {\bibfnamefont {T.~W.}\ \bibnamefont {Watts}},
  \bibinfo {author} {\bibfnamefont {M.}~\bibnamefont {Moosa}}, \bibinfo
  {author} {\bibfnamefont {Y.}~\bibnamefont {Liu}},\ and\ \bibinfo {author}
  {\bibfnamefont {P.~L.}\ \bibnamefont {McMahon}},\ }\bibfield  {title}
  {\bibinfo {title} {Quantum variational solving of nonlinear and
  multidimensional partial differential equations},\ }\href
  {https://link.aps.org/doi/10.1103/PhysRevA.109.062616} {\bibfield  {journal}
  {\bibinfo  {journal} {Phys. Rev. A}\ }\textbf {\bibinfo {volume} {109}},\
  \bibinfo {pages} {062616} (\bibinfo {year} {2024})}\BibitemShut {NoStop}%
\bibitem [{\citenamefont {Over}\ \emph {et~al.}(2025)\citenamefont {Over},
  \citenamefont {Bengoechea}, \citenamefont {Rung}, \citenamefont {Clerici},
  \citenamefont {Scandurra}, \citenamefont {{de Villiers}},\ and\ \citenamefont
  {Jaksch}}]{Over2024}%
  \BibitemOpen
  \bibfield  {author} {\bibinfo {author} {\bibfnamefont {P.}~\bibnamefont
  {Over}}, \bibinfo {author} {\bibfnamefont {S.}~\bibnamefont {Bengoechea}},
  \bibinfo {author} {\bibfnamefont {T.}~\bibnamefont {Rung}}, \bibinfo {author}
  {\bibfnamefont {F.}~\bibnamefont {Clerici}}, \bibinfo {author} {\bibfnamefont
  {L.}~\bibnamefont {Scandurra}}, \bibinfo {author} {\bibfnamefont
  {E.}~\bibnamefont {{de Villiers}}},\ and\ \bibinfo {author} {\bibfnamefont
  {D.}~\bibnamefont {Jaksch}},\ }\bibfield  {title} {\bibinfo {title} {Boundary
  treatment for variational quantum simulations of partial differential
  equations on quantum computers},\ }\href
  {https://www.sciencedirect.com/science/article/pii/S0045793024003396}
  {\bibfield  {journal} {\bibinfo  {journal} {Comput. Fluids}\ }\textbf
  {\bibinfo {volume} {288}},\ \bibinfo {pages} {106508} (\bibinfo {year}
  {2025})}\BibitemShut {NoStop}%
\bibitem [{\citenamefont {Schuld}\ and\ \citenamefont
  {Petruccione}(2021)}]{Schuld2021}%
  \BibitemOpen
  \bibfield  {author} {\bibinfo {author} {\bibfnamefont {M.}~\bibnamefont
  {Schuld}}\ and\ \bibinfo {author} {\bibfnamefont {F.}~\bibnamefont
  {Petruccione}},\ }\bibinfo {title} {Representing data on a quantum
  computer},\ in\ \href {https://doi.org/10.1007/978-3-030-83098-4_4} {\emph
  {\bibinfo {booktitle} {Machine Learning with Quantum Computers}}}\ (\bibinfo
  {publisher} {Springer International Publishing},\ \bibinfo {address} {Cham},\
  \bibinfo {year} {2021})\ pp.\ \bibinfo {pages} {147--176}\BibitemShut
  {NoStop}%
\bibitem [{\citenamefont {Ben-Dov}\ \emph {et~al.}(2024)\citenamefont
  {Ben-Dov}, \citenamefont {Shnaiderov}, \citenamefont {Makmal},\ and\
  \citenamefont {Dalla~Torre}}]{Ben-Dov2024}%
  \BibitemOpen
  \bibfield  {author} {\bibinfo {author} {\bibfnamefont {M.}~\bibnamefont
  {Ben-Dov}}, \bibinfo {author} {\bibfnamefont {D.}~\bibnamefont {Shnaiderov}},
  \bibinfo {author} {\bibfnamefont {A.}~\bibnamefont {Makmal}},\ and\ \bibinfo
  {author} {\bibfnamefont {E.~G.}\ \bibnamefont {Dalla~Torre}},\ }\bibfield
  {title} {\bibinfo {title} {Approximate encoding of quantum states using
  shallow circuits},\ }\href {https://doi.org/10.1038/s41534-024-00858-1}
  {\bibfield  {journal} {\bibinfo  {journal} {npj Quantum Information}\
  }\textbf {\bibinfo {volume} {10}},\ \bibinfo {pages} {65} (\bibinfo {year}
  {2024})}\BibitemShut {NoStop}%
\bibitem [{\citenamefont {Kazeev}\ and\ \citenamefont
  {Khoromskij}(2012)}]{Kazeev2012}%
  \BibitemOpen
  \bibfield  {author} {\bibinfo {author} {\bibfnamefont {V.~A.}\ \bibnamefont
  {Kazeev}}\ and\ \bibinfo {author} {\bibfnamefont {B.~N.}\ \bibnamefont
  {Khoromskij}},\ }\bibfield  {title} {\bibinfo {title} {Low-rank explicit qtt
  representation of the laplace operator and its inverse},\ }\href
  {https://doi.org/10.1137/100820479} {\bibfield  {journal} {\bibinfo
  {journal} {SIAM J. Matrix Anal. Appl.}\ }\textbf {\bibinfo {volume} {33}},\
  \bibinfo {pages} {742} (\bibinfo {year} {2012})}\BibitemShut {NoStop}%
\bibitem [{\citenamefont {Oseledets}(2010)}]{Oseledets2010}%
  \BibitemOpen
  \bibfield  {author} {\bibinfo {author} {\bibfnamefont {I.~V.}\ \bibnamefont
  {Oseledets}},\ }\bibfield  {title} {\bibinfo {title} {Approximation of
  $2^d\times2^d$ matrices using tensor decomposition},\ }\href
  {https://doi.org/10.1137/090757861} {\bibfield  {journal} {\bibinfo
  {journal} {SIAM J. Matrix Anal. Appl.}\ }\textbf {\bibinfo {volume} {31}},\
  \bibinfo {pages} {2130} (\bibinfo {year} {2010})}\BibitemShut {NoStop}%
\bibitem [{\citenamefont {Zhao}\ \emph {et~al.}(2023)\citenamefont {Zhao},
  \citenamefont {Bukov}, \citenamefont {Heyl},\ and\ \citenamefont
  {Moessner}}]{Zhao2023}%
  \BibitemOpen
  \bibfield  {author} {\bibinfo {author} {\bibfnamefont {H.}~\bibnamefont
  {Zhao}}, \bibinfo {author} {\bibfnamefont {M.}~\bibnamefont {Bukov}},
  \bibinfo {author} {\bibfnamefont {M.}~\bibnamefont {Heyl}},\ and\ \bibinfo
  {author} {\bibfnamefont {R.}~\bibnamefont {Moessner}},\ }\bibfield  {title}
  {\bibinfo {title} {Making trotterization adaptive and energy-self-correcting
  for nisq devices and beyond},\ }\href
  {https://doi.org/10.1103/PRXQuantum.4.030319} {\bibfield  {journal} {\bibinfo
   {journal} {PRX Quantum}\ }\textbf {\bibinfo {volume} {4}},\ \bibinfo {pages}
  {030319} (\bibinfo {year} {2023})}\BibitemShut {NoStop}%
\bibitem [{\citenamefont {Puig}\ \emph {et~al.}(2025)\citenamefont {Puig},
  \citenamefont {Drudis}, \citenamefont {Thanasilp},\ and\ \citenamefont
  {Holmes}}]{Puig2025}%
  \BibitemOpen
  \bibfield  {author} {\bibinfo {author} {\bibfnamefont {R.}~\bibnamefont
  {Puig}}, \bibinfo {author} {\bibfnamefont {M.}~\bibnamefont {Drudis}},
  \bibinfo {author} {\bibfnamefont {S.}~\bibnamefont {Thanasilp}},\ and\
  \bibinfo {author} {\bibfnamefont {Z.}~\bibnamefont {Holmes}},\ }\bibfield
  {title} {\bibinfo {title} {Variational quantum simulation: A case study for
  understanding warm starts},\ }\href
  {https://doi.org/10.1103/PRXQuantum.6.010317} {\bibfield  {journal} {\bibinfo
   {journal} {PRX Quantum}\ }\textbf {\bibinfo {volume} {6}},\ \bibinfo {pages}
  {010317} (\bibinfo {year} {2025})}\BibitemShut {NoStop}%
\bibitem [{\citenamefont {Mitarai}\ \emph {et~al.}(2018)\citenamefont
  {Mitarai}, \citenamefont {Negoro}, \citenamefont {Kitagawa},\ and\
  \citenamefont {Fujii}}]{Mitarai2018}%
  \BibitemOpen
  \bibfield  {author} {\bibinfo {author} {\bibfnamefont {K.}~\bibnamefont
  {Mitarai}}, \bibinfo {author} {\bibfnamefont {M.}~\bibnamefont {Negoro}},
  \bibinfo {author} {\bibfnamefont {M.}~\bibnamefont {Kitagawa}},\ and\
  \bibinfo {author} {\bibfnamefont {K.}~\bibnamefont {Fujii}},\ }\bibfield
  {title} {\bibinfo {title} {Quantum circuit learning},\ }\href
  {https://doi.org/10.1103/PhysRevA.98.032309} {\bibfield  {journal} {\bibinfo
  {journal} {Phys. Rev. A}\ }\textbf {\bibinfo {volume} {98}},\ \bibinfo
  {pages} {032309} (\bibinfo {year} {2018})}\BibitemShut {NoStop}%
\bibitem [{\citenamefont {Schuld}\ \emph {et~al.}(2019)\citenamefont {Schuld},
  \citenamefont {Bergholm}, \citenamefont {Gogolin}, \citenamefont {Izaac},\
  and\ \citenamefont {Killoran}}]{Schuld2018}%
  \BibitemOpen
  \bibfield  {author} {\bibinfo {author} {\bibfnamefont {M.}~\bibnamefont
  {Schuld}}, \bibinfo {author} {\bibfnamefont {V.}~\bibnamefont {Bergholm}},
  \bibinfo {author} {\bibfnamefont {C.}~\bibnamefont {Gogolin}}, \bibinfo
  {author} {\bibfnamefont {J.}~\bibnamefont {Izaac}},\ and\ \bibinfo {author}
  {\bibfnamefont {N.}~\bibnamefont {Killoran}},\ }\bibfield  {title} {\bibinfo
  {title} {Evaluating analytic gradients on quantum hardware},\ }\href
  {https://doi.org/10.1103/PhysRevA.99.032331} {\bibfield  {journal} {\bibinfo
  {journal} {Phys. Rev. A}\ }\textbf {\bibinfo {volume} {99}},\ \bibinfo
  {pages} {032331} (\bibinfo {year} {2019})}\BibitemShut {NoStop}%
\bibitem [{\citenamefont {Spall}(1998)}]{Spall1998}%
  \BibitemOpen
  \bibfield  {author} {\bibinfo {author} {\bibfnamefont {J.}~\bibnamefont
  {Spall}},\ }\href@noop {} {\emph {\bibinfo {title} {An overview of the
  simultaneous perturbation method for efficient optimization}}}\ (\bibinfo
  {publisher} {Johns Hopkins apl technical digest},\ \bibinfo {year} {1998})\
  p.\ \bibinfo {pages} {482–492}\BibitemShut {NoStop}%
\bibitem [{\citenamefont {Chorin}\ \emph {et~al.}(1979)\citenamefont {Chorin},
  \citenamefont {Marsden},\ and\ \citenamefont {Leonard}}]{Chorin1979}%
  \BibitemOpen
  \bibfield  {author} {\bibinfo {author} {\bibfnamefont {A.~J.}\ \bibnamefont
  {Chorin}}, \bibinfo {author} {\bibfnamefont {J.~E.}\ \bibnamefont
  {Marsden}},\ and\ \bibinfo {author} {\bibfnamefont {A.}~\bibnamefont
  {Leonard}},\ }\bibfield  {title} {\bibinfo {title} {A mathematical
  introduction to fluid mechanics}\ }(\bibinfo {year} {1979})\BibitemShut
  {NoStop}%
\bibitem [{\citenamefont {Peric}\ and\ \citenamefont
  {Abdel-Maksoud}(2015)}]{Peric2015}%
  \BibitemOpen
  \bibfield  {author} {\bibinfo {author} {\bibfnamefont {R.}~\bibnamefont
  {Peric}}\ and\ \bibinfo {author} {\bibfnamefont {M.}~\bibnamefont
  {Abdel-Maksoud}},\ }\bibfield  {title} {\bibinfo {title} {Reliable damping of
  free surface waves in numerical simulations},\ }\href
  {https://doi.org/10.1080/09377255.2015.1119921} {\bibfield  {journal}
  {\bibinfo  {journal} {Sh. Technol. Res.}\ }\textbf {\bibinfo {volume} {63}}
  (\bibinfo {year} {2015})}\BibitemShut {NoStop}%
\bibitem [{\citenamefont {Harlow}\ and\ \citenamefont
  {Welch}(1965)}]{harlow1965}%
  \BibitemOpen
  \bibfield  {author} {\bibinfo {author} {\bibfnamefont {F.~H.}\ \bibnamefont
  {Harlow}}\ and\ \bibinfo {author} {\bibfnamefont {J.~E.}\ \bibnamefont
  {Welch}},\ }\bibfield  {title} {\bibinfo {title} {Numerical {{Calculation}}
  of {{Time-Dependent Viscous Incompressible Flow}} of {{Fluid}} with {{Free
  Surface}}},\ }\href {https://doi.org/10.1063/1.1761178} {\bibfield  {journal}
  {\bibinfo  {journal} {Phys. Fluids}\ }\textbf {\bibinfo {volume} {8}},\
  \bibinfo {pages} {2182} (\bibinfo {year} {1965})}\BibitemShut {NoStop}%
\bibitem [{\citenamefont {Nielsen}\ and\ \citenamefont
  {Chuang}(2010)}]{nielsen_chuang2010}%
  \BibitemOpen
  \bibfield  {author} {\bibinfo {author} {\bibfnamefont {M.}~\bibnamefont
  {Nielsen}}\ and\ \bibinfo {author} {\bibfnamefont {I.}~\bibnamefont
  {Chuang}},\ }\href@noop {} {\emph {\bibinfo {title} {Quantum Computation and
  Quantum Information: 10th Anniversary Edition}}}\ (\bibinfo  {publisher}
  {Cambridge University Press},\ \bibinfo {year} {2010})\BibitemShut {NoStop}%
\bibitem [{\citenamefont {Haghshenas}\ \emph {et~al.}(2022)\citenamefont
  {Haghshenas}, \citenamefont {Gray}, \citenamefont {Potter},\ and\
  \citenamefont {Chan}}]{Haghshenas2022}%
  \BibitemOpen
  \bibfield  {author} {\bibinfo {author} {\bibfnamefont {R.}~\bibnamefont
  {Haghshenas}}, \bibinfo {author} {\bibfnamefont {J.}~\bibnamefont {Gray}},
  \bibinfo {author} {\bibfnamefont {A.~C.}\ \bibnamefont {Potter}},\ and\
  \bibinfo {author} {\bibfnamefont {G.~K.-L.}\ \bibnamefont {Chan}},\
  }\bibfield  {title} {\bibinfo {title} {Variational power of quantum circuit
  tensor networks},\ }\href {https://doi.org/10.1103/PhysRevX.12.011047}
  {\bibfield  {journal} {\bibinfo  {journal} {Phys. Rev. X}\ }\textbf {\bibinfo
  {volume} {12}},\ \bibinfo {pages} {011047} (\bibinfo {year}
  {2022})}\BibitemShut {NoStop}%
\bibitem [{\citenamefont {Wang}\ \emph {et~al.}(2019)\citenamefont {Wang},
  \citenamefont {Higgott},\ and\ \citenamefont {Brierley}}]{Wang2019}%
  \BibitemOpen
  \bibfield  {author} {\bibinfo {author} {\bibfnamefont {D.}~\bibnamefont
  {Wang}}, \bibinfo {author} {\bibfnamefont {O.}~\bibnamefont {Higgott}},\ and\
  \bibinfo {author} {\bibfnamefont {S.}~\bibnamefont {Brierley}},\ }\bibfield
  {title} {\bibinfo {title} {Accelerated variational quantum eigensolver},\
  }\href {https://doi.org/10.1103/PhysRevLett.122.140504} {\bibfield  {journal}
  {\bibinfo  {journal} {Phys. Rev. Lett.}\ }\textbf {\bibinfo {volume} {122}},\
  \bibinfo {pages} {140504} (\bibinfo {year} {2019})}\BibitemShut {NoStop}%
\bibitem [{\citenamefont {Grinko}\ \emph {et~al.}(2021)\citenamefont {Grinko},
  \citenamefont {Gacon}, \citenamefont {Zoufal},\ and\ \citenamefont
  {Woerner}}]{Grinko2021}%
  \BibitemOpen
  \bibfield  {author} {\bibinfo {author} {\bibfnamefont {D.}~\bibnamefont
  {Grinko}}, \bibinfo {author} {\bibfnamefont {J.}~\bibnamefont {Gacon}},
  \bibinfo {author} {\bibfnamefont {C.}~\bibnamefont {Zoufal}},\ and\ \bibinfo
  {author} {\bibfnamefont {S.}~\bibnamefont {Woerner}},\ }\bibfield  {title}
  {\bibinfo {title} {Iterative quantum amplitude estimation},\ }\href
  {https://doi.org/10.1038/s41534-021-00379-1} {\bibfield  {journal} {\bibinfo
  {journal} {npj Quantum Information}\ }\textbf {\bibinfo {volume} {7}},\
  \bibinfo {pages} {52} (\bibinfo {year} {2021})}\BibitemShut {NoStop}%
\bibitem [{\citenamefont {Li}\ \emph {et~al.}(2008)\citenamefont {Li},
  \citenamefont {Perlman}, \citenamefont {Wan}, \citenamefont {Yang},
  \citenamefont {Meneveau}, \citenamefont {Burns}, \citenamefont {Chen},
  \citenamefont {Szalay},\ and\ \citenamefont {Eyink}}]{John_Hopkins_Turb}%
  \BibitemOpen
  \bibfield  {author} {\bibinfo {author} {\bibfnamefont {Y.}~\bibnamefont
  {Li}}, \bibinfo {author} {\bibfnamefont {E.}~\bibnamefont {Perlman}},
  \bibinfo {author} {\bibfnamefont {M.}~\bibnamefont {Wan}}, \bibinfo {author}
  {\bibfnamefont {Y.}~\bibnamefont {Yang}}, \bibinfo {author} {\bibfnamefont
  {C.}~\bibnamefont {Meneveau}}, \bibinfo {author} {\bibfnamefont
  {R.}~\bibnamefont {Burns}}, \bibinfo {author} {\bibfnamefont
  {S.}~\bibnamefont {Chen}}, \bibinfo {author} {\bibfnamefont {A.}~\bibnamefont
  {Szalay}},\ and\ \bibinfo {author} {\bibfnamefont {G.}~\bibnamefont
  {Eyink}},\ }\bibfield  {title} {\bibinfo {title} {A public turbulence
  database cluster and applications to study lagrangian evolution of velocity
  increments in turbulence},\ }\href@noop {} {\bibfield  {journal} {\bibinfo
  {journal} {Journal of Turbulence}\ }\textbf {\bibinfo {volume} {9}},\
  \bibinfo {pages} {031015} (\bibinfo {year} {2008})}\BibitemShut {NoStop}%
\bibitem [{\citenamefont {Perlman}\ \emph {et~al.}(2007)\citenamefont
  {Perlman}, \citenamefont {Burns}, \citenamefont {Li},\ and\ \citenamefont
  {Meneveau}}]{John_Hopkins_Turb2}%
  \BibitemOpen
  \bibfield  {author} {\bibinfo {author} {\bibfnamefont {E.}~\bibnamefont
  {Perlman}}, \bibinfo {author} {\bibfnamefont {R.}~\bibnamefont {Burns}},
  \bibinfo {author} {\bibfnamefont {Y.}~\bibnamefont {Li}},\ and\ \bibinfo
  {author} {\bibfnamefont {C.}~\bibnamefont {Meneveau}},\ }\bibfield  {title}
  {\bibinfo {title} {Data exploration of turbulence simulations using a
  database cluster},\ }\href@noop {} {\bibfield  {journal} {\bibinfo  {journal}
  {Supercomputing SC07, ACM, IEEE}\ } (\bibinfo {year} {2007})}\BibitemShut
  {NoStop}%
\bibitem [{Joh()}]{John_Hopkins_Turb3}%
  \BibitemOpen
  \href {https://doi.org/https://doi.org/10.7281/T1KK98XB} {\bibinfo {title}
  {John hopkins turbulence database - isotropic turbulence}}\BibitemShut
  {NoStop}%
\bibitem [{\citenamefont {Schollwöck}(2011)}]{Schollw_ck_2011}%
  \BibitemOpen
  \bibfield  {author} {\bibinfo {author} {\bibfnamefont {U.}~\bibnamefont
  {Schollwöck}},\ }\bibfield  {title} {\bibinfo {title} {The density-matrix
  renormalization group in the age of matrix product states},\ }\href
  {http://dx.doi.org/10.1016/j.aop.2010.09.012} {\bibfield  {journal} {\bibinfo
   {journal} {Ann. Phys.}\ }\textbf {\bibinfo {volume} {326}},\ \bibinfo
  {pages} {96–192} (\bibinfo {year} {2011})}\BibitemShut {NoStop}%
\bibitem [{\citenamefont {Brandão}\ \emph {et~al.}(2016)\citenamefont
  {Brandão}, \citenamefont {Harrow},\ and\ \citenamefont
  {Horodecki}}]{Brandao2016}%
  \BibitemOpen
  \bibfield  {author} {\bibinfo {author} {\bibfnamefont {F.~G. S.~L.}\
  \bibnamefont {Brandão}}, \bibinfo {author} {\bibfnamefont {A.~W.}\
  \bibnamefont {Harrow}},\ and\ \bibinfo {author} {\bibfnamefont
  {M.}~\bibnamefont {Horodecki}},\ }\bibfield  {title} {\bibinfo {title} {Local
  random quantum circuits are approximate polynomial-designs},\ }\href
  {https://doi.org/10.1007/s00220-016-2706-8} {\bibfield  {journal} {\bibinfo
  {journal} {Communications in Mathematical Physics}\ }\textbf {\bibinfo
  {volume} {346}},\ \bibinfo {pages} {397} (\bibinfo {year}
  {2016})}\BibitemShut {NoStop}%
\bibitem [{\citenamefont {Goswami}\ \emph {et~al.}(2025)\citenamefont
  {Goswami}, \citenamefont {Schmelcher},\ and\ \citenamefont
  {Mukherjee}}]{Goswami2025-arxiv}%
  \BibitemOpen
  \bibfield  {author} {\bibinfo {author} {\bibfnamefont {K.}~\bibnamefont
  {Goswami}}, \bibinfo {author} {\bibfnamefont {P.}~\bibnamefont
  {Schmelcher}},\ and\ \bibinfo {author} {\bibfnamefont {R.}~\bibnamefont
  {Mukherjee}},\ }\href@noop {} {\bibinfo {title} {Qudit-based scalable quantum
  algorithm for solving the integer programming problem}} (\bibinfo {year}
  {2025}),\ \Eprint {https://arxiv.org/abs/2508.13906} {arXiv:2508.13906}
  \BibitemShut {NoStop}%
\bibitem [{\citenamefont {Shende}\ \emph {et~al.}(2004)\citenamefont {Shende},
  \citenamefont {Markov},\ and\ \citenamefont {Bullock}}]{Shende2004}%
  \BibitemOpen
  \bibfield  {author} {\bibinfo {author} {\bibfnamefont {V.~V.}\ \bibnamefont
  {Shende}}, \bibinfo {author} {\bibfnamefont {I.~L.}\ \bibnamefont {Markov}},\
  and\ \bibinfo {author} {\bibfnamefont {S.~S.}\ \bibnamefont {Bullock}},\
  }\bibfield  {title} {\bibinfo {title} {Minimal universal two-qubit
  controlled-not-based circuits},\ }\href
  {https://doi.org/10.1103/PhysRevA.69.062321} {\bibfield  {journal} {\bibinfo
  {journal} {Phys. Rev. A}\ }\textbf {\bibinfo {volume} {69}},\ \bibinfo
  {pages} {062321} (\bibinfo {year} {2004})}\BibitemShut {NoStop}%
\bibitem [{\citenamefont {Lubasch}\ \emph {et~al.}(2018)\citenamefont
  {Lubasch}, \citenamefont {Moinier},\ and\ \citenamefont
  {Jaksch}}]{Lubasch2018}%
  \BibitemOpen
  \bibfield  {author} {\bibinfo {author} {\bibfnamefont {M.}~\bibnamefont
  {Lubasch}}, \bibinfo {author} {\bibfnamefont {P.}~\bibnamefont {Moinier}},\
  and\ \bibinfo {author} {\bibfnamefont {D.}~\bibnamefont {Jaksch}},\
  }\bibfield  {title} {\bibinfo {title} {Multigrid renormalization},\ }\href
  {https://doi.org/https://doi.org/10.1016/j.jcp.2018.06.065} {\bibfield
  {journal} {\bibinfo  {journal} {Journal of Computational Physics}\ }\textbf
  {\bibinfo {volume} {372}},\ \bibinfo {pages} {587} (\bibinfo {year}
  {2018})}\BibitemShut {NoStop}%
\bibitem [{\citenamefont {Slattery}\ \emph {et~al.}(2022)\citenamefont
  {Slattery}, \citenamefont {Villalonga},\ and\ \citenamefont
  {Clark}}]{Slattery2022}%
  \BibitemOpen
  \bibfield  {author} {\bibinfo {author} {\bibfnamefont {L.}~\bibnamefont
  {Slattery}}, \bibinfo {author} {\bibfnamefont {B.}~\bibnamefont
  {Villalonga}},\ and\ \bibinfo {author} {\bibfnamefont {B.~K.}\ \bibnamefont
  {Clark}},\ }\bibfield  {title} {\bibinfo {title} {Unitary block optimization
  for variational quantum algorithms},\ }\href
  {https://doi.org/10.1103/PhysRevResearch.4.023072} {\bibfield  {journal}
  {\bibinfo  {journal} {Phys. Rev. Res.}\ }\textbf {\bibinfo {volume} {4}},\
  \bibinfo {pages} {023072} (\bibinfo {year} {2022})}\BibitemShut {NoStop}%
\bibitem [{\citenamefont {Kitaev}(1995)}]{Kitaev1995}%
  \BibitemOpen
  \bibfield  {author} {\bibinfo {author} {\bibfnamefont {A.~Y.}\ \bibnamefont
  {Kitaev}},\ }\href@noop {} {\bibinfo {title} {Quantum measurements and the
  abelian stabilizer problem}} (\bibinfo {year} {1995}),\ \Eprint
  {https://arxiv.org/abs/quant-ph/9511026} {arXiv:quant-ph/9511026}
  \BibitemShut {NoStop}%
\bibitem [{\citenamefont {Fomichev}\ \emph {et~al.}(2024)\citenamefont
  {Fomichev}, \citenamefont {Hejazi}, \citenamefont {Zini}, \citenamefont
  {Kiser}, \citenamefont {Fraxanet}, \citenamefont {Casares}, \citenamefont
  {Delgado}, \citenamefont {Huh}, \citenamefont {Voigt}, \citenamefont
  {Mueller},\ and\ \citenamefont {Arrazola}}]{Fomichev2023}%
  \BibitemOpen
  \bibfield  {author} {\bibinfo {author} {\bibfnamefont {S.}~\bibnamefont
  {Fomichev}}, \bibinfo {author} {\bibfnamefont {K.}~\bibnamefont {Hejazi}},
  \bibinfo {author} {\bibfnamefont {M.~S.}\ \bibnamefont {Zini}}, \bibinfo
  {author} {\bibfnamefont {M.}~\bibnamefont {Kiser}}, \bibinfo {author}
  {\bibfnamefont {J.}~\bibnamefont {Fraxanet}}, \bibinfo {author}
  {\bibfnamefont {P.~A.~M.}\ \bibnamefont {Casares}}, \bibinfo {author}
  {\bibfnamefont {A.}~\bibnamefont {Delgado}}, \bibinfo {author} {\bibfnamefont
  {J.}~\bibnamefont {Huh}}, \bibinfo {author} {\bibfnamefont {A.-C.}\
  \bibnamefont {Voigt}}, \bibinfo {author} {\bibfnamefont {J.~E.}\ \bibnamefont
  {Mueller}},\ and\ \bibinfo {author} {\bibfnamefont {J.~M.}\ \bibnamefont
  {Arrazola}},\ }\bibfield  {title} {\bibinfo {title} {Initial state
  preparation for quantum chemistry on quantum computers},\ }\href
  {https://link.aps.org/doi/10.1103/PRXQuantum.5.040339} {\bibfield  {journal}
  {\bibinfo  {journal} {PRX Quantum}\ }\textbf {\bibinfo {volume} {5}},\
  \bibinfo {pages} {040339} (\bibinfo {year} {2024})}\BibitemShut {NoStop}%
\bibitem [{\citenamefont {Siegl}\ \emph {et~al.}(2025)\citenamefont {Siegl},
  \citenamefont {Reese}, \citenamefont {Hashizume}, \citenamefont {van
  Hülst},\ and\ \citenamefont {Jaksch}}]{dataset}%
  \BibitemOpen
  \bibfield  {author} {\bibinfo {author} {\bibfnamefont {P.}~\bibnamefont
  {Siegl}}, \bibinfo {author} {\bibfnamefont {G.~S.}\ \bibnamefont {Reese}},
  \bibinfo {author} {\bibfnamefont {T.}~\bibnamefont {Hashizume}}, \bibinfo
  {author} {\bibfnamefont {N.-L.}\ \bibnamefont {van Hülst}},\ and\ \bibinfo
  {author} {\bibfnamefont {D.}~\bibnamefont {Jaksch}},\ }\href
  {https://doi.org/10.25592/uhhfdm.18010} {\bibinfo {title} {{Dataset
  Tensor-Programmable Quantum Circuits for Solving Differential Equations
  (Version 2)}}} (\bibinfo {year} {2025})\BibitemShut {NoStop}%
\bibitem [{\citenamefont {Bergholm~et al.}(2018)}]{Pennylane}%
  \BibitemOpen
  \bibfield  {author} {\bibinfo {author} {\bibfnamefont {V.}~\bibnamefont
  {Bergholm~et al.}},\ }\href@noop {} {\bibinfo {title} {Pennylane: Automatic
  differentiation of hybrid quantum-classical computations}} (\bibinfo {year}
  {2018}),\ \Eprint {https://arxiv.org/abs/1811.04968} {arXiv:1811.04968}
  \BibitemShut {NoStop}%
\bibitem [{\citenamefont {Paszke}\ \emph {et~al.}(2019)\citenamefont {Paszke},
  \citenamefont {Gross}, \citenamefont {Massa}, \citenamefont {Lerer},
  \citenamefont {Bradbury}, \citenamefont {Chanan}, \citenamefont {Killeen},
  \citenamefont {Lin}, \citenamefont {Gimelshein}, \citenamefont {Antiga},
  \citenamefont {Desmaison}, \citenamefont {K{\"o}pf}, \citenamefont {Yang},
  \citenamefont {DeVito}, \citenamefont {Raison}, \citenamefont {Tejani},
  \citenamefont {Chilamkurthy}, \citenamefont {Steiner}, \citenamefont {Fang},
  \citenamefont {Bai},\ and\ \citenamefont {Chintala}}]{Paszke2019}%
  \BibitemOpen
  \bibfield  {author} {\bibinfo {author} {\bibfnamefont {A.}~\bibnamefont
  {Paszke}}, \bibinfo {author} {\bibfnamefont {S.}~\bibnamefont {Gross}},
  \bibinfo {author} {\bibfnamefont {F.}~\bibnamefont {Massa}}, \bibinfo
  {author} {\bibfnamefont {A.}~\bibnamefont {Lerer}}, \bibinfo {author}
  {\bibfnamefont {J.}~\bibnamefont {Bradbury}}, \bibinfo {author}
  {\bibfnamefont {G.}~\bibnamefont {Chanan}}, \bibinfo {author} {\bibfnamefont
  {T.}~\bibnamefont {Killeen}}, \bibinfo {author} {\bibfnamefont
  {Z.}~\bibnamefont {Lin}}, \bibinfo {author} {\bibfnamefont {N.}~\bibnamefont
  {Gimelshein}}, \bibinfo {author} {\bibfnamefont {L.}~\bibnamefont {Antiga}},
  \bibinfo {author} {\bibfnamefont {A.}~\bibnamefont {Desmaison}}, \bibinfo
  {author} {\bibfnamefont {A.}~\bibnamefont {K{\"o}pf}}, \bibinfo {author}
  {\bibfnamefont {E.}~\bibnamefont {Yang}}, \bibinfo {author} {\bibfnamefont
  {Z.}~\bibnamefont {DeVito}}, \bibinfo {author} {\bibfnamefont
  {M.}~\bibnamefont {Raison}}, \bibinfo {author} {\bibfnamefont
  {A.}~\bibnamefont {Tejani}}, \bibinfo {author} {\bibfnamefont
  {S.}~\bibnamefont {Chilamkurthy}}, \bibinfo {author} {\bibfnamefont
  {B.}~\bibnamefont {Steiner}}, \bibinfo {author} {\bibfnamefont
  {L.}~\bibnamefont {Fang}}, \bibinfo {author} {\bibfnamefont {J.}~\bibnamefont
  {Bai}},\ and\ \bibinfo {author} {\bibfnamefont {S.}~\bibnamefont
  {Chintala}},\ }\href@noop {} {\bibinfo {title} {Pytorch: An imperative style,
  high-performance deep learning library}} (\bibinfo {year} {2019}),\ \Eprint
  {https://arxiv.org/abs/1912.01703} {arXiv:1912.01703} \BibitemShut {NoStop}%
\bibitem [{\citenamefont {Bradbury}\ \emph {et~al.}(2018)\citenamefont
  {Bradbury}, \citenamefont {Frostig}, \citenamefont {Hawkins}, \citenamefont
  {Johnson}, \citenamefont {Leary}, \citenamefont {Maclaurin}, \citenamefont
  {Necula}, \citenamefont {Paszke}, \citenamefont {Vander{P}las}, \citenamefont
  {Wanderman-{M}ilne},\ and\ \citenamefont {Zhang}}]{jax2018github}%
  \BibitemOpen
  \bibfield  {author} {\bibinfo {author} {\bibfnamefont {J.}~\bibnamefont
  {Bradbury}}, \bibinfo {author} {\bibfnamefont {R.}~\bibnamefont {Frostig}},
  \bibinfo {author} {\bibfnamefont {P.}~\bibnamefont {Hawkins}}, \bibinfo
  {author} {\bibfnamefont {M.~J.}\ \bibnamefont {Johnson}}, \bibinfo {author}
  {\bibfnamefont {C.}~\bibnamefont {Leary}}, \bibinfo {author} {\bibfnamefont
  {D.}~\bibnamefont {Maclaurin}}, \bibinfo {author} {\bibfnamefont
  {G.}~\bibnamefont {Necula}}, \bibinfo {author} {\bibfnamefont
  {A.}~\bibnamefont {Paszke}}, \bibinfo {author} {\bibfnamefont
  {J.}~\bibnamefont {Vander{P}las}}, \bibinfo {author} {\bibfnamefont
  {S.}~\bibnamefont {Wanderman-{M}ilne}},\ and\ \bibinfo {author}
  {\bibfnamefont {Q.}~\bibnamefont {Zhang}},\ }\href
  {http://github.com/jax-ml/jax} {\bibinfo {title} {{JAX}: composable
  transformations of {P}ython+{N}um{P}y programs}} (\bibinfo {year}
  {2018})\BibitemShut {NoStop}%
\bibitem [{\citenamefont {Luchnikov}\ \emph {et~al.}(2021)\citenamefont
  {Luchnikov}, \citenamefont {Krechetov},\ and\ \citenamefont
  {Filippov}}]{Luchnikov2021}%
  \BibitemOpen
  \bibfield  {author} {\bibinfo {author} {\bibfnamefont {I.~A.}\ \bibnamefont
  {Luchnikov}}, \bibinfo {author} {\bibfnamefont {M.~E.}\ \bibnamefont
  {Krechetov}},\ and\ \bibinfo {author} {\bibfnamefont {S.~N.}\ \bibnamefont
  {Filippov}},\ }\bibfield  {title} {\bibinfo {title} {Riemannian geometry and
  automatic differentiation for optimization problems of quantum physics and
  quantum technologies},\ }\href {https://doi.org/10.1088/1367-2630/ac0b02}
  {\bibfield  {journal} {\bibinfo  {journal} {New Jour. Phys.}\ }\textbf
  {\bibinfo {volume} {23}},\ \bibinfo {pages} {073006} (\bibinfo {year}
  {2021})}\BibitemShut {NoStop}%
\bibitem [{\citenamefont {Li}\ \emph {et~al.}(2020)\citenamefont {Li},
  \citenamefont {Li},\ and\ \citenamefont {Todorovic}}]{Li2020}%
  \BibitemOpen
  \bibfield  {author} {\bibinfo {author} {\bibfnamefont {J.}~\bibnamefont
  {Li}}, \bibinfo {author} {\bibfnamefont {F.}~\bibnamefont {Li}},\ and\
  \bibinfo {author} {\bibfnamefont {S.}~\bibnamefont {Todorovic}},\ }\href@noop
  {} {\bibinfo {title} {Efficient riemannian optimization on the stiefel
  manifold via the cayley transform}} (\bibinfo {year} {2020}),\ \Eprint
  {https://arxiv.org/abs/2002.01113} {arXiv:2002.01113} \BibitemShut {NoStop}%
\bibitem [{\citenamefont {Strang}(2016)}]{strang2016}%
  \BibitemOpen
  \bibfield  {author} {\bibinfo {author} {\bibfnamefont {G.}~\bibnamefont
  {Strang}},\ }\href@noop {} {\emph {\bibinfo {title} {Introduction to Linear
  Algebra}}}\ (\bibinfo  {publisher} {Wellesley},\ \bibinfo {year}
  {2016})\BibitemShut {NoStop}%
\end{thebibliography}%
\clearpage
\newpage

\onecolumngrid
\appendix
\renewcommand{\thesection}{\Alph{section}}
\renewcommand{\thesubsection}{\Roman{subsection}}
\pagestyle{plain}
\setcounter{page}{1}

\begin{center}
{\Large\textbf{Appendix}}
 \end{center}
First, we present the parameters employed for our use case of the linearized Euler equations with a periodic source in \autoref{subsec:Parameter}. 
\change{Next, we explain the construction of the MPOs from the differential operators and providing especially details on the differential operators and the sponge MPO, which facilitates non-reflective boundary conditions in \autoref{subsec:MPO_representation}.}
In \autoref{subsec:MPO-to-UNITARIES}, we explain how we compile the necessary quantum operations from the MPOs.
\change{Next, we show the performance of the algorithm for large qubit numbers for the advection-diffusion equation in \autoref{sec:app-AdvDiff}. In  \autoref{subsec:scaling_succ_prob} we show the scaling of the success probability for different qubit numbers. Next, in  \autoref{subsec:cost-landscape} we analyse the trainability and the sensitivity of the loss landscape to shot noise.}
Next, the computation of the normalization constant $f_{\hat{O},j}$, the derivation of the angle $\varphi$ used in the adapted Hadamard test, as well as the derivation of the convergence measure, i.e. the fidelity, in \autoref{subsec:adaptedH}. 
\change{Next, we present in \autoref{subsec:Euler} the cost functions used for Euler time stepping of the Burgers' and the advection-diffusion equation.}
Finally, we present the cost functions and quantum circuits obtained using the 4th order Runge-Kutta  time stepping scheme in \autoref{subsec:Cost}. 
We use the same notation and definitions as in the main text.

\section{System and Training Details}\label{subsec:Parameter}
Here, we shortly outline the specific parameters that describe the system as well as the choose circuit sizes and trainining details.
\paragraph{Euler equation}
The specific parameters used in our example of the linear Euler equation with a periodic point source are density $\bar{\rho}=1.225~\frac{\text{kg}}{\text{m}^3}$,  the frequency and amplitude of the point source $\omega=100~\text{Hz}$ and $A_0=0.4c$ , and the sound of speed $c=340.2~\frac{\text{m}}{\text{s}}$.
We study a spatial domain of size $x \in [-4,4]$.
The ansatz is encoded into $6$ qubits, corresponding to a discretization of the domain into $N_x=2^6=64$ data points.
This domain includes the inner zone $x_{\text{inner}} \in [-2,2]$ of unperturbed spatial evolution as well as the outer zones $x_{\text{outer}} \in [-4,-2]$ and $\in [2,4]$, where the sponge damps the signal to implement non-reflective boundary conditions. 
We discretize the space with first order finite differences and a use a 4th order Runga Kutta time stepping scheme with a stepsize of $dt=2.5\cdot10^{-4}$~s. The corresponding cost function are detailed in  \autoref{subsec:Cost}. We use the expression of the bounded sponge operator explained in \autoref{subsec:MPO_representation}, with $\kappa=0.13$, $\Tilde{n}=4$, and $\gamma_{max}=1500$.

We perform the optimization using an Adam Optimizer, followed by additional training epochs with a Limited-memory Broyden–Fletcher–Goldfarb–Shanno algorithm (LBFGS). 
For the simulation of the results shown in \autoref{fig:circuit-schemes} (b) and (c), we used a brickwall ansatz with 14 layers. We trained it using a learning rate of $\text{lr}_{\text{Adam}}=0.05$ and $\text{lr}_{\text{LBFGS}}=0.5$ and a number of epochs of $\text{n}_{\text{epochs,Adam}}=751$ and $\text{n}_{\text{epochs,LBFGS}}=75$.
All runs are performed using the quantum computing software framework PennyLane \cite{Pennylane} together with pyTorch for the parameter optimization \cite{Paszke2019}.

\paragraph{Burgers' equation}
\change{For the Burgers' equation we consider a spatial domain of size $x \in [0,2\pi]$ and discretize it on a uniform grid with $N_x=2^6=64$  lattice points with lattice spacing $dx=2\pi/N_x$. We choose $\nu=0.001$ and $dt=0.5dx$, such that a shock evolves within the first $60$ Euler time steps. The choice of such a large $dt$ leads to a small time evolution error, but which is smaller than the introduced grid error. For the simulation of the results, we use a circuit with $10$ layers, a learning rate of $\rm{lr}=0.005$
 and $1000$ training epochs using the Adam optimizer. All runs are performed using the quantum computing software framework PennyLane \cite{Pennylane} together with Jax \cite{jax2018github} for the parameter optimization.
 The weights that encode the initial conditions are trained prior to the the application of the tensor-programmable quantum scheme by miimizing the relative error $\epsilon_u(t=0)\approx 1.5\times10^{-6}$. In principle, it is also possible to start from a delta peak, and evolve this with the diffusion-equation until the correct gauss width is reached.}

\section{\change{Matrix Product Operator Representation of Differential Operators}}\label{subsec:MPO_representation}

\change{We introduce the Matrix-Product Operator (MPO) of matrix $O$ in its generic form \cite{Schollw_ck_2011}:
\begin{align}
   \mathcal{O}  = \sum_{\boldsymbol{\zeta},\boldsymbol{\sigma},\boldsymbol{\sigma}'}
    O[1]_{\zeta_0, \zeta_1}^{\sigma_1,\sigma_1'}
    \ldots O[n]_{\zeta_{n-1}, \zeta_n}^{\sigma_n,\sigma_n'}
      \ket{\boldsymbol{\sigma}}
      \bra{\boldsymbol{\sigma}'},
\end{align}
where $\boldsymbol{\zeta}$ ($\boldsymbol{\sigma}$) denotes the virtual (physical) indices, respectively, and $O[\cdot]$ the MPO cores, which are 4-dimensional tensors. We assume $\zeta_0=\zeta_n=1$ and define the maximal bond dimension $\zeta_{\mathcal{O}}=\text{max}(\text{dim}(\zeta_j))$ with $j=0,\dots,n$. Whenever \(O\) admits a sum of simple tensor products, it can be written in a factorized “matrix-of-operators” form
\(O = A[1]\Join \cdots \Join A[n]\), where each block entry of \(A[j]\) is a local \(2\times 2\) operator \(B_{\sigma_j,\sigma_j'}\). The symbol \(\Join\) indicates that block multiplication is a tensor product on the physical legs and an ordinary matrix product on the virtual legs. We use the elementary \(2\times 2\) projectors \(I_1, I_2\) and the shift \(J\) and \(J^T\) as a convenient local operator basis:
\begin{align}
    I_1 = \begin{pmatrix}
        1 & 0 \\
        0 & 0 \\
    \end{pmatrix} \quad I_2 = \begin{pmatrix}
        0 & 0 \\
        0 & 1 \\
    \end{pmatrix} \quad J = \begin{pmatrix}
        0 & 1 \\
        0 & 0 \\
    \end{pmatrix} \quad 
    J^T = \begin{pmatrix}
        0 & 0 \\
        1 & 0 \\
    \end{pmatrix}.
\end{align}
These generators yield compact MPOs for common gates and for banded matrices, while keeping the MPO bond dimension small.
For two qubits, \(\mathrm{CNOT}\) decomposes into a sum of two product terms and thus admits an MPO with bond \(\zeta_{\mathcal O}=2\). The first core selects the control state via \((I_1,I_2)\), the second core applies either \(I\) or \(\sigma_x\) on the target conditioned on the virtual index, now written as
\begin{align}
    \mathrm{CNOT} = \begin{pmatrix}
        1 & 0 & 0 & 0 \\
        0 & 1 & 0 & 0 \\
        0 & 0 & 0 & 1 \\
        0 & 0 & 1 & 0 \\
    \end{pmatrix}= \underset{\zeta_0}{\left\lbrace
  \vphantom{(I_1,\; I_2)}
\right.}
\smash[t]{\overbrace{(I_1,\; I_2)}^{\zeta_1}} 
 \Join
 \underset{\zeta_1}{\left\lbrace
  \vphantom{\begin{pmatrix} I \\ \sigma_x \end{pmatrix}}
\right.}
\smash[t]{\overbrace{\begin{pmatrix} I \\ \sigma_x \end{pmatrix}}^{\zeta_2}}
    =  I_1  \otimes I + I_2 \otimes \sigma_x,
\end{align}
with Pauli matrix $\sigma_x$. Similarly, tridiagonal Toeplitz matrices \(\mathrm{TriDiag}(\alpha,\beta,\gamma)\) have a standard three–block MPO with constant bond \(\zeta_{\mathcal O}=3\) \cite{Kazeev2012}
\begin{align}
    \text{TriDiag}(\alpha, \beta, \gamma)=
\begin{pmatrix}
\alpha & \beta  &        &        &        \\
\gamma & \alpha & \beta  &        &        \\
       & \gamma & \alpha & \beta  &        \\
       &        & \ddots & \ddots & \ddots \\
       &        &        & \gamma & \alpha
\end{pmatrix} = (\alpha I+\beta J + \gamma J^T, \gamma J, \beta J^T) \Join \begin{pmatrix}
    I & 0 & 0\\
    J^T & J & 0 \\
    J & 0 & J^T
\end{pmatrix}^{n-2} \Join \begin{pmatrix}
    I \\
    J^T \\
    J \\
\end{pmatrix}.
\end{align}
In this MPO form, standard finite-difference operators with at most nearest–neighbor coupling admit compact representations. For instance, the central, second–order accurate first–derivative stencil on a uniform grid with spacing \(\Delta x\) corresponds to the tridiagonal Toeplitz choice
\[
\alpha=0,\qquad
\beta=\frac{1}{2\Delta x},\qquad
\gamma=-\frac{1}{2\Delta x},
\]
i.e.\ \(\mathrm{TriDiag}\bigl(\alpha=0,\beta=\tfrac{1}{2\Delta x},\gamma = -\tfrac{1}{2\Delta x}\bigr)\).
The block structure above directly encodes operators under Dirichlet boundary conditions. If different boundary conditions are required (e.g.~ periodic or Neumann) the needed corrections can be added as low–rank modifications to the base \(\mathrm{TriDiag}(\alpha,\beta,\gamma)\). Each added nonzero matrix element (such as a periodic link between the first and last grid point) increases the MPO bond dimension by at most one, so boundary corrections preserve a small bond dimension \cite{Kiffner2023}.
\newline
\newline
The respective decomposition for a bounded 1D sponge operator, with maximal factor of 1, are given by
\begin{align}
A[j] &= \frac{1}{e^{(2^{\tilde{n}}-1) \kappa}-1}(
I_2, \ I_1 \ I_2, \ I_1) \text{ for } j = 1 \\
A[j] &=\begin{pmatrix}
I_2 & & & \\
& I_1 & & \\
&  & I_2 &  \\
& & & I_1
\end{pmatrix}
2 \leq j \leq n -\tilde{n} \\
A[j] &= \begin{pmatrix}
J_1(j) & & & \\
& J_1^{\mathfrak{t}}(j) & & \\
&  & I &  \\
& & & I
\end{pmatrix}, \text{ for } n -\tilde{n} < j \leq n -1 \\
A[j] &= \begin{pmatrix}
J_1(j)\\
J_1^{\mathfrak{t}}(j)\\
I\\
I
\end{pmatrix}, \text{ for } j=n
\end{align}
where $\mathfrak{t}$ refers to a mirroring with respect to the anti-diagonal.
}

\section{Matrix Product Operators to Quantum Gates}\label{subsec:MPO-to-UNITARIES}

\begin{figure}
    \centering
    \includegraphics{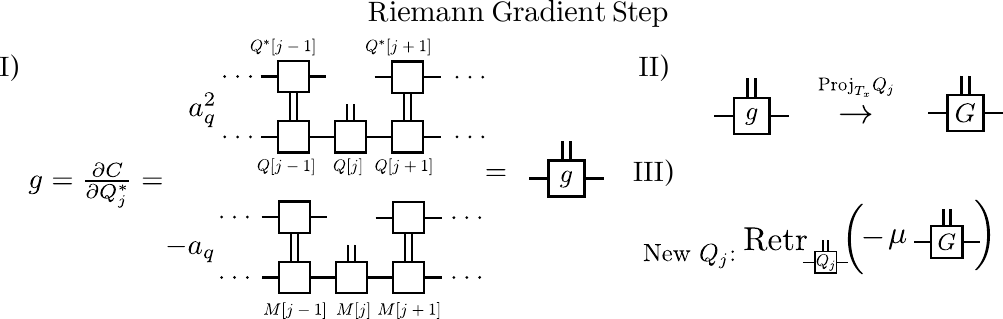}
    \caption{
    Sketch of the execution of a single Riemannian gradient step on the tensors of a unitary MPO $\mathcal{Q}$ that approximates a target MPO $\mathcal{M}$. It can be divided into three sub-steps: I) for each core $Q_j$ the gradient is computed by deriving the cost function $C$ with respect to its complex conjugate $Q_j^*$, we denote the result by $g$, II) the gradient $g$ is projected onto the tangent space of $Q_j$ via $g - \frac{1}{2} Q_j (Q_j^T g + g^T Q_j) \vcentcolon= G $ \cite{Luchnikov2021}, where we have defined the Riemannian gradient $G$, III) the new $Q_j$ is found by a retraction antiparallel to the Riemannian gradient, where the magnitude of the update step is controlled by the learning rate $\mu$. Using the QR decomposition as a retraction map, this last step has the form $\text{Retr}_{Q_j}^{QR}(- \mu G) = QR \left(Q_j - \mu G \right)$. We note that for the core $Q_1$, which is not in the Stiefel manifold, special measures must be taken. After step (I), $Q_j$ and the gradient $g$ must be transposed, then (II) and (III) are carried out, and the transposed result then gives the new $Q_j$.
}
    \label{fig:RiemannianGD}
\end{figure}
The algorithm for determining quantum gates that prepare an arbitrary MPS is well known \cite{Ran2020,Smith2024, Malz2024}. This approach yields an exact encoding and provides an upper bound on the circuit depth for generating a certain amount of entanglement \cite{Lubasch2020}. Recently, also the translation of MPOs into quantum gates has been reported \cite{Nibbi2024, Termanova2024}. 
The latter work by Termanova et al. will be outlined in the following. Its reduced requirements in qubit numbers made it a promising candidate for integration into the VQA framework.
Let us start by introducing the MPO as \cite{Schollw_ck_2011}
\begin{align}
   \mathcal{O}  = 
    \sum_{\boldsymbol{\zeta},\boldsymbol{\sigma},\boldsymbol{\sigma}'}
    O[1]_{\zeta_0, \zeta_1}^{\sigma_1,\sigma_1'}
    \ldots O[n]_{\zeta_{n-1}, \zeta_n}^{\sigma_n,\sigma_n'}
      \ket{\boldsymbol{\sigma}}
      \bra{\boldsymbol{\sigma}'},
\end{align}
where $\boldsymbol{\zeta}$ ($\boldsymbol{\sigma}$) denotes the virtual (physical) indices, respectively, and $O[\cdot]$ the MPO cores, which are 4-dimensional tensors. We assume $\zeta_0=\zeta_n=1$ and define the maximal bond dimension $\zeta_{\mathcal{O}}=\text{max}(\text{dim}(\zeta_j))$ with $j=0,\dots,n$.

We introduce the MPO $\mathcal{Q}$, which shall approximate target MPO $\mathcal{M}$ while satisfying isometric constraints. To account for the limitations on dimensionality and degrees of freedom imposed by these constraints, we expand the search for $\mathcal{Q}$ to encompass a larger Hilbert space. This is done implicitly by setting its bond dimension $Z_{\mathcal{Q}}=2^{\ell}$ where $\ell$ is a positive integer and $Z_{\mathcal{Q}}>\zeta_{\mathcal{M}}$.

Following this, the search procedure is then formulated as a constrained optimization problem, which reads as \cite{Termanova2024}
\begin{align}
    C=\ &\underset{c_{\text{MPO}},\hat{\mathcal{Q}}}{\text{min}}
    \  \lVert c_{\text{MPO}} \mathcal{Q} -\mathcal{M} \rVert^2 \label{eq:minimization_mpo}\\
    &\text{subject to} \ \ Q[1]^{\dagger}  \in \text{St}(r, s) \notag\\  & \hspace{51pt}Q[j] \hspace{1pt} \ \in \text{St}(r, s)  \quad \forall j = 2, \ldots, n, \notag
\end{align}
where $c_{\text{MPO}}$ is a normalization constant, $\lVert \cdot \rVert$ the Frobenius norm and $\text{St}(r, s)$ the Stiefel manifold, which is the set of all $r \times s$ matrices with orthonormal columns, where $r\geq s$ \cite{Li2020}. Here, we introduced $Q[j]:=Q[j]_{(Z_\mathcal{Q}, \sigma_j), (\sigma_j', Z_\mathcal{Q})}$ as the reshaped, isometric cores for $1<j<n$, and $Q[1]:=Q[1]_{\sigma_j, (\sigma_j', Z_\mathcal{Q})}$ and $Q[n]:=Q[n]_{(Z_\mathcal{Q}, \sigma_j), \sigma_j'}$ respectively. The normalization constant $c_{\text{MPO}}$ can be determined via \cite{Termanova2024} 
\begin{align}
    c_{\text{MPO}} = \text{Re} \frac{\text{tr}\left[\mathcal{Q}^\dagger  \mathcal{M}  \right]}{\lVert \mathcal{M} \rVert^2}.
\end{align}
With all of this in place, we briefly outline the constraint minimization of \autoref{eq:minimization_mpo}, graphically depicted in \autoref{fig:RiemannianGD}: (i) Initialize the isometric cores of $\mathcal{Q}$ (ii) Compute normalization $c_{\text{MPO}}$ (iii) Perform a single Riemannian gradient step on all tensors $Q_j$ \cite{Li2020} (iv) Repeat (ii) and (iii) in an alternating manner until the error measure $\epsilon = \frac{\norm{c_{\text{MPO}}\mathcal{Q}-\mathcal{M}}^2}{\norm{\mathcal{M}}^2} $ reaches the set tolerance.
In step (iii) the gradient $g=\frac{\partial C}{\partial{Q[j]^*}}$  is projected  onto the tangent space of the core, resulting in $G$. A retraction is then performed in this direction, scaled by the learning rate, i.e., $-\mu G$. This retraction can be performed, for example, by a QR decomposition or a Cayley transformation \cite{Li2020, Luchnikov2021}. 

As soon as the algorithm has reached the set convergence criterion, the boundary isometric cores $Q[1]$ and $Q[n]$ with the shapes $2 \times (2 Z_\mathcal{Q})$ and $ (2 Z_\mathcal{Q}) \times 2$, respectively, need to be raised to unitaries. To do so, the remaining columns (rows) are filled using Gram-Schmidt orthonormalization procedure \cite{strang2016}, respectively. This results in the target matrix only being applied probabilistically, as the padding also enables a trajectory within the nullspace.

\section{Advection-Diffusion equation} \label{sec:app-AdvDiff}

\change{
Next to the examples given in the main text, we implemented the advection-diffusion equation 
\begin{equation}
    \frac{\partial\phi}{\partial t} = \nu\Delta\phi -c\nabla\phi
\end{equation}
with $\nu=0.1$, $c=20$ and periodic boundary conditions using the proposed quantum-tensor scheme. Its simple structure and reduced qubit requirements compared to the non-linear use case makes it a suitable candidate for larger-scale simulations and the analysis of the cost function given in \autoref{subsec:cost-landscape}.
We consider the same initial condition as for the Burgers' equation, a Gauss peak with $\sigma=0.5$ resolved with $N_x=2^{10}=1024$ lattice points and a lattice spacing of $dx=2\pi/N_x$. 
Figure~\ref{fig:AdvDiff} depicts the field evolution and the relative error over time, showing a good agreement with the classical solution. 
\begin{figure}
    \centering
    \includegraphics[width=0.8\linewidth]{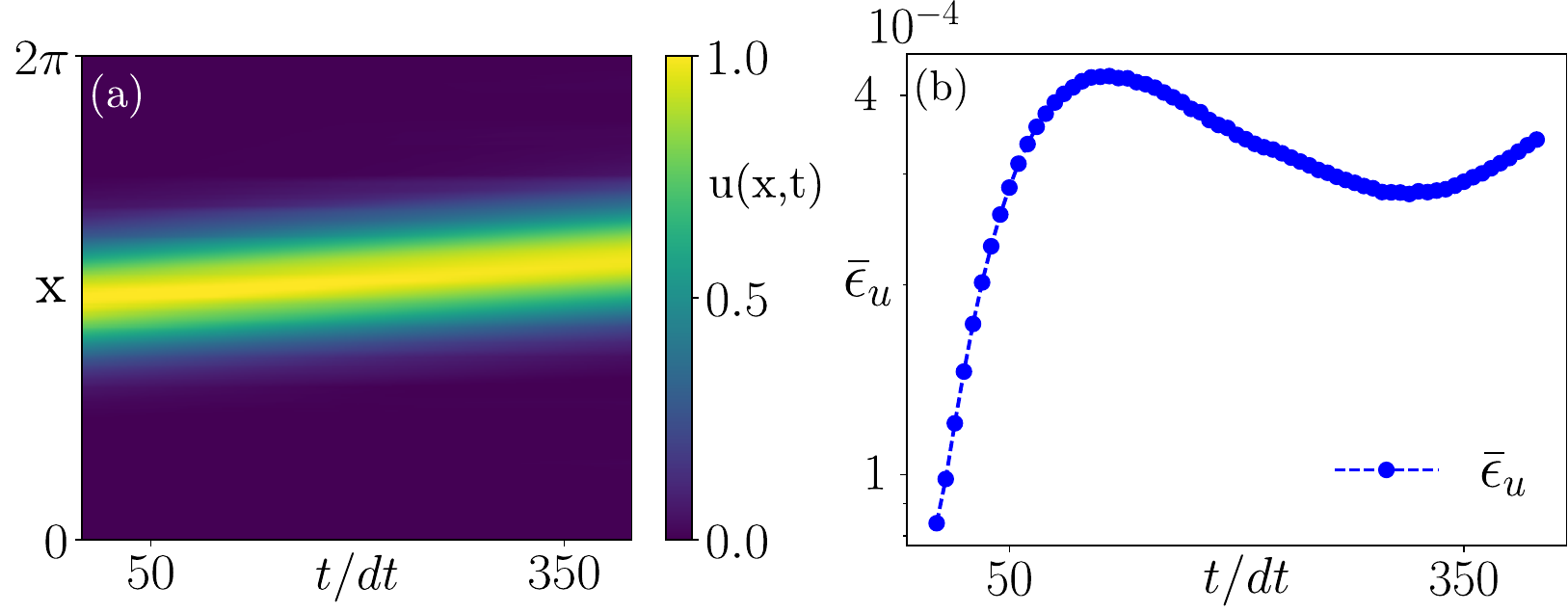}
    \caption{\change{Time evolution of the field according to the advection-diffusion equation. 
    (a) Evolution of the initial Gauss peak $u(x,t=0)$ over 400 evolution steps and 
    (b) the relative error $\bar \epsilon_u$ over time.}}
    \label{fig:AdvDiff}
\end{figure}
}

\section{\change{Scaling of the success probability with system size}}\label{subsec:scaling_succ_prob}

\change{The value of the success probability $\alpha_{\textrm{succ}}$ plays a crucial role in the determination of the norm correction $f_{\hat{O},j}$. As it depends on both the operator and the field it acts on, it is a use-case dependent quantity. Here, we look at the average success probability of all differential operators implemented for this work, following the procedure introduced by Termanova et al. in \cite{Termanova2024}.
They recognize, that the average success probability $\bar{\alpha}_{\rm{succ}}$ of a matrix $M$ is directly connected to its frobenius norm as 
\begin{equation}
    \bar{\alpha}_{\rm{succ}} = \frac{\norm{A}}{2^n},
\end{equation}
where $A = \frac{1}{c_{\rm{MPO}}}M$ and $c_{\rm{MPO}}$ is the multiplicative factor from the mapping of the MPO to a Unitary. }
\change{
To study the scaling of the success probability over the system size, we consider the optimal $c$, being the leading singular value of $M$. Then, we find constant or converging average success probabilities for all differential operators used within this work as shown in \autoref{fig:app_probs}~(a).}  

\change{Next, we are interested in the scaling of the success probability of the non-linear multiplication implemented by the $\mathrm{CNOT}$ gates.} \change{While for random vectors, the point-wise multiplication might result in an exponential decay of the success probability with system size, for the discussed use-case of the Burgers' equation the situation is more subtle. 
Figure~\ref{fig:app_probs}~(b) shows $\alpha_{\rm{succ}}$ over the system size for different widths of the gauss peak. We observe, that its value remains constant until the peak is resolved, and only then decays with system size. This is a behavior also observed by Lubasch et al. \cite{Lubasch2020} for their example of the non-linear Schrödinger equation. The implications of that get clearer in \autoref{fig:app_probs}~(c) where, we consider the success probability over system size, considering different gauss widths $\sigma$ and choosing the number of required qubits by restricting the grid-error below a threshold: considering a constant grid error, the quantum circuit is capable to resolve increasingly  narrow peaks with an increasingly better resolution without gaining any loss in the success probability.
}
\change{Furthermore, we expect techniques as phase estimation and amplitude amplification to mitigate the challenges that arise for use cases that result in $\alpha_{\rm{succ}}$ and will investigate their potential in future works.}


\begin{figure}
    \centering
    \includegraphics[width=\linewidth]{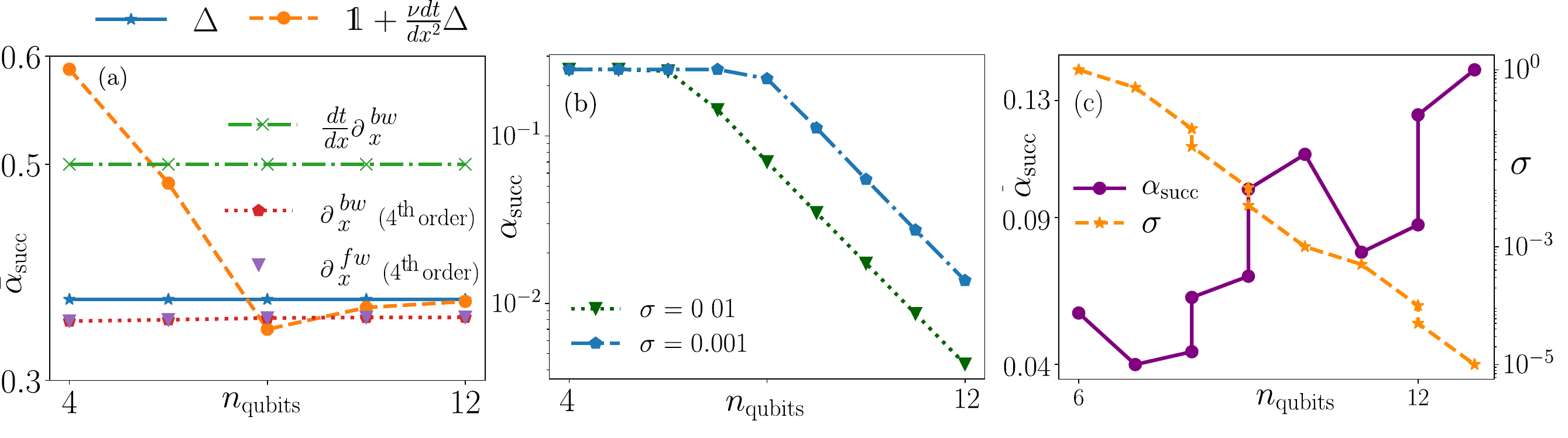}
    \caption{\change{(a) Scaling of the maximal average success probability $\bar{\alpha}_{\textrm{succ}}$ for all differential operators used throughout this work. The success probability of the sponge operator highly depends on its width compared to the system size. (b) Scaling of the success probability $\alpha_{\textrm{succ}}$ the point-wise multiplication $u\nabla u$, as present in the Burgers' equation for a Gauss peak of width $\sigma$.
    (c) Scaling of  $\alpha_{\textrm{succ}}$ when considering different Gauss peak widths and choosing the number of qubits according to a maximal allowed root-mean-square-error $<0.001$ between the solution and the next better resolution.}  }
    \label{fig:app_probs}
\end{figure}

\section{\change{Analysis of the Trainability}} \label{subsec:cost-landscape}
\change{In this section we study the shape of the loss landscape of the training parameters and its sensitivity to shot noise.
We consider the first time step of the advection-diffusion equation, as it allows us to go to larger qubit number as the non-linear use case with significantly reduced resource requirements. Please note, that we observe qualitatively the same cost landscapes for the other use cases.
We where mostly concerned with the following questions: First, considering initializations of the weights from the previous time step, what is the shape of the loss landscape for the single parameters. Second,  how is this loss landscape affected by noise? Third, how does this loss landscape change with system size. The results in \autoref{fig:ap-loss} show clearly that weight-reinitialization leads to a sine-like loss landscape for each parameter, which roughens in the presence of shots. Importantly the impact of shots on the loss landscape seems to remain constant over the system size. Furthermore, the magnitude of the sine-like loss function remains large for increasing system sizes, ensuring well trainability even for larger scale problems.
We tested the trainability of the next time-step for the advection-diffusion equation with shots using the exact gradient computed with parameter shift rule or a gradient approximation obtained from the "simultaneous perturbation stochastic approximation" (SPSA) algorithm \cite{Spall1998,Kandala2017} combined with an Adam optimization. With both strategies, we reach the best possible resolved loss for different shot numbers.}
\begin{figure}[bt] 
    \centering
    \includegraphics[width=\linewidth]{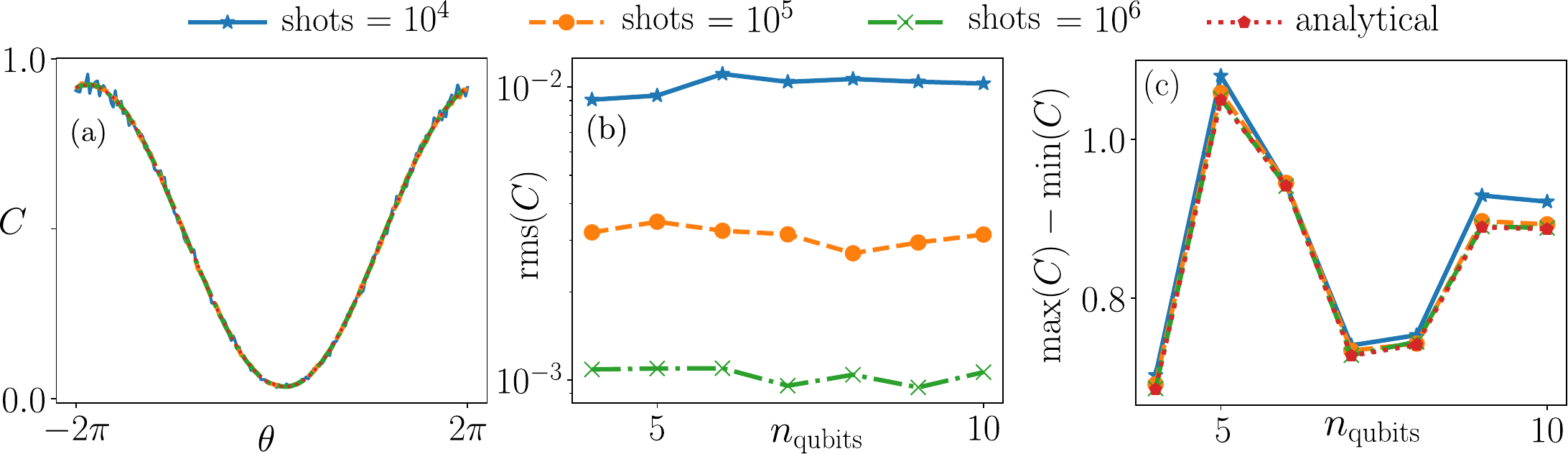}
    \caption{
    \change{Analysis of the cost landscape $C(\theta^i_{j+1}) = 1-\braket{\sigma_z}_{\textrm{anc}}$ of one representative training parameter $i$ for different shot numbers and the shot-free analytical solution.  
    (a) Cost landscape of a gaussian peak with $\sigma=0.5$ evolved according to the advection-diffusion equation and resolved with $10$ qubits.  We note that, the biggest deviations in the cost landscape for small shot numbers appear far from the optimum.
    (b) Root-mean-square-error between the shot-based  and the analytical cost-landscape over the system size.
    (c) Difference between the maximum and the minimum value in the cost landscape over the system size, showing that the cost landscape remains pronounced when increasing the system size.
    All quantities where computed with $dx=2\pi/N_x$ and a time step of $0.1\nu dx$, such that a stable simulation is possible.
    All cost landscapes are plotted using $\varphi=0.6126$.}
    }
    \label{fig:ap-loss}
\end{figure}

\section{Deriving the norm constant $f_{\hat{O},j}$, the optimal angle $\varphi_{\textrm{opt}}$ and the convergence measure $\mathcal{F}$}\label{subsec:adaptedH}

In the following we present the calculation of the norm constant $f_{\hat{O},j}$ for the standard and adapted Hadamard test. Furthermore we determine the optimal rotation angle $\varphi$
\change{and derive the convergence measure accessible via the adapted Hadamard test}

As explained in the main text, applying the operator $\hat{O}$ with help of the unitaries $\hat{U}_{\text{MPO}}$  on the quantum computer, implements the correct operation up to the factor $f_{\hat{O},j}c_{\text{MPO}}$
\begin{equation}
\begin{aligned}
    \braket{\hat{O}}&=c_{\textit{MPO}}\cdot \mathrm{Re}\bra{\Psi}P_{\ket{0}_{\mathrm{aux}}\bra{0}_{\mathrm{aux}}}\hat{U}_{Q}\ket{\Psi}
    =c_{\textit{MPO}}\cdot f_{\hat{O},j} \cdot \braket{\sigma_z}_\text{anc}.\\
    \label{eq:calculateO}
\end{aligned}
\end{equation}
Here, the operator $\hat{U}_Q$ summarizes all controlled unitaries, $\ket{0}_{\text{anc}}\ket{\Psi}$ is the initial state. 
\change{For linear differential equations}, we have $\hat{U}_Q=\hat{U}(\boldsymbol{\theta}^{\dagger}_{j+1})\hat{U}_\textit{MPO}\hat{U}(\boldsymbol{\theta}_{j})$ and $\ket{0}_{\text{anc}}\ket{\Psi}=\ket{0}_{\text{anc}}\ket{\mathbf{0}}$. 
While $c_{\text{MPO}}$ is known from the algorithm which determines the unitaries from the initial MPO, $f_{\hat{O},j}$ needs to be determined from the success probability.

The quantum circuit of the adapted Hadamard test produces prior to measurement the state
\change{
\begin{equation}\label{eq:full_state}
\begin{aligned}
    & \hat{R}_Y(-\varphi)P_{\ket{0}_{\mathrm{aux}}\bra{0}_{\mathrm{aux}}} \hat{U}_{Q}\hat{H}\ket{1}_{\text{anc}}\ket{\mathbf{0}}\\
    &=\frac{1}{\sqrt{2}}\frac{1}{1+\alpha_{\text{succ}}}P_{\ket{0}_{\mathrm{aux}}\bra{0}_{\mathrm{aux}}}
    \left(\ket{0}_{\text{anc}}(-\cos(\frac{\varphi}{2})\ket{\mathbf{0}}+\sin(\frac{\varphi}{2}) \hat{U}_{Q}\ket{\mathbf{0}})+\ket{1}_{\text{anc}}(\sin(\frac{\varphi}{2})\ket{\mathbf{0}}+\cos(\frac{\varphi}{2}) \hat{U}_{Q}\ket{\mathbf{0}})\right),
\end{aligned}
\end{equation}
}
\change{
For optimally trained $\boldsymbol{\theta}_{j+1}$
measuring the global ancilla qubit yields the expectation value 
\begin{equation}\label{eq:exp-val}
\begin{aligned}
\braket{\sigma_\text{z}}_{\text{anc}}&=\frac{2\sin(\varphi)-(\sqrt{\alpha_{\text{succ}}}-\frac{1}{\sqrt{\alpha_{\text{succ}}}})\cos(\varphi)}{1+\alpha_{\text{succ}}}\mathrm{Re}\bra{\mathbf{0}}\hat{U}_{Q}\ket{\mathbf{0}}
=\frac{1}{f_{\hat{O},j}}\mathrm{Re}\bra{\mathbf{0}}\hat{U}_{Q}\ket{\mathbf{0}}.
\end{aligned}
\end{equation}
}
This yields 
\change{
\begin{equation}
f_{\hat{O},j} =\frac{1+\alpha_{\text{succ}}}{2\sin(\varphi)-(\sqrt{\alpha_{\text{succ}}}-\frac{1}{\sqrt{\alpha_{\text{succ}}}})\cos(\varphi)} 
\end{equation}
}
for the norm constant $f_{\hat{O},j}$ of the adapted Hadamard test with the special case of the standard Hadamard test ($\varphi=\pi/2$)
\begin{equation}
    f_{\hat{O},j}(\varphi=\pi/2)=\frac{1+\alpha_{\text{succ}}}{2}.
\end{equation}

With this norm constant $f_{\hat{O},j}$, which depends on the success probability $\alpha_{\text{succ}}$ and the rotation angle $\varphi$ of the adapted Hadamard test, $\braket{\hat{O}}$ can then be calculated according to \autoref{eq:calculateO}.

\change{
If all operators of a differential equation are summarizes within $\hat{O}$ and assuming optimally trained $\boldsymbol{\theta}_{j+1}$, there is an optimal angle $\varphi_{\textrm{opt}}=2\arctan(\sqrt{\alpha_{\text{succ}}})$ that leads to $\braket{\sigma_z}_{\text{anc}}=1$ and $f_{\hat{O},j}=\sqrt{\alpha}$.
}

\change{
Furthermore, we can derive a measure of convergence  from \autoref{eq:full_state}, that allows to compute the fidelity $\mathcal{F}=||\braket{\mathbf{0}|P_{\ket{0}_{\mathrm{aux}}\bra{0}_{\mathrm{aux}}}\hat{U}_Q\|\mathbf{0}}||^2$ between the normalized trained solution and the real solution from $\braket{\sigma_\text{z}}_{\text{anc}}$. For simplicity, we assume optimal $\varphi = \varphi_{\textrm{opt}}$}. \change{For general $\boldsymbol{\theta}_{j+1}$, $\norm{\bra{\mathbf{0}}\hat{U}_{Q}\ket{\mathbf{0}}}^2=\alpha_{\textrm{succ}}\mathcal{F}$ and the global expectation value can be computed from \autoref{eq:full_state}, reading
\begin{equation}
\braket{\sigma_z}_{\textrm{anc}} = \frac{2\sqrt{\alpha_{\text{succ}}\mathcal{F}}\sin(\varphi_{\textrm{opt}})-(\alpha_{\text{succ}}-1)\cos(\varphi_{\textrm{opt}})}{1+\alpha_{\text{succ}}},
\end{equation}
which allows us to infer the fidelity of the trained solution during the training as
\begin{equation}
    \mathcal{F} = \frac{(\alpha_{\text{succ}}\braket{\sigma_z}_{\textrm{anc}}+\alpha_{\text{succ}}\cos(\varphi_{\textrm{opt}})+\braket{\sigma_z}_{\textrm{anc}}-\cos(\varphi_{\textrm{opt}}))^2}{4\alpha_{\text{succ}}\sin^2(\varphi_{\textrm{opt}})}.
\end{equation}
}

\section{\change{Cost functions}}\label{sec:cost_functions}
\change{In this section we introduce the utilized cost functions for the euler time stepping and the 4th order Runge Kutta scheme.}
\subsection{\change{Euler Time Stepping}}\label{subsec:Euler}
\change{To simulate the time evolution of the Burgers' equation and the advection-diffusion equation, we employ the explicit Euler time stepping and summarize both PDEs within one quantum circuit.
Hence, the equation we need to solve can be summarized as
\begin{equation}
\left(\mathbf{1}+dt\frac{\partial}{\partial t}\right)\phi(x,t)=\hat{O}\phi(x,t).
\end{equation}
When the solution at time step $t_j=n\cdot dt$ is known, $\phi(t_{j+1})$ is computed according to
\begin{equation}
\begin{aligned}
    & \phi(x, t_{j+1}) = (\mathbf{1} +dt\frac{\partial}{\partial t})\phi(x,t_j)=\hat{O}_{\textrm{full}}\phi(x,t_j) ,\\
    & t_{j+1} = t_j + dt.
    \end{aligned}
\end{equation}
}

\change{
This results in a cost function 
\begin{equation}
\begin{aligned}
    C(\theta_{j+1}) 
    &= \norm{\ket{\phi_{t_{j+1}}} -\hat{O}_{\textrm{full}}\ket{\phi_{t_{j}}}}^2\\
    &\propto -\Re\bra{\phi_{t_{j+1}}}\hat{O}_{\textrm{full}}\ket{\phi_{t_j}} + \textrm{constant}\\
\end{aligned}
\end{equation}
}
\subsection{4th Order Runge Kutta Time Stepping}\label{subsec:Cost}
To perform the time evolution of the linearized Eulers equation shown in the main text, we implemented a 4th order Runge Kutta (RK4) scheme. Consider a differential equation of form 
\begin{equation}
\frac{\partial\phi(x,t)}{\partial t} =g(t, \phi(x,t_j)).
\end{equation}
When the solution at time step $t_j=n\cdot dt$ is known, $\phi(t_{j+1})$ is computed according to
\begin{equation}
\begin{aligned}
& \phi(x, t_{j+1}) = \phi(x,t_j) + \frac{dt}{6}(k_1 +2k_2 +2k_3 +k_4)\\
& t_{j+1} = t_j + dt,
\end{aligned}
\end{equation}
where
\begin{equation} \label{eq:rk_classic}
\begin{aligned}
& k_1 = g(t_j, \phi(x,t_j)) \\
& k_2 = g(t_j + \frac{dt}{2}, \phi(x,t_j) + \frac{dt}{2}k_1)\\
& k_3 = g(t_j + \frac{dt}{2}, \phi(x,t_j) + \frac{dt}{2}k_2)\\
& k_4 = g(t_j + dt, \phi(x,t_j) + dt k_3).
\end{aligned}
\end{equation}
For the quantum solver, we need to translate this procedure into cost functions that are to be minimized.
According to the five equations required to compute the final $\phi(x, t_{j,m+1})$, we define five different optimization steps. The steps do not compute the $k_m$ directly, but instead the sum of $\phi^*_{j,m} = c^{st}_m\phi_{j} + c^{rk}_m k_m$, choosing $c^{st}_m$ and $c^{rk}_m$ such that $\phi^*$ corresponds to the right hand side of \autoref{eq:rk_classic} for each RK4 step:
\begin{equation} \label{eq:cost_rk4}
\begin{aligned}
    &C^{\phi}_1 =\norm{\ket{\phi_1^*}- \ket{\phi_j} - \frac{1}{2}dt \ket{k_1}}^2, \\
    &C^{\phi}_2 =\norm{\ket{\phi_2^*} -\ket{\phi_j} - \frac{1}{2}dt \ket{k_2}}^2 \\
    &C^{\phi}_3 =\norm{\ket{\phi_3^*} -\ket{\phi_j} - dt \ket{k_3}}^2, \\
    &C^{\phi}_4 =\norm{\ket{\phi_4^*} + \frac{1}{3}\ket{\phi_n} - \frac{1}{6}dt \ket{k_4}}^2, \\
    &C^{\phi}_5 =\norm{\ket{\phi_\text{final}} -\frac{1}{3}\ket{\phi_1^*} - \frac{2}{3}\ket{\phi_2^*} -\frac{2}{3}\ket{\phi_3^*} -\ket{\phi_4^*}}^2 ,
\end{aligned}
\end{equation}
where $\ket{\phi_m^*} = \theta^0_{j,m}\hat{U}(\boldsymbol{\theta}_{j,m})\ket{0}$ and $\ket{\phi_\text{final}}=\theta^0_{j+1}\hat{U}(\boldsymbol{\theta}_{j+1})\ket{0}$. The indices $j$ and $m$ refer to the index of the time step and the Runge Kutta step respectively.
In the considered example, the linear Euler equation, pressure and velocity are coupled. Hence, the right hand side of \autoref{eq:rk_classic} depends of both the pressure and velocity field, i.e., $f(t, \ket{\phi(x,t_j)})\rightarrow f(t, \ket{p(x,t_j)}, \ket{u(x,t_j)})$. In the quantum register the pressure and velocity field are encoded as $\ket{p(x,t_j)} = \theta^{0}_n\hat{P}(\boldsymbol{\theta}_j)\ket{0}$ and $\ket{u(x,t_j)} = \theta^{0}_j\hat{U}(\boldsymbol{\theta}_j)\ket{0}$
The first four cost functions $C_m$ can be constructed from \autoref{eq:Euler_eqs} and follow the scheme
\begin{equation}
\begin{aligned}
C_m^p(\boldsymbol{\theta}_{j, m+1}) = &\norm{\ket{p_{j,m+1}^*}-c^{st}_m\ket{p_{j,m}}+c^{rk}_m\cdot dt\left(\rho\cdot c^2\hat{\nabla} \ket{u_{j,m}}+\hat{\gamma(x)}\ket{p_{j,m}} -A_0\sin(\omega t)\ket{\delta(x)}\right)}^2 \\
=&~(\theta^0_{j,m+1})^2
-2\theta^0_{j,m+1}\theta^0_{j,m}\Re\bra{0}\hat{P}^\dag(\boldsymbol{\theta}_{j,m+1})\hat{P}(\boldsymbol{\theta}_{j,m})\ket{0} \\
&+2dt\rho c^2 f_{\hat{\nabla},j}^{u}\theta_{j,m+1}^0\theta_n^0\Re\bra{0}\hat{P}^\dag(\boldsymbol{\theta}_{j,m+1})\hat{\nabla}_{MPO}\hat{U}(\boldsymbol{\theta}_{j,m})\ket{0} \\
&+ 2 f_{\hat{\gamma},j,m}^{p} \Re\theta^0_{j,m+1}\theta^0_{j,m}\bra{0}\hat{P}^\dag(\boldsymbol{\theta}_{j,m+1})\hat{\gamma}_{MPO}\hat{P}(\boldsymbol{\theta}_{j,m})\ket{0} \\
&-2  A_0\sin(\omega t)\Re\theta^0_{j,m+1}\theta^0_{j,m}\bra{0}\hat{P}^\dag(\boldsymbol{\theta}_{j,m+1})\ket{\delta(x)} \\
&+\mathrm{const.}
\end{aligned}
\end{equation}
and
\begin{equation}
\begin{aligned}
C_{m}^{u}(\boldsymbol{\theta}_{j,m+1}) = &\norm{\ket{u_{j,m+1}}-c^{st}_{m}\ket{u_{j,m}}+c^{rk}_{m}\cdot dt\left(\frac{1}{\rho}\cdot\hat{\nabla} \ket{p_{j,m}}+\hat{\gamma}(x)\ket{u_{j,m}}\right)}^2\\
=&~(\theta^0_{j,m+1})^2
-2\theta^0_{j,m+1}\theta^0_{j,m}\Re\bra{0}\hat{U}^\dag(\boldsymbol{\theta}_{j,m+1})\hat{U}(\boldsymbol{\theta}_{j,m})\ket{0}\\
&+2\frac{dt}{\bar\rho}f_{\hat{\nabla},j,m}^{p}\theta_{j,m+1}^0\theta_{j,m}^0\Re\bra{0}\hat{U}^\dag(\boldsymbol{\theta}_{j,m+1})\hat{\nabla}_{MPO}\hat{P}(\boldsymbol{\theta}_{j,m})\ket{0} \\
&+ 2f_{\hat{\gamma},j,m}^{u}  \theta^0_{j,m+1}\theta^0_{j,m}\Re\bra{0}\hat{U}^\dag(\boldsymbol{\theta}_{j,m+1})\hat{\gamma}_{MPO}\hat{U}(\boldsymbol{\theta}_{j,m})\ket{0}\\
&+\mathrm{const.}
\end{aligned}
\end{equation}
The last cost functions $C_5^{u/p}$ only depend on one field, and hence don't differ from \autoref{eq:cost_rk4}.

\end{document}